%% file: main.tex
\documentclass[sigconf, 10pt, screen]{acmart}
\usepackage{blindtext}

\input{0_packages}
\input{0_config}
\input{0_macros}

\renewcommand\footnotetextcopyrightpermission[1]{} 
\setcopyright{none}

\definecolor{mycitegreen}{RGB}{0,90,0}
\definecolor{myrefblue}{RGB}{90,0,0}
\definecolor{forestgreen}{rgb}{0.13, 0.65, 0.13}

\acmDOI{}

\acmISBN{}

\acmPrice{}

\makeatletter
\def\author@bx@sep{0pc}
\makeatother

\begin{document}

\pagestyle{plain}
\title{Rethinking Machine Learning Collective Communication \hspace{-0.15em}as\hspace{-0.15em} a\hspace{-0.15em} Multi-Commodity \hspace{-0.15em}Flow \hspace{-0.15em}Problem}

\author{Behnaz Arzani}
\authornote{both authors contributed equally to the paper}
\affiliation{\institution{Microsoft Research}}
\email{bearzani@microsoft.com}

\author{Siva Kesava Reddy  Kakarla}
\authornotemark[1]
\affiliation{\institution{Microsoft Research}}
\email{sivakakarla@microsoft.com}

\author{Miguel Castro}
\affiliation{\institution{Microsoft}}
\email{mcastro@microsoft.com}

\author{Srikanth Kandula}
\affiliation{\institution{Microsoft Research}}
\email{Srikanth@microsoft.com}

\author{Saeed Maleki}
\affiliation{\institution{Microsoft Research}}
\email{saemal@microsoft.com}

\author{Luke Marshall}
\affiliation{\institution{Microsoft Research}}
\email{luke.marshall@microsoft.com}

\begin{abstract}
We show communication schedulers' recent work proposed for ML collectives does not scale to the increasing problem sizes that arise from training larger models. These works also often produce suboptimal schedules. We make a connection with similar problems in traffic engineering and propose a new method, {\sysname}, that finds better quality schedules (\eg finishes collectives faster and/or while sending fewer bytes) and does so more quickly on larger topologies. We present results on many different GPU topologies that show substantial improvement over the state-of-the-art.
\end{abstract}

\maketitle

\input{intro}
\input{motivation}
\input{solution}
\input{eval}

\input{relatedwork}
\input{conclusion}

\noindent{\bf This work does not raise any ethical concerns.}

\bibliographystyle{ACM-Reference-Format}
\begin{small}
\bibliography{SIGCOMM23}
\end{small}

\clearpage
\input{appendix}

\end{document}

%% file: 0_packages.tex
\usepackage{amsmath,amsfonts}
\usepackage[T1]{fontenc}
\usepackage[utf8]{inputenc}

\usepackage{upquote}

\usepackage{microtype}
\UseMicrotypeSet[protrusion]{basicmath}

\usepackage{hyperref}
\usepackage{graphicx,grffile}

\usepackage{times} 
\usepackage{tikz}
\usepackage{tikz-network}
\usetikzlibrary{arrows.meta,arrows,backgrounds,patterns,positioning,shadings, decorations.markings, shapes.multipart, shapes, decorations.pathreplacing,calligraphy}
\usepackage{endnotes,epsfig}
\usepackage{float}
\usepackage{multirow}
\usepackage{xspace}
\usepackage{tabularx}
\usepackage{ragged2e}
\usepackage{booktabs}
\usepackage{paralist}
\usepackage[american]{babel}
\usepackage[shortlabels]{enumitem}
\usepackage{courier,color,wrapfig}
\usepackage{xspace}
\usepackage{balance}
\usepackage{epstopdf}
\usepackage{multirow}
\usepackage{booktabs}
\usepackage{url}
\usepackage{pifont}
\usepackage{tikz}
\usepackage{comment}
\usepackage{mathtools}
\usepackage[normalem]{ulem}
\usepackage{xkeyval}
\usepackage[textsize=small]{todonotes}
\usepackage{balance}
\usepackage{ifthen}
\usepackage{etoolbox}
\usepackage{subcaption}

\usepackage{sepfootnotes}
\usepackage[compact]{titlesec}

\usepackage{lscape}
\usepackage{makecell}
\usepackage{longtable}

\usepackage{array}
\newcolumntype{L}[1]{>{\raggedright\let\newline\\\arraybackslash\hspace{0pt}}m{#1}}
\newcolumntype{C}[1]{>{\centering\let\newline\\\arraybackslash\hspace{0pt}}m{#1}}
\newcolumntype{R}[1]{>{\raggedleft\let\newline\\\arraybackslash\hspace{0pt}}m{#1}}
\usepackage{amsmath,amsthm}
\usepackage{mismath}
\usepackage{bbm}
\usepackage[linesnumbered,boxed]{algorithm2e}

\makeatletter
\def\BState{\State\hskip-\ALG@thistlm}
\makeatother

\usepackage[capitalise,nameinlink]{cleveref}
\crefname{algorithm}{Algorithm}{Algorithms}
\Crefname{algorithm}{Algorithm}{Algorithms}
\crefname{section}{\S}{Sections}
\Crefname{section}{\S}{Sections}
\crefname{figure}{Figure}{Figures}
\Crefname{figure}{Figure}{Figures}
\crefname{equation}{Equation}{Equations}
\Crefname{equation}{Equation}{Equations}
\crefname{listing}{Listing}{Listings}
\Crefname{listing}{Listing}{Listings}
\crefname{defn}{definition}{definitions}


%% file: 0_config.tex
\PassOptionsToPackage{usenames,dvipsnames}{color} 
\hypersetup{unicode=true,
	colorlinks=true,
	linkcolor=blue,
	citecolor=blue,
	anchorcolor=blue,
	urlcolor=blue,
	breaklinks=true}
\makeatletter
\def\maxwidth{\ifdim\Gin@nat@width>\linewidth\linewidth\else\Gin@nat@width\fi}
\def\maxheight{\ifdim\Gin@nat@height>\textheight\textheight\else\Gin@nat@height\fi}
\makeatother
\setkeys{Gin}{width=\maxwidth,height=\maxheight,keepaspectratio}
\newcommand\paraspace{\vspace*{1ex}}

\providecommand\parab[1]{\paraspace\noindent\textbf{#1}}
\providecommand\parae[1]{\paraspace\textbf{\textit{#1}}}

\apptocmd\normalsize{%
	\abovedisplayskip=0pt
	\abovedisplayshortskip=0pt
	\belowdisplayskip=5pt
	\belowdisplayshortskip=5pt
}{}{}


%% file: 0_macros.tex
\newcommand{\sysname}{\textsc{TE-CCL}\xspace}
\newcommand{\alltoall}{\textsc{AllToAll}\xspace}
\newcommand{\allgather}{\textsc{AllGather}\xspace}
\newcommand{\scattergather}{\textsc{ScatterGather}\xspace}

\newcommand{\dgx}{{\small\textsf{DGX1}}\xspace}
\newcommand{\dgxtwo}{{\small\textsf{DGX2}}\xspace}
\newcommand{\internalone}{{\small\textsf{Internal 1}}\xspace}
\newcommand{\internaltwo}{{\small\textsf{Internal 2}}\xspace}
\newcommand{\ndv}{{\small\textsf{NDv2}}\xspace}

\newcommand{\ie}{\emph{i.e.,}\xspace}
\newcommand{\eg}{\emph{e.g.,}\xspace}

\newcommand{\squishlist}{
	\begin{list}{$\bullet$}
		{ \setlength{\itemsep}{0pt}
			\setlength{\parsep}{3pt}
			\setlength{\topsep}{3pt}
			\setlength{\partopsep}{0pt}
			\setlength{\leftmargin}{1.5em}
			\setlength{\labelwidth}{1em}
			\setlength{\labelsep}{0.5em} } }
	
	\newcommand{\pgftextcircled}[1]{
		\setbox0=\hbox{#1}%
		\dimen0\wd0%
		\divide\dimen0 by 2%
		\begin{tikzpicture}[baseline=(a.base)]%
			\useasboundingbox (-\the\dimen0,0pt) rectangle (\the\dimen0,1pt);
			\node[circle,draw,outer sep=0pt,inner sep=0.1ex] (a) {#1};
		\end{tikzpicture}
	}

	\newcommand{\squishend}{
\end{list}  }

\makeatletter
\newcommand{\customlabel}[3]{%
   \protected@write \@auxout {}{\string \newlabel {#1}{{#2}{\thepage}{#2}{#1}{}} }%
   \hypertarget{#1}{#3}
}
\makeatother

%% file: intro.tex
\section{Introduction}
\label{sec::intro}

Near-optimal collective communication optimizers~\cite{TACCL, SCCL, Blink}~---~that optimize the communication routes and schedules of distributed training~---~cannot scale to what cloud operators need. This is because cloud operators run large multi-tenant GPU clusters where they schedule distributed training jobs over many GPUs. Tools that find optimum topologies, hardware architectures, or co-optimize various aspects of distributed training~\cite{wang23topoopt, mahajanbetter,zhao2022optimal} also rely on these optimizers and call them multiple times during their search.

Without communication optimizers GPU clusters spend a significant amount of time with idle GPUs: prior work reports the GPUs in BERT~\cite{devlin2018bert} and DeepLight~\cite{deng2021deeplight} spent $11\%$ and $63\%$ of the time idle respectively~\cite{TACCL}. The problem becomes worse as we move to faster GPUs. Current communication optimizers leave significant room for improvement: for example, we show we can improve upon state-of-the-art solutions such as TACCL~\cite{TACCL} {\em by over} $2\times$ on its two chassis \ndv topology~\cite{ndv2} (\cref{fig:ndv2_topo}).

We scale near-optimal collective communication optimizers (\eg SCCL~\cite{SCCL})~---~that model the problem imperfectly but optimally solve their model~---~to enable cloud operators to use them for today's large GPU collectives and improve their runtime to make them more usable as part of other collective optimizers such as~\cite{mahajanbetter, wang23topoopt,zhao2022optimal}~---~our goal is to improve the solution quality of state-of-the-art {\em heuristics} (\eg TACCL~\cite{TACCL}) while maintaining the same ability to scale.

The input to a collective communication optimizer is a {\em demand} (\eg \alltoall, \allgather, \textsc{AllReduce}): a set of interconnected GPUs where each GPU has a certain amount of data to send to other GPUs in the interconnect. The goal of the optimizer is to produce routes and schedules that either maximize bandwidth utilization~\cite{Blink} or minimize job completion time~\cite{TACCL, SCCL} for the input demand or both.

Near-optimal optimizers (\eg~\cite{SCCL}) apply to a single chassis~\cite{SCCL}. In contrast, operators require solutions that scale to topologies with 30-60 chassis (and project larger topologies)~\cite{chatgpt}. Heuristic scale but often produce highly sub-optimal solutions~\cite{Blink, TACCL}. This is becoming a problem as topologies grow and more users share the same underlying network.

SCCL cannot scale because it uses SMT solvers~\cite{campion}. The heuristics avoid using SMT solvers and scale better but fail to account for one or more factors (\eg identifying where traffic should be copied inside the network, enforcing synchronization barriers, and properly accounting for latency of transfers) and produce sub-optimal solutions as a result.

We propose an alternate solution: \sysname. Our insight is that we can model the problem of collective communication optimization through techniques from a class of problems known as multi-commodity flow.

Operators use multi-commodity flow problems in traffic engineering (TE) and use flow conservation constraints to model the flow of traffic --- they assign paths to maximize a cost function~\cite{bertsimas1997introduction}. They, too, take a set of demands as input and produce routes and schedules that optimize various objectives. But the collective problem has nuances that are not present in a traditional multi-commodity flow model:

\noindent \textbf{Temporal variations.} Multi-commodity flow problems assume ``sustained demand'': such problems rely on a continuous flow of data between a source and destination (for several minutes), and this is why the demand in these problems is a bandwidth request (with units such as bits/sec). But GPUs in a collective have finite data to send~--~the demand in these problems is a transfer request (with units such as bits). 

This means we can no longer minimize the delay on the longest path to minimize the transfer time as traditional flow problems do: we can no longer assume an uninterrupted flow of traffic to approximate the delay cost of transfers (see~\cref{sec::background}).

\noindent \textbf{Support for store and forward.} Traditional flow problems~\cite{bertsimas1997introduction} do not model caches. We show in \cref{sec:eval} that we can speed up the solver to find schedules faster if we use the available memory in GPUs.

\noindent\textbf{Supporting copy.} Unlike typical use-cases of the network flow formulation (\eg in the TE context~\cite{SWAN, B4}), collective communication often multicasts the same data to multiple parties, which requires the model to appropriately copy data within the network (and adjust the traditional flow conservation constraints accordingly).

Some prior works do extend multi-commodity flow problems to incorporate these concerns: \eg Calendaring~\cite{kandula2014calendaring} supports deadlines on fixed-size transfers, NetStitcher~\cite{netstitcher} allows for store-and-forward, and several multicast TE works~\cite{noronha1994optimum,doar1993multicast} support copying (see~\cref{sec:relatedwork}). But, it is non-trivial to combine these techniques to add support for all three dimensions {\em simultaneously} without affecting scalability. 

We adapt multi-commodity flow problems to model all three behaviors and solve the general collective communication optimization problem. Our solution is a scalable mixed-integer linear program with optimality gap guarantees (based on the primal-dual theorem~\cite{bertsimas1997introduction}). We show that this solution scales to much larger collectives than techniques such as TACCL~\cite{TACCL} and SCCL~\cite{SCCL} and improves the solution quality.

For certain collectives we can scale this solution even further by converting the MILP into an LP by removing all integer variables. In the general case, we improve scalability by partitioning the problem in time~---~a technique 
 inspired by the $A^{*}$ ~\cite{Astar} from robotics.

{\sysname}'s solutions match the solution quality of SCCL and outperform the quality of state-of-the-art solutions such as TACCL~\cite{TACCL}~---~we show {\em a minimum of $2\times$} performance improvement on the same 2 chassis \ndv topology TACCL uses~---~and shortest path schedules~\cite{zhao2022optimal} because the optimization models the end-to-end problem (whereas these works contain consecutive optimizations that only see a partial view of the problem at each stage), and adds support for copy and store-and-forward. 
As part of \sysname we are also able to account for multi-tenant, heterogeneous topologies where links have different latency, and bandwidth costs and tenants have different priorities to support cloud-scale GPU clusters better.

Our contributions are as follows:
\begin{compactitem}
	\item We present a novel, scalable, solution to the collective communication optimization problem. To the best of our knowledge, this is the first multi-commodity based solution to this problem. This new mode of thinking provides an opportunity to improve other aspects of machine learning collectives such as topology design and adapting to failures.
	\item We show how to scale this solution to larger topologies through a linear program for \alltoall-like demands and a technique inspired by $A^{*}$ in the general case.
	\item We evaluate \sysname both on popular topologies and on the proprietary, large-scale topologies from a large public cloud. We show our solution improves the solution quality of TACCL~\cite{TACCL} by a minimum of $2\times$ in many scenarios. We find TACCL's heuristic is unreliable (produces different solutions in each run) and cannot find a feasible solution in many cases. In contrast, \sysname is reliable, produces the same solution in each run, and finds a feasible solution in instances where TACCL was infeasible. \sysname and \sysname have similar abilities to scale although \sysname was able to run on much larger topologies.
\end{compactitem}

%% file: motivation.tex
\section{Background and Motivation}
\label{sec::background}

We present the necessary background on collective communication and motivate the need for scalable communication schedules for ML collectives. We then describe the multi-commodity flow formulation, how it relates to collective communication optimization, and show why we should modify them to model delay, store-and-forward, and copy.

\subsection{The need for fast collective scheduling}

ML collectives have pronounced communication patterns with flavors of multicast aggregation trees: \eg \allgather, \alltoall, \scattergather (Figure 2 in TACCL~\cite{TACCL} illustrates these communication patterns and how they differ).

These communication patterns constitute a {\em demand} on the network where each GPU wants to send data to other GPUs. For example, in an \allgather demand, each source GPU intends to send all of its data to all other GPUs, and in an \alltoall demand, each GPU wants to send data to all other GPUs, but the data it sends to each GPU is different.

Collective communication optimizers take these demands as input and find solutions that route and schedule them efficiently to minimize transfer time. Operators use these optimizers in their multi-tenant GPU clusters and as part of solutions that help improve their offerings~\cite{zhao2022optimal, mahajanbetter,wang23topoopt}.

Most optimizers use the $\alpha-\beta$ cost model~\cite{alphabeta}. $\beta$ is the transmission time of bytes on a link (how long it takes for the NIC to get the bytes on the wire): if we send $\mathcal{B}$ bytes on a link with capacity $C$ bytes per second, it takes $\frac{\mathcal{B}}{C}$ seconds for the bytes to cross that link and $\beta = \frac{1}{C}$. $\alpha$ is the constant delay of a link. In its simplest form, we can think of it as the propagation delay over a link, but it can also include other factors such as the fixed compute cost of consolidating the data and making the call to the network stack to transmit it. It takes $\alpha + \beta S$ seconds to send a chunk of size $S$ over a link. 

Most existing optimizers fail to scale to large topologies (\eg SCCL~\cite{SCCL}) or produce sub-optimal schedules (\eg NCCL~\cite{SCCL, zhao2022optimal}, TACCL~\cite{TACCL}). SCCL uses SMT solvers and does not scale. TACCL separates the routing and scheduling problems and fails to co-optimize the two. The shortest path first algorithm in~\cite{zhao2022optimal} fails to leverage copy.  

\SetVertexStyle[LineWidth=0.75pt, LineColor=brown!80!black, TextColor=brown!80!black]

\newcommand{\vertex}[4]{
    \Vertex[size=.5,
            shape=circle,
            fontsize=\small,
            fontscale=1,
            label=#3,
            x=#1,y=#2,
            color=brown!15]{#4}
}

\newcommand{\tedge}[3]{
    \Edge[fontsize=\small, fontscale=1, lw=0.75pt, label=#3,,position={above=0.5mm}](#1)(#2)
}

\begin{figure*}[ht]
     \centering
    \begin{subfigure}[b]{0.4\textwidth}
      \centering
        \begin{tikzpicture}
            \node[circle, draw=OliveGreen, fill=OliveGreen!5, inner sep=0pt, minimum size=14pt, line width=0.75pt] (A) at (0,0) {\color{OliveGreen}$s_1$};
            \node[circle, draw=brown!80!black, fill=brown!15, inner sep=0pt, minimum size=14pt, line width=0.75pt] (B) at (1.25,0) {\color{brown!80!black}$h_1$};
            \node[circle, draw=brown!80!black,fill=brown!15, inner sep=0pt, minimum size=14pt, line width=0.75pt] (C) at (2.5,0) {\color{brown!80!black}$h_2$};
            \node[circle, draw=brown!80!black,fill=brown!15, inner sep=0pt, minimum size=14pt, line width=0.75pt] (D) at (3.75,0) {\color{brown!80!black}$h_3$};
            \node[circle, draw=brown!80!black, fill=brown!15, inner sep=0pt, minimum size=14pt, line width=0.75pt] (E) at (5,0) {\color{brown!80!black}$d$};
            \node[circle, draw=blue!70, fill=blue!5, inner sep=0pt, minimum size=14pt, line width=0.75pt] (F) at (2.75,-1) {\color{blue!70}$s_2$};
            
            \draw[pattern=north west lines, pattern color=OliveGreen] (0.15,0.25) rectangle (0.35,0.45);
            \draw[pattern=crosshatch, pattern color=blue!70] (3.05,-0.9) rectangle (3.25,-1.1);
            
            \tedge{A}{B}{$\alpha_1$}
            \tedge{B}{C}{$\alpha_1$}
            \tedge{C}{D}{$\alpha_1$}
            \tedge{D}{E}{0}
            \Text[x=2.5,y=-1.65, fontsize=\small]{$\alpha_2 = 2\beta + 3\alpha_1$}
            \Text[x=2.5,y=-2, fontsize=\small]{TE solutions' cost estimate: $\alpha_2 + 4\beta$}
            \Text[x=2.5,y=-2.35, fontsize=\small]{Correct cost estimate: $\alpha_2 + 3\beta$}
            \Edge[fontsize=\small, fontscale=1, lw=0.75pt, label=$\alpha_2$,position={xshift=-2.5mm, yshift=1mm}](D)(F)
        \end{tikzpicture}
	\caption{Proper modeling of $\alpha$ delay}
	\label{fig:modeling:delay}
     \end{subfigure}
     \begin{subfigure}[b]{0.3\textwidth}
      \centering
        \begin{tikzpicture}
            \node[circle, draw=blue!70, fill=blue!5, inner sep=0pt, minimum size=14pt, line width=0.75pt] (A) at (0,0) {\color{blue!70}$s_2$};
            \node[circle, draw=brown!80!black, fill=brown!15, inner sep=0pt, minimum size=14pt, line width=0.75pt] (B) at (1.25,0) {\color{brown!80!black}$h$};
            \node[circle, draw=brown!80!black, fill=brown!15, inner sep=0pt, minimum size=14pt, line width=0.75pt] (C) at (2.5,0) {\color{brown!80!black}$d$};
            \node[circle, draw=OliveGreen, fill=OliveGreen!5, inner sep=0pt, minimum size=14pt, line width=0.75pt] (D) at (0.25,1) {\color{OliveGreen}$s_1$};
            \node[circle, draw=darkgray, fill=gray!5, inner sep=0pt, minimum size=14pt, line width=0.75pt] (E) at (0.25,-1) {\color{darkgray}$s_3$};

            \draw[pattern=crosshatch, pattern color=blue!70] (0.15,0.25) rectangle (0.35,0.45);
            \draw[pattern=north west lines, pattern color=OliveGreen] (0.55,0.9) rectangle (0.75,1.1);
            \draw[pattern=horizontal lines, pattern color=darkgray] (0.55,-0.9) rectangle (0.75,-1.1);
            
            \tedge{A}{B}{$1$}
            \tedge{B}{C}{$2$}
            \tedge{B}{D}{$1$}
            \tedge{B}{E}{1}
            \Text[x=2.65,y=-0.65, fontsize=\small]{$s_i - h$ are 1 unit/s capacity}
            \Text[x=2.65,y=-1, fontsize=\small]{$h - d$ is 2 units/s capacity}
            \Text[x=2.25,y=-1.5, fontsize=\small]{Without store and forward: 3 solutions}
            \Text[x=2.25,y=-1.85, fontsize=\small]{With store and forward: 3 + 3 solutions}
        \end{tikzpicture}
	\caption{Modeling store and forward}
	\label{fig:modeling:stfwd}
     \end{subfigure}
     \begin{subfigure}[b]{0.28\textwidth}
      \centering
        \begin{tikzpicture}
            \node[circle, draw=OliveGreen, fill=OliveGreen!5, inner sep=0pt, minimum size=14pt, line width=0.75pt] (A) at (0,0) {\color{OliveGreen}$s$};
            \node[circle, draw=brown!80!black, fill=brown!15, inner sep=0pt, minimum size=14pt, line width=0.75pt] (B) at (1.25,0) {\color{brown!80!black}$h$};
            \node[circle, draw=brown!80!black, fill=brown!15, inner sep=0pt, minimum size=14pt, line width=0.75pt] (C) at (2.5,0) {\color{brown!80!black}$d_2$};
            \node[circle, draw=brown!80!black, fill=brown!15, inner sep=0pt, minimum size=14pt, line width=0.75pt] (D) at (2.5,1) {\color{brown!80!black}$d_1$};
            \node[circle, draw=brown!80!black, fill=brown!15, inner sep=0pt, minimum size=14pt, line width=0.75pt] (E) at (2.5,-1) {\color{brown!80!black}$d_3$};
            \draw[pattern=north west lines, pattern color=OliveGreen] (0.15,0.25) rectangle (0.35,0.45);
            
            \tedge{A}{B}{$1$}
            \tedge{B}{C}{$1$}
            \tedge{B}{D}{$1$}
            \tedge{B}{E}{1}
            \Text[x=2.25,y=-1.5, fontsize=\small]{Without copy: 4 sec}
            \Text[x=2.25,y=-1.85, fontsize=\small]{With copy: 2 sec}
        \end{tikzpicture}
	\caption{Modeling copy}
	\label{fig:modeling:copy}
     \end{subfigure}
     \caption{Examples that show why we should model properly: (a) $\alpha$-delay: the maximum delay across all the paths is an incorrect estimate; (b) store-and-forward: buffers improve the solver time as there are more solutions (c) Copy: we can leverage copy to use the available bandwidth more efficiently.}
     \label{fig:modeling}
     \vspace*{-\baselineskip}
\end{figure*}
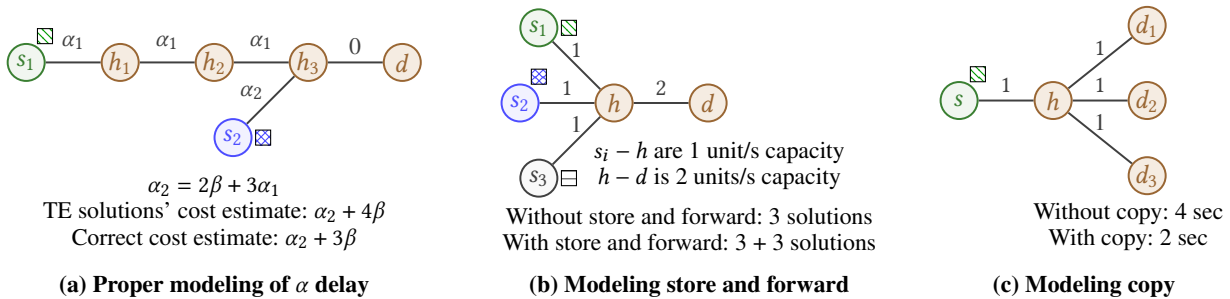

\subsection{Background on network flow solutions}

Many works find optimal routes for wide area traffic engineering (WAN-TE) and for multicast networks (\eg~\cite{SWAN, B4, NCFlow, PoP, netstitcher, kandula2014calendaring,doar1993multicast,Sharp, noronha1994optimum}). These problems also take as input a set of {\em demands}: ``rate requests'' between a source-destination pair. The solutions aim to meet these demands and maximize the total flow the network carries, or the network utilization, or maintain fairness without violating capacity constraints.

Although these formulations take different forms (most notable of these is the path-formulation which takes a set of paths as input and only allows flows to go over the input paths~\cite{SWAN, B4}) they share the following key components:

\noindent{\textbf{Capacity constraints.}} Ensure that the traffic the solution allocates on a link never exceeds its capacity.

\noindent{\textbf{Flow conservation constraints.}} Ensure that the solution does not create traffic ``out of thin air'' but that each non-source node forwards what it receives or consumes it.

\noindent{\textbf{An objective.}} The objective encodes what the optimization is trying to minimize or maximize: the cost model. The most common TE objectives include max-min fair rate allocations, total satisfied demand, or the link utilization.

We observe that the multi-commodity flow and the collective communication optimization problems have many commonalities: both take a set of demands and a topology as input and produce routes (and schedules) to optimize an objective. 

But the two are different as the collective optimizer requires we account for: copy, store-and-forward, and temporal behavior (and the impact on the latency cost as a result). We next discuss each of these in detail:

\noindent{\textbf{Temporal behavior.}} In the collective problem, the source wants to transfer a fixed number of bits~---~once the network satisfies the demand from that source, the demand goes to zero and frees up capacity. The network can then re-assign this capacity to other demands. This happens in the traditional TE as well but at larger time-scales and most deployed TE solvers periodically re-solve the optimization to handle it. 
This is not a problem at face-value --- after all we solve the problem offline --- but it impacts the scalability of the solution in the collective setting.  Calendaring~\cite{kandula2014calendaring} and Netstitcher~\cite{netstitcher} both model this, but they do not model propagation delay and hence fail to address an important side-effect:

\noindent \textit{Modeling delay (the $\alpha$-cost)}. Most TE solutions (\eg~\cite{noronha1994optimum,doar1993multicast}) compute the delay-cost as the maximum delay across all paths where the delay of a path is the sum of the delay on each of its links. These models assume the total time needed to fulfill a demand is the transmission delay (or $\beta$-cost) + this delay-cost.

We show why this model breaks through an example (\cref{fig:modeling:delay}). Here, two sources ($s_1$ and $s_2$) want to send a unit of traffic
(\raisebox{2.3pt}{\tikz[baseline=(char.base)]{\node[shape=rectangle,draw, pattern=north west lines, pattern color=OliveGreen, minimum size=1mm] (char) {};}} 
and 
\raisebox{2.3pt}{\tikz[baseline=(char.base)]{\node[shape=rectangle,draw, pattern=crosshatch, pattern color=blue!70, minimum size=1mm] (char) {};}}) 
to destination $d$. The links on the path from $s_1$ to $h_3$ have a propagation delay $\alpha_1$ and those on $s_2$ to $h_3$ have a propagation delay of $\alpha_2$ where $\alpha_2 = 2\beta + 3\alpha_1$. If we take the traditional TE approach to model the delay, the path with the maximum delay is the one between $S_2$ and $d$ which has a propagation delay of $\alpha_2$. It also takes an additional $4\beta$ for the traffic to get from both $S_1$ and $S_2$ to $d$: the TE solutions estimate $\alpha_2 + 4\beta$ as the completion time.

            

        
\begin{figure}[t]
	\centering
	\includegraphics[width=0.38\textwidth]{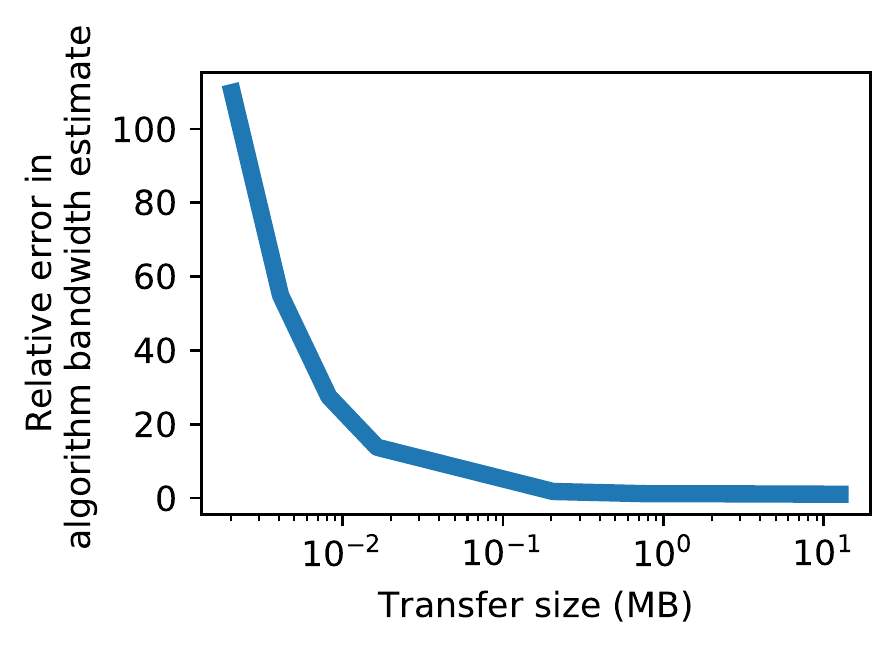}
	\caption{The relative error in the algorithm bandwidth estimate (the output buffer size / transmission time) of a collective schedule that does not model alpha compared to one that does. We use a proprietary topology from a public cloud with $2$ chassis, $8$ GPUs, and $40$ edges where the $\alpha$ of inter-GPU and the GPU to the switch links is $0.6$ and $0.75$ microseconds respectively. \label{fig:alpha}
	}
\end{figure}

But because of the higher propagation delay on the link $s_2$-$h_3$ the data from $s_1$ and $s_2$ both arrive at $h_3$ at the same time ($t = \beta + \alpha_2$), since the propagation delay on the link $h_3$-$d$ is zero, the total time to complete the transfer is $\alpha_2 + 3\beta$. 

The impact of $\alpha$ is greater for smaller transfers (\cref{fig:alpha}): the error in our estimate of algorithm bandwidth for a schedule where we do not model $\alpha$ to one where we do goes up to $100\times$.

\noindent \textbf{Store-and-forward.} Most nodes in a collective topology can buffer incoming traffic before sending it out. We can use this to improve solver time (\cref{fig:modeling:stfwd}) as the number (space) of optimal solutions increases. In \cref{fig:modeling:stfwd}, without store and forward, in the first second, any two nodes (3 schedules) can send their chunks to $h$. With store and forward, we can have three additional schedules where all three sources send to $h$ in the first second, and we then choose in which order to send them to the destination in the next. The solution quality is the same in both cases (we satisfy the demand in 3s). WE confirmed this in our experiments in~\cref{subsec:microbenchmarks} across all the scenarios we considered. For some collective demands store-and-forward may also help with transfer time (though it did not in our experiments). 

But traditional TE does not model buffering~\cite{SWAN, B4}. NetStitcher~\cite{netstitcher} models store and forward but assumes flows do not compete for bandwidth and solves a separate optimization for each flow: it is sub-optimal and does not scale. Some multi-cast TE solutions model intermediate caches~\cite{doar1993multicast}, but they fail to account for the delay, and it is difficult to modify them to do so. 

\noindent \textbf{Copy.} Some collective demands (\eg \allgather) consist of sources that send the same data to multiple destinations (\ie multicast). Traditional TE does not model copy (\eg SWAN and B4~\cite{SWAN, B4}) and produces sub-optimal solutions (see \cref{fig:modeling:copy}). Multi-cast TE~\cite{doar1993multicast,noronha1994optimum} addresses this problem but fails to model delay (these works assume sustained demands) and, in some instances~\cite{noronha1994optimum}, store-and-forward.

We formulate the collective communication optimization problem as a TE problem that supports these elements. The challenge is to maintain scalability. We show our model, as-is, outperforms current state-of-the-art solutions such as SCCL~\cite{SCCL} in its ability to scale and TACCL~\cite{TACCL} in its solution quality. We further improve it's scalability through a technique inspired by $A^*$ from robotics.

%% file: solution.tex
\section{Solution}
\label{sec::solution}

We next describe how we model the collective communication problem as a multi-commodity flow problem. We build on the ideas in Calendaring~\cite{kandula2014calendaring} and Netstitcher~\cite{netstitcher} to model delay, model store-and-forward, and copy. 
 
But this solution does not scale to topologies with more than 64 GPUs. We scale it by changing our mixed integer program (MILP) into a linear program (LP) for demands such as \alltoall where sources send different data to each destination and do not benefit from copy (\cref{sec::partial}); and through a more general solution we call $A^*$(\cref{sec:Astar}).  
\begin{table*}[h]
	\centering
	\begin{tabularx}{\textwidth}{ l X  }
		\textbf{Variable} & \textbf{Description}  \\ 
		\hline
		\hline
		$N$ & Set of nodes in the graph \\
		$S$ & Set of nodes in the graph that are switches ($S \subset N$) \\
		$E$ & Set of edges in the graph ($E \subseteq 2^{N \times N}$). Edges are unidirectional. \\
		$C$ & Chunk IDs ($C = \{0, 1, 2, \ldots, \mathtt{C}\}$). Each node has $ \le \mathtt{C} + 1$ number of chunks.\\
		$D$ & Demand function ($N \times C \times N \to \{0, 1\}$) where $D_{s,c,d}$ is whether destination $d$ wants chunk with id $c$ from node $s$\\
		$\tau$ & Epoch duration\\
		$K$ & The set of epochs ($K = \{0, 1, 2, \ldots, \mathtt{K}\}$) \\
		$F_{s, i, j, k, (c)}$ & Amount of source $s$ chunks that are going over link $(i,j) \in E$ at epoch $k \in K$  \\
		$B_{s, i,k,(c)}$ &  Amount of source $s$ chunks that are in node $i$'s buffer at the \textit{start} of epoch $k$ \\
		$T_{ij}$ & Capacity of link $(i,j) \in E$ \\
		$\alpha_{ij}$ & Fixed latency associated with link $(i,j) \in E$\\
		$\delta_{ij}$ &  Number of epochs contained within an $\alpha_{ij}$ for each link $(i,j) \in E$ \\
		$\mathsf{R}_{s,d,k}$ & Source $s$ chunks that node $d$ \textit{read} off of the network \textit{in} epoch $k$\\
		$\mathcal{R}_{s,d,k, (c)}$ & Source $s$ chunks read off the network by $d$ \textit{up to} epoch $k$.\\
	\end{tabularx}
	\caption{Our notation. We put in parentheses the index (c) because we only use it when demands benefit from copy. When we model copy the values of $F$ and $B$ are integers. We show for some demands (where copy is not useful) we can use real variables instead in \cref{sec::partial}.\label{table:notation}}
\end{table*}

\subsection{The general model}
\label{subsec::generalform}

We describe our notation in \cref{table:notation}. Like any other multi-commodity flow problem we need to specify: capacity and flow conservation constraints, and an objective.

But, to model delay, store-and-forward, and copy we need to introduce a few new concepts: chunks, epochs, and buffers. 

Our notion of chunks is similar to prior work (\eg SCCL): a chunk (like packets) is a block of bytes\footnote{We allow our solution to split chunks into smaller blocks when we move to the linear program form.}. 

We use epochs (similar to how SCCL uses rounds) to make time discrete: epochs are fixed periods of time --- our solution produces a schedule that tells the user in which epoch they should send a chunk and on which link.

We discuss chunk sizes and epoch durations in detail in \cref{subsec::consider}. For now, we assume $\tau$ is the epoch duration and $T_{ij}$ is the capacity of a link (where the units are chunks per second), and that epoch is sufficient for at least one chunk to traverse any link.

We use buffers to model store-and-forward. To simplify the explanation we assume each node has enough buffer to store the entire network demand if it needs to (we show how to remove this assumption in \cref{sec:limitedbuffers}).

To model copy, we need to track each chunk: we use $F_{s,i,j,k,c}$ and $B_{s,i,k,c}$ to track whether chunk $c$ from source $s$ is going over link $(i,j)$ or is in node $i$'s buffer at epoch $k$ respectively. 

\newcommand{\north}{\raisebox{2.3pt}{\protect\tikz[baseline=(char.base)]{\protect\node[shape=rectangle,draw, pattern=crosshatch, pattern color=brown!80!black, minimum size=1mm] (char) {};}}}

\newcommand{\cross}{\raisebox{2.3pt}{\protect\tikz[baseline=(char.base)]{\protect\node[shape=rectangle,draw, pattern=north west lines, pattern color=brown!80!black, minimum size=1mm] (char){};}}}

\begin{figure}[t]
	\centering
		\begin{tikzpicture}
			\vertex{0}{0}{$s$}{A}
			\vertex{1.5}{1}{$d_1$}{B}
			\vertex{1.5}{-1}{$d_2$}{C}
			\vertex{3}{0}{$d_3$}{D}
			
			\draw[pattern=north west lines, pattern color=brown!80!black] (-0.35,0.05) rectangle (-0.55,0.25);
			\draw[pattern=crosshatch, pattern color=brown!80!black] (-0.35,-0.15) rectangle (-0.55,0.05);

			\draw[pattern=north west lines, pattern color=brown!80!black] (1.4,1.3) rectangle (1.6,1.5);
			\draw[pattern=north west lines, pattern color=brown!80!black] (1.4,-1.3) rectangle (1.6,-1.5);
			
			\draw[pattern=north west lines, pattern color=brown!80!black] (3.35,0.05) rectangle (3.55,0.25);
			\draw[pattern=north west lines, pattern color=brown!80!black] (3.35,-0.15) rectangle (3.55,0.05);

			\Edge[fontsize=\small, fontscale=1, lw=0.75pt, label=$0.5$,position={above=0.2mm}](C)(D)
			\Edge[fontsize=\small, fontscale=1, lw=0.75pt, label=$0.5$,position={above=0.2mm}](A)(B)
			\Edge[fontsize=\small, fontscale=1, lw=0.75pt, label=$0.5$,position={above=0.2mm}](A)(C)
			\Edge[fontsize=\small, fontscale=1, lw=0.75pt, label=$0.5$,position={above=0.2mm}](B)(D)
			\Text[x=-2,y=0, fontsize=\small]{Epoch 0 one chunk}
            \Text[x=-2,y=-0.35, fontsize=\small]{(as two halves)}
			\Text[x=.75,y=-1.42, fontsize=\small]{Epoch 1}
            \Text[x=.75,y=1.38, fontsize=\small]{Epoch 1}
            \Text[x=4.15,y=0, fontsize=\small]{Epoch 2}
		\end{tikzpicture}
	\protect\caption{An example of why we need integer variables to track each chunk. If we allow partial chunks (\north,\;\cross)  and copy at the same time, we run into a situation where the optimization can send the same copy of part of a chunk (\cross) to two neighboring nodes (in this case $d_1$ and $d_2$) and they can forward it along to the destination ($d_3$). Since the formulation has no way of knowing these two halves are the same, it thinks $d_3$ has received the full chunk.}
	\label{fig:example_copy}
\end{figure}
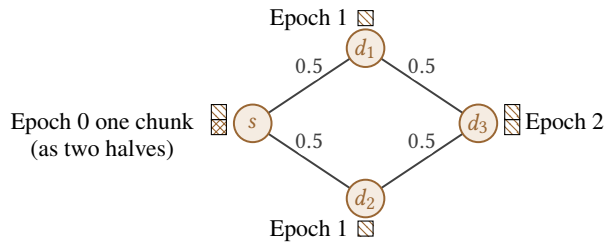

We need to use integer variables for $F_{s,i,j,k,c}$ and $B_{s,i,k,c}$ to model copy~---~we cannot allow chunks to be split into smaller pieces. We use the example in \cref{fig:example_copy} to explain why. Source $s$ sends the first half of a chunk (\cross) to both destinations $d_1$ and $d_2$. These nodes then both forward it to $d_3$: they have no way of knowing this is the same half. The optimization now thinks it has delivered the full chunk to $d_3$ while it has only delivered one half of it twice: it will send the second half of the chunk to both $d_1$ and $d_2$ but not to $d_3$. Using integers for $F_{s,i,j,k,c}$ and $B_{s,i,k,c}$ allows us to avoid this problem (we do not need this for demands that do not benefit from copy \cref{sec::partial}). We can increase the number of chunks to decrease the size of each individual chunk and support smaller transmission blocks (the optimization automatically consolidates them to bigger transmission units if needed) --- but this increases the number of variables and slows down the optimization.

We now have everything we need:

\noindent \textbf{Capacity constraints.} Capacity constraints ensure we do not send more data than the link can carry in an epoch. We have:

\begin{align*}
&\text{Capacity Constraint} \big(i,j,k\big)  \triangleq \nonumber\\
&\sum_{s \in N}\sum_{c \in C} F_{s,i,j,k,c} \le T_{ij} \tau
\end{align*}

\noindent \textbf{Flow conservation constraints.} The purpose of these constraints is to ensure the network does not create or lose traffic. The traditional form of these constraints specifies: a node should either consume or forward all of the traffic it receives. Here, we need to change these constraints to account for: (a) copy --- nodes can create new traffic; (b) delay. 

To model delay, we need to ensure a node does not forward a chunk if it has not received it. We first compute $\delta_{ij} = \frac{\alpha_{ij}}{\tau}$: number of epochs it takes for a chunk to traverse a link. Traffic that node $i$ sends to node $j$ at the beginning of epoch $k$ arrives at node $j$ by the end of epoch $k + \lceil \delta_{ij} \rceil$. Node $j$ can forward a chunk it receives from node $i$ if node $i$ sent it $\lceil \delta_{ij} \rceil$ ago.

Copy, by definition, violates traditional flow conservation constraints: it creates traffic where it didn't exist before. But, the node does not need to copy the chunk on the same link in the same epoch. We use this, along with $\delta_{ij}$ to rewrite the flow conservation constraints as follows:  

\begin{align*}
& \text{Flow conservation constraints} \big(s, n, k, c \big) \triangleq \nonumber \\
& B_{s,n,k, c} + \sum_{\forall j \mid (j,n) \in E} F_{s,j,n,k - \lceil\delta_{jn}\rceil, c} \ge \max_{\forall j \mid (n,j) \in E} F_{s,n,j, k + 1,c}
\end{align*}

This constraint encodes that what the node $n$ has in its buffer along with what it receives in epoch $k$ has to be larger than what it sends out in the next epoch on \textit{each} of its outgoing links. We track the buffer contents as follows:

\begin{align*}
& \text{Buffer constraints} \big(s,n,k,c) \triangleq \nonumber \\
& B_{s,n, k,c } = B_{s,n,k-1,c} + \sum_{\forall j \mid (j,n) \in E} F_{s,j,n,k - \lceil\delta_{jn}\rceil - 1, c}
\end{align*}

The buffers accumulate all traffic the GPU has received up to that point. Nodes have enough memory for this: for collective demands such as \allgather each GPU needs all the chunks that are sent over the network and stores them anyway. But it is straight-forward to model limited buffers as well if we track what we should remove from the buffer in each epoch (see \cref{sec:limitedbuffers}). We evaluate the benefit of buffers using an \allgather demand in \cref{sec:eval}.

The first and last epoch's flow conservation constraints are slightly different from the above: a node does not receive anything in the first epoch and doesn't send anything in the last. We refer the reader to the appendix for details due to space constraints (see \cref{sec::initialization}).  

We next need to account for demands: we need to make sure all demands are met at the end.

\noindent\textbf{Destination constraints.} These constraints ensure each node receives its full demand by the end: 

\begin{align*}
& \text{Destination constraints} \big(s, d, k, c\big) \triangleq \nonumber\\
&\mathcal{R}_{s,d,k,c} = \min(D_{s,d,c}, B_{s,d,k+1,c} ) \quad \& \nonumber \\
& \mathcal{R}_{s,d,\mathtt{K},c} = D_{s,d,c}
\end{align*}

where $\mathcal{R}_{s,d,k,c}$ is whether $d$ has received chunk $c$ of source $s$ by epoch $k$. These destination constraints are different from their counterparts in traditional TE models. This is because of copy: $d$ may want a chunk and also relay the chunk to others. Hence, we cannot assume $d$ wants to consume everything in its buffers. This is why we take the minimum of $D_{s,d,c}$ and $B_{s,d,k+1,c}$. We ensure $d$ eventually receives its full demand by the last epoch $\mathtt{K}$ by setting $\mathcal{R}_{s,d,\mathtt{K}, c}$ to $D_{s,d,c}$.

\parab{Modeling switches.} So far, we have only modeled the behavior of GPU nodes. While some topologies (\eg within a single \dgx node~\cite{SCCL}) only consist of GPUs, almost all larger topologies use switches to connect GPU blocks. We have to model switches differently because they have limited memory: we cannot buffer chunks at the switch. Hence, we set the buffer at each switch to zero.

Traffic needs to pay the $\alpha$ delay cost of two links to cross a switch: one from the node to the switch and one from the switch to the node.

Most of today's switches support copy~\cite{Sharp}, and so we model switches with this assumption (switches have the same flow conservation constraint as other nodes). But we can also model switches without this capability to support legacy hardware. One way is to replace the flow conservation constraints at the switch with the traditional TE flow conservation constraints (what comes into the switch must go out). 

Another option is to use the approach from TACCL~\cite{TACCL}: replace switches with {\em hyper-edges} and allow the user to choose which hyper-edges to allow. For this second model we need to add additional constraints and due to limited space we refer the reader to \cref{sec:legacy} for the details. 

The former two approaches are easier to use in practice: the user does not need to specify a sketch (which is a crucial in TACCL) or pick which GPU communicates with which other GPU~---~when we looked at the TACCL code we found the authors used their \texttt{uc-min} and \texttt{uc-max} strategy along with the user-specified sketch to automatically find which links to enable for switches within the node, but for cross-node links they pre-identified which links perform best manually. We need to understand the topologies well to write such sketches and we found it difficult when we evaluated new topologies with TACCL. In contrast, our solution requires no human in the loop~---~the user only needs to specify the topology and the demand matrix~---~but the solver is slightly slower.

\noindent \textbf{The objective.} Our optimization objective is to finish the transfer as quickly as possible. We can encode this as follows:

\begin{align*}
&\text{Objective function} \triangleq \sum_{\forall k \in K, \forall s,d \in N:s\not=d} \frac{1}{k + 1} \mathcal{R}_{s, d, k}
\end{align*}

Notice how the objective gives fewer rewards as $k$ increases: the objective improves if the schedule satisfies the demand as soon as possible.
If we combine the objectives with our constraints we arrive at an optimization that maximizes the objective subject to all of the above constraints.


One nuance here is that the optimization has multiple optima: the objective does not discourage solutions where we send flows that do not satisfy any demand (as long as the schedule satisfies all demands as quickly as possible the solution is optimal). Such solutions are clearly wasteful.

To avoid such {\em silly} cases, we can do one of two things: (a) we can either add a term to the objective to discourage unnecessary flows; or (b) we can zero out those flows in post-processing the solutions. The first results in higher solver run-times as it becomes harder for the solver to prove optimality. 

We use the latter approach where we run an algorithm similar to a reverse DFS. We start from each destination, and track the flows from that destination to the source until we account for its entire demand. We then remove (zero-out) all remaining flows as there is no demand corresponding to them. This takes $\bigO(\abs{N} + \abs{E})$ time where $N$ is the number of nodes in the graph and $E$ is the number of edges.

\section{Scaling}

Our formulation is general and pushes beyond the scale boundaries of SCCL and outperforms the solution quality of TACCL. But it is slow for topologies with more than $32$ chassis. We next show two methods to scale this solution. 
The first works in situations where copy is not useful (\eg \alltoall) and preserves optimality. The second is general (\ie supports copy): it solves the problem by partitioning it in time (its goal, in each time partition, is to make as much progress as it can towards finishing the transfer). This later model is sub-optimal, but outperforms the TACCL heuristic (see \cref{sec:eval}) as it more accurately captures the optimization incentives and constraints. Its formulation allows users to trade-off optimality and speed by changing the number of partitions (smaller partitions increase sub-optimality but improve scalability).

\subsection{Scaling by converting to a linear program}
\label{sec::partial}

There is only one reason we needed integer variables for our model: copy! But some demands do not benefit from copy --- this is when each destination wants a unique segment of information from each source. In these scenarios we can change our formulation into a linear program (LP). LPs are convex optimization programs which we can solve in polynomial time and scale much better than MILPs. 

We remove support for copy and modify the flow conservation constraints back to their traditional form.
The following constraint dictates: a node either buffers a chunk it received, forwards it in the next epoch, or consumes it. Notice a node can consume a chunk it received at the end of an epoch. We do not track individual chunks since we no longer need to worry about duplicates. This reduces the number of variables. 

\begin{align}
	\text{Flow conservation constraints} \big(s, n, k \big) \triangleq \nonumber\\
	\sum_{\{j \mid (j,n) \in E\}} F_{s, j, n,  k - \lceil\delta_{jn} \rceil} + B_{s, n, k} =\nonumber\\ 
	B_{s, n, k + 1} + \mathsf{R}_{s, n, k} + \sum_{\{j \mid (n,j) \in E\}} F_{s,n, j, k+1} \nonumber
\end{align}

The flow conservation constraints for switches are different: a switch does not consume chunks and does not buffer them --- we remove those terms from the flow conservation equations.

Since destinations no longer need to both consume {\em and} forward chunks, we can modify the destination constraints:

\begin{align}
&\text{Destination constraint} \big(s, d, k\big) \triangleq \nonumber\\
& \mathcal{R}_{s, d, k} = \sum_{r = 0}^k \mathsf{R}_{s, d, r} \quad \& \nonumber\\
& \mathcal{R}_{s,d, \mathtt{K}} = \sum_{\forall c} D_{s,d,c} \nonumber
\end{align}

Our LP produces a {\em rate allocation} to demands that originate from each source on each link. From this we generate a schedule that we then execute in hardware (we translate these rates to paths for each chunk through the same DFS-like solution we described earlier). This is a straight-forward algorithm --- TE solutions also use similar algorithms that we can adopt~\cite{kandula2014calendaring, netstitcher} --- and we omit it due to space constraints. 

\subsection{Scaling using the $A^{*}$ technique}

The LP form allows us to scale the solution to large topologies, but it does not permit copy. Copy is important for demands such as \allgather (see \cref{sec::background}). We also provide a second scaling method inspired by the $A^{*}$ technique in robotics~\cite{Astar}. 

We partition the problem into multiple rounds. In each round we no longer find a solution that satisfies all demands but instead motivate the solver to make as much progress towards this goal as it can. These optimizations have fewer variables and are faster. We sequentially solve them one after the other until we reach a round where we meet all demands.

Here we need to address two new modeling challenges:

\noindent{\textbf{Encoding the right incentives.}} We need to remove the constraint that required the optimization to meet all demands by the last epoch --- otherwise the optimization in each round may become infeasible. This means our objective function is no longer sufficient: it only says {\em if} it is feasible to satisfy a demand do so as fast as possible, but it does not reward incremental progress~---~we need to augment our objective with a term that rewards the optimization for moving data closer to the destinations in each round. But how to do this in a way that preserves the MILP format?

We augment our topology with logical links that allow us to compute this reward function: we add logical edges to the graph that connect each node to all the destinations and add weights to each of these logical edges that correspond to the minimum distance --- we compute these weights using the Floyd Warshall algorithm~\cite{FW} and the $\alpha$-delay cost of each edge --- from the node to each destination. We can now use these edges to encode a viable cost function which we can add to our original objective. Due to space constraints we refer the reader to the \cref{sec:Astar} for the details.

\noindent{\textbf{Modeling delay.}} Chunks that we send on any link $(i,j)$ my not reach $j$ by the end of the round (because of the $\alpha_{ij}$-delay on that link) but instead arrive in a future round. We therefore need to maintain state from one round to the next and incorporate these late arrivals in our formulation.

We refer the reader to the appendix for the full formulation.

\section{Important Considerations}
\label{subsec::consider}	

Earlier we described how to formulate collective communication optimization using a TE approach. All three formulations (the general MILP form, the LP form, and $A^{*}$) find solutions for any input demand but only the general MILP form and the $A^{*}$ model support copy. There are a number of parameters in these formulations we need to choose carefully:

\noindent \textbf{Epoch durations and chunk sizes.} A side-effect of using integer variables in the MILP formulation and the $A^{*}$-based technique is that the choice of chunk-size and epoch duration is important (the LP is not sensitive to these settings)~---~smaller epochs allow for finer-grained schedules that better leverage the available network capacity. To find the best chunk size we can sweep a range of values to find the best one quickly. We can also take this as an input~---~smaller chunks allow for finer grained schedules but can increase the resource usage on a node. Users can also utilize solutions such as~\cite{mahajanbetter} to pick what is optimum for their work-flow.

To set the epoch duration we can do one of two things: (a) to get the best schedule from the vanilla MILP formulation we can set the epoch duration to the time it takes the slowest link to transmit a chunk~---~ the MILP cannot send anything if we use smaller epochs because of the capacity constraints; (b) we can set the epoch duration based on the time it takes the {\em fastest} link to transmit a chunk. Option (b) enables the MILP to produce finer grained schedules but to use it we have to modify the capacity constraints and the flow conservation constraints: the capacity constraints ensure we don't exceed the capacity constraint on the slowest link and the flow conservation constraints ensure we do not forward a chunk before receiving it. Due to space constraints we refer the reader to the appendix for the details (see \cref{sec:fastlink}). We compare the two approaches in \cref{sec:eval}. Option (b) produces better schedules which is why we use it for most of our evaluations. 

\noindent \textbf{Number of epochs.} We need to input an upper bound on the number of epochs which estimates how many epochs it may take to fully satisfy the demand: pick too small a number and the optimization will be infeasible, pick too large of a number and the MILP will be too large and too slow. To streamline finding the right number of epochs --- and to not burden the user with having to identify what numbers to use --- we develop a simple algorithm which finds a loose upper bound on how long we need to satisfy all the demands.

To find this number, we quickly sweep a range of transmission times: for each transmission time, we use coarser grain epoch durations (very large epochs) and run the optimization. Because we use large epoch sizes, we have fewer variables, which allows us to solve the optimization quickly. The solution of these runs is not optimal (because the epochs are too coarse), but it gives us an idea of how long we need when we switch to the optimal epoch duration. We describe the process in detail in \cref{algo:finding_num_epochs} in the \cref{sec:num_epochs}. We use the output to initialize the optimization which automatically identifies if a lower number of epochs is sufficient. 

\noindent \textbf{Number of epochs in a round in $A^{*}$.} 
We solve round after round of $A^{*}$ until we deliver all the demands. Users can choose how many epochs to use in each round. The smaller the number of epochs in a round, the faster the optimization and the higher the optimality gap. Picking a small number of epochs per round also impacts the state we need to maintain. In our experiments, we set the number of epochs such that chunks do not arrive later than one round in the future.

\noindent \textbf{The topology, $\alpha$, and $\beta$ inputs.} \sysname takes the topology and the values for $\alpha$ and $\beta$ as input. We do not provide an independent method for computing these values. 

\noindent \textbf{Which switch model to use.} We provide two switch models: one that allows the switch to copy chunks (to model networks with the SHArP protocol~\cite{Sharp} enabled) and one which does not (the latter is similar to TACCL's hyper-edge model). It is up to the user to decide which model to use in the optimizer.

\noindent \textbf{Modeling variable bandwidth.} Our model supports networks with variable bandwidth. To add support for this, we have to assume bandwidth only changes from one epoch to the next. We can then take the capacity matrix for each epoch and use that in our capacity constraints.

\noindent \textbf{Use in multi-tenant clusters.} TE-CCL supports multi-tenant communication optimization: all our models accept a network demand as input~---~to model a multi-tenant environment we have to change the demand matrix to the sum of the demands across all collectives. The capacity constraints will ensure we do not exceed network capacity and the objective ensures we minimize the total completion time across all tenants.

We can also support priorities across tenants (\ie prioritizing one tenant's completion time over the others) if we add a separate buffer and read variable for each tenant: we can then add the priorities the objective function. This change increases the number of variables in the MILP which slow it down~---~we may have to use $A^{*}$ in this case but this does not impact the quality of the solution compared to when we solve a single tenant problem at the same scale.

\noindent \textbf{Scaling through intermediate solutions.} The solver we use, Gurobi~\cite{gurobi}, often finds an optimal solution and then spends a long time proving it is optimal~---~ often the solution does not improve even after the solver runs for an additional $10$ hours. We therefore apply a timeout and stop the solver after $2$ hours and use the solution at that point. Gurobi reports its progress through the primal-dual gap~\cite{boyd_co}.

%% file: eval.tex
\vspace{-0.15cm}
\section{Evaluation}
\label{sec:eval}
We implement our solution in Python. We use Gurobi~\cite{gurobi} to solve the optimizations. We convert our solution into MSCCL~\cite{SCCL}, which can then port it into a schedule that runs on the hardware. We plan to release our code.


The goal in this evaluation is to:

\begin{compactitem}
	\item Compare \sysname to state-of-the art: both in scale and in terms of solution quality.
	\item Show \sysname scales to the large topologies.
	\item Show the impact of each of our different design choices.
\end{compactitem}

\noindent \textbf{Metrics.} We use the following metrics to evaluate \sysname:

\parae{Solver time.} The time it takes~---~which includes the time to setup the variables and constraints in the solver~---~ to solve the collective optimization problem. 

\parae{Transfer time.} The time it takes for the transfer to complete: for all the nodes to receive their full demand.

\parae{Output buffer size.} The data each GPU receives once we satisfy the demand (we borrow this from TACCL~\cite{TACCL}).

\parae{Transfer size.} The amount of data each GPU sends to others: for example, a GPU in an \allgather demand with a transfer size of $1$ GB sends $1$ GB of data to {\em each} other GPU. 

\parae{Algorithmic bandwidth.} The output buffer size divided by the transfer time (this metric is from TACCL~\cite{TACCL}).

\parab{Topologies and workloads.} We evaluate \sysname using the topologies in \cref{table:topologies}. We use common topologies such as \dgx, \dgxtwo~\cite{dgx2}, and \ndv~\cite{ndv2} as well as two proprietary topologies from a public cloud provider.

\parab{\sysname variants.} We use three variants of \sysname in our evaluations: the optimal (where we use the vanilla MILP for \allgather and LP for \alltoall), the early-stop version for \allgather (where we use Gurobi's ability to find a good solution~--~ which is at most 30\% away from optimal~--~quickly), and $A^{*}$ for \allgather.

Gurobi runs into numerical issues with \alltoall on large topologies (more than 64 nodes): we need to run it with a different configuration (\texttt{method = 2}~\cite{method2}) which causes it to produce a feasible (but not optimal) solution. In those cases, we run the solver in a loop and do a binary search (on the number of epochs) to find the optimal solution.

We set the epoch duration based on the bandwidth of the fastest link. In the cases where $\alpha > 200\times\tau$ we increase the epoch duration by $5\times$ to avoid large models (since $\alpha$ dominates this does not materially impact the solution).

\sysname solves optimization problems to produce a schedule, and the optimization is deterministic, outputting the same number of epochs to meet the demand every time we run it. 
The solver times also do not vary significantly for a given optimization across runs.
\noindent\textbf{Baselines.} We compare our solution to two state-of-the-art solutions: TACCL~\cite{TACCL} and SCCL~\cite{SCCL}. 

\parae{TACCL.} We obtained the TACCL code from the authors and track and report the solver time. 
\sysname takes an additional $\beta$ compared to TACCL to route chunks through a switch: TACCL replaces the switch with direct edges between the nodes and only pays one transmission delay to cross that link whereas \sysname models the switch itself and pays two transmission delays~---~one from the node to the switch and one from the switch to the node. To compare fairly against TACCL we change our model of the switch to do the same when comparing with TACCL.

\parae{SCCL.} We compare to SCCL using the public SCCL code-base~\cite{msccl} and also re-ran our experiments using the SCCL artifact from their submission (which the authors gave us). We verified and confirmed with the authors we used SCCL correctly and that our numbers are correct. 

\begin{table}[t]
	{\small
		\centering
		\begin{tabularx}{\linewidth}{ l c c }
			\textbf{Topology} & \textbf{\# of GPUs per chassis} & \textbf{\# of edges per chassis}\\ 
			\hline
			\hline
			\internalone & 4 & 8\\
			\internaltwo & 2 & 2\\
			\dgx & 8 & 32\\
			\ndv & 8 & 32 \\
			\dgxtwo & 17 & 32\\
		\end{tabularx}
	}
	\caption{Our topologies. The internal topologies are from a large public cloud and are proprietary: $\alpha$ is $0.6 \mu s$ and $0.75 \mu s$ on their GPU to GPU and GPU to switch links.}\label{table:topologies}
\end{table}

\parab{Platform.} We use the solvers and the schedules they produce to compute the transfer times and algorithmic bandwidth for SCCL, TACCL, and \sysname.
We checked using a single $8$ GPU \dgx node that these estimates match what we get from running on hardware for both \sysname and TACCL.

We report the capacity and delay for the public topologies in the \cref{sec:topology details}.

\parab{Unexplored avenues.} 
We show from testing on a \dgx that \sysname's estimates of collective latency match the actual runtimes on prototype hardware. We do not have access and the budget to run hardware experiments at scale on different kinds of GPUs. Thus, the effect of factors such as congestion, message batch sizes and other GPU implementation artefacts on the collective latency remains an unknown. But our results on all of the other metrics such as solver times and our ability to scale to large topologies hold regardless.

\subsection{Comparison to SCCL and TACCL} 

\parab{SCCL.} SCCL has two modes: one minimizes latency (\texttt{least-steps}) and one produces an instance solution (\texttt{instance}) with the number of chunks, rounds, and steps as input.

Our solution is equivalent to the former but the SCCL \texttt{least-steps} command took over a day to produce a solution for \allgather demands with more than $3$ chunks and \alltoall demands with more than $1$ chunk on a \dgx topology (the SCCL paper does not evaluate this mode). In contrast, we ran \sysname with $\max K = \mathtt{K} = 10$ (the maximum number of epochs the optimization can use to satisfy the demand) and $25KB$ chunks, and it finished in $\le 0.65s$ for all \allgather demands and $\le 0.97s$ for \alltoall demands with less than $5$ chunks.

\begin{table}[t]
	{\small
		\centering
		\begin{tabularx}{\linewidth}{ l c c }
			\textbf{Collective, \# chunks} & \textbf{SCCL ($\mu$s)} & \textbf{\sysname ($\mu$s)}\\ 
			\hline
			\hline
			\allgather, 1 & 3.4 & 4\\
			\allgather, 2 & 5.1 & 5\\
			\allgather, 3 & 8 & 6.1\\
			\alltoall, 1 & 3.4 & 4 \\
		\end{tabularx}
	}
	\caption{Comparing the transfer time from SCCL \texttt{least-steps} with \sysname ($\mathtt{K}=10$ and chunk size = $25$ KB). \sysname can better pipeline chunks and so pays less $\alpha$ cost with larger transfers.}\label{table:SCCL_vs_TECCL_opt}
\end{table}

We used $25KB$ chunks to capture the impact of $\alpha$ ($\alpha = 0.7\mu$s) on the solutions (\cref{table:SCCL_vs_TECCL_opt}): for all $> 1$ chunk cases \sysname outperforms SCCL as it models the $\alpha$ delay better~---~ it ensures a node receives a chunk before forwarding it but pipelines traffic; SCCL enforces a barrier instead. SCCL performs better in the $1$ chunk case as \sysname cannot leverage its ability to pipeline.

\begin{figure*}[th]
	\centering
	\begin{subfigure}[b]{0.23\textwidth}
		\centering
		{\label{fig:ndv2_allgather_taccl_bw}\includegraphics{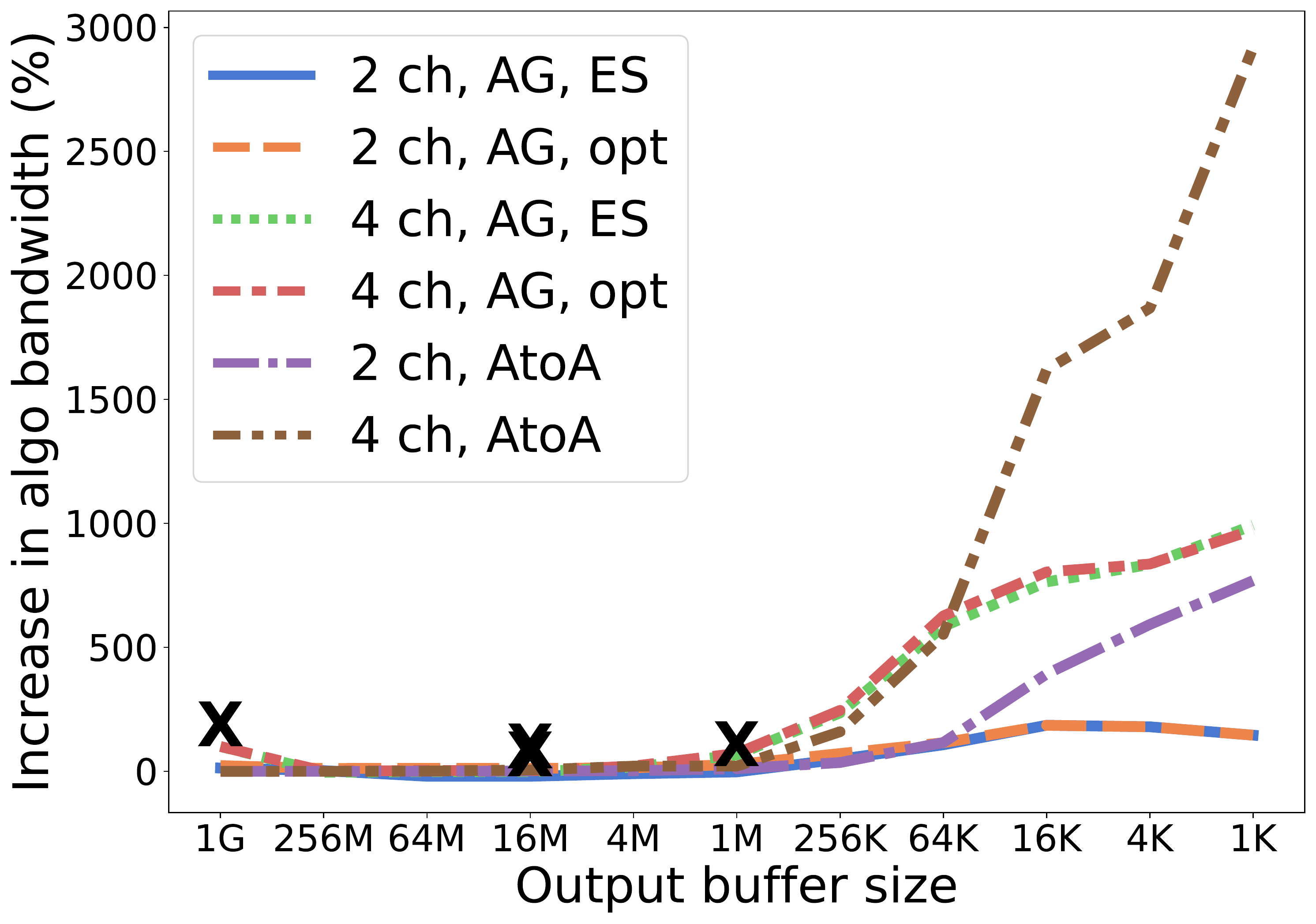}}
		\caption{\ndv}
	\end{subfigure}
	\begin{subfigure}[b]{0.23\textwidth}
		\centering
		{\label{fig:dgx2_allgather_taccl_bw}\includegraphics{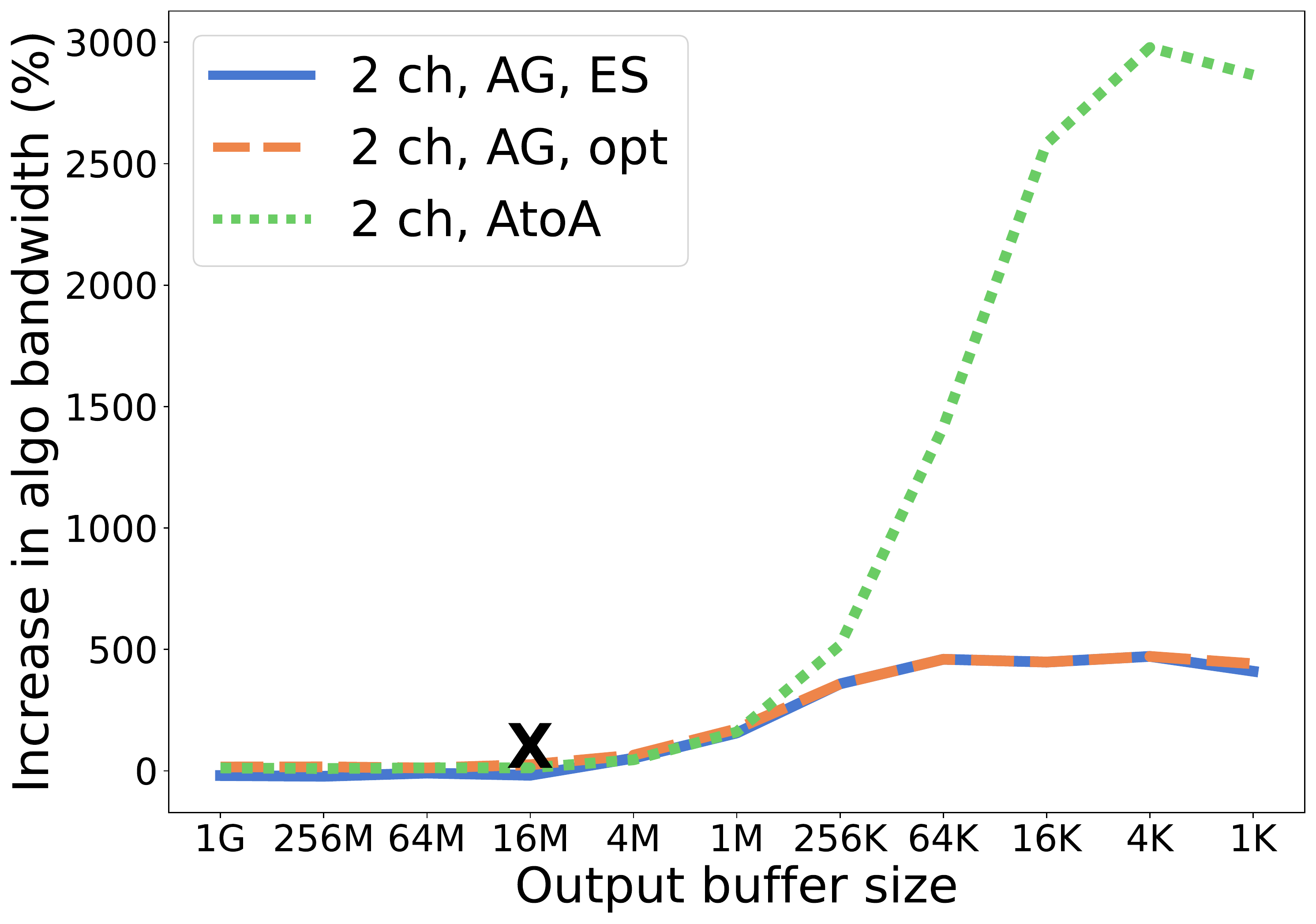}}
		\caption{\dgxtwo}
	\end{subfigure}
	\begin{subfigure}[b]{0.23\textwidth}
		\centering
		{\label{fig:internal1_allgather_taccl_bw}\includegraphics{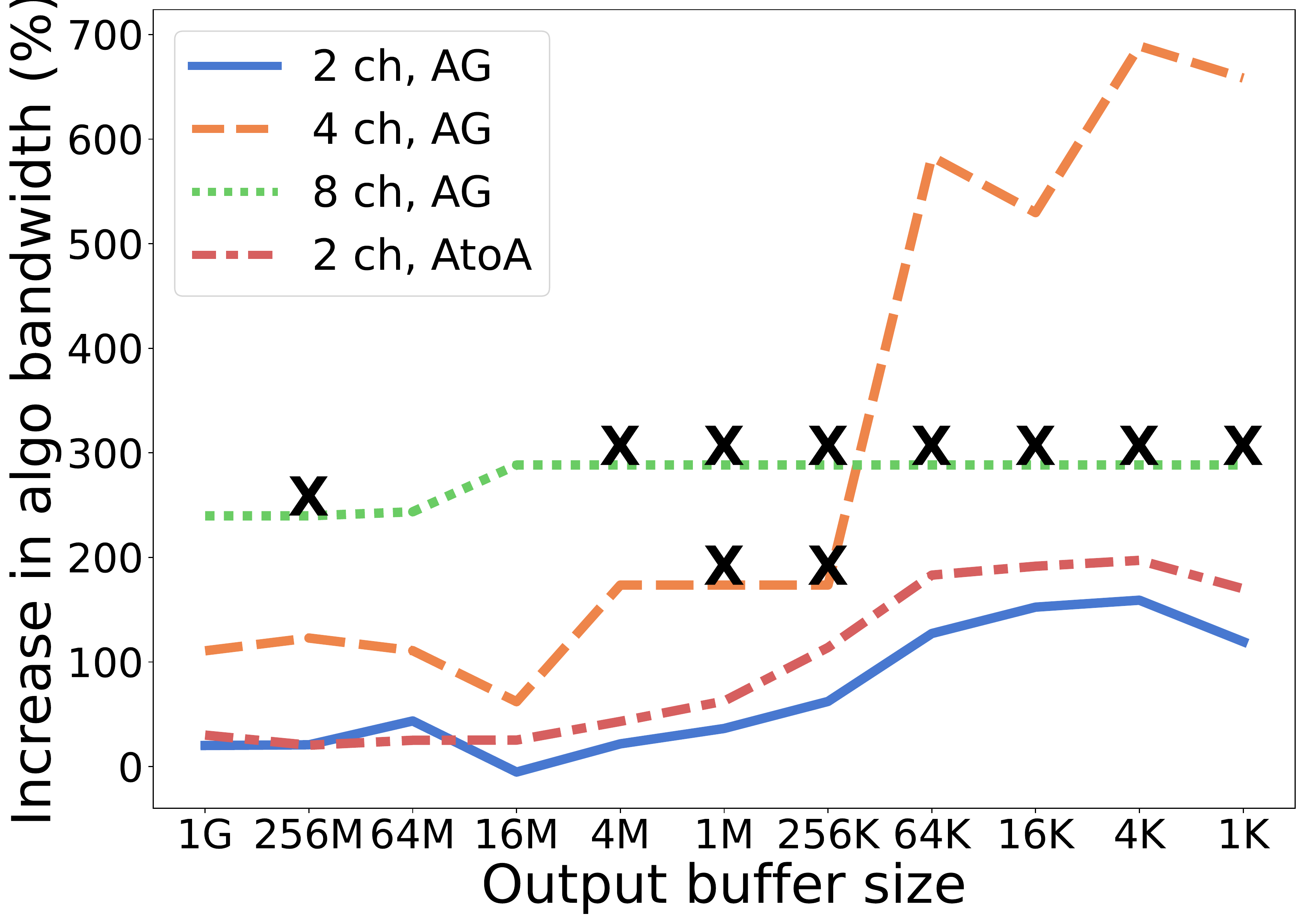}}
		\caption{\internalone}
	\end{subfigure}
	\begin{subfigure}[b]{0.23\textwidth}
		\centering
		{\label{fig:internal2_allgather_taccl_bw}\includegraphics{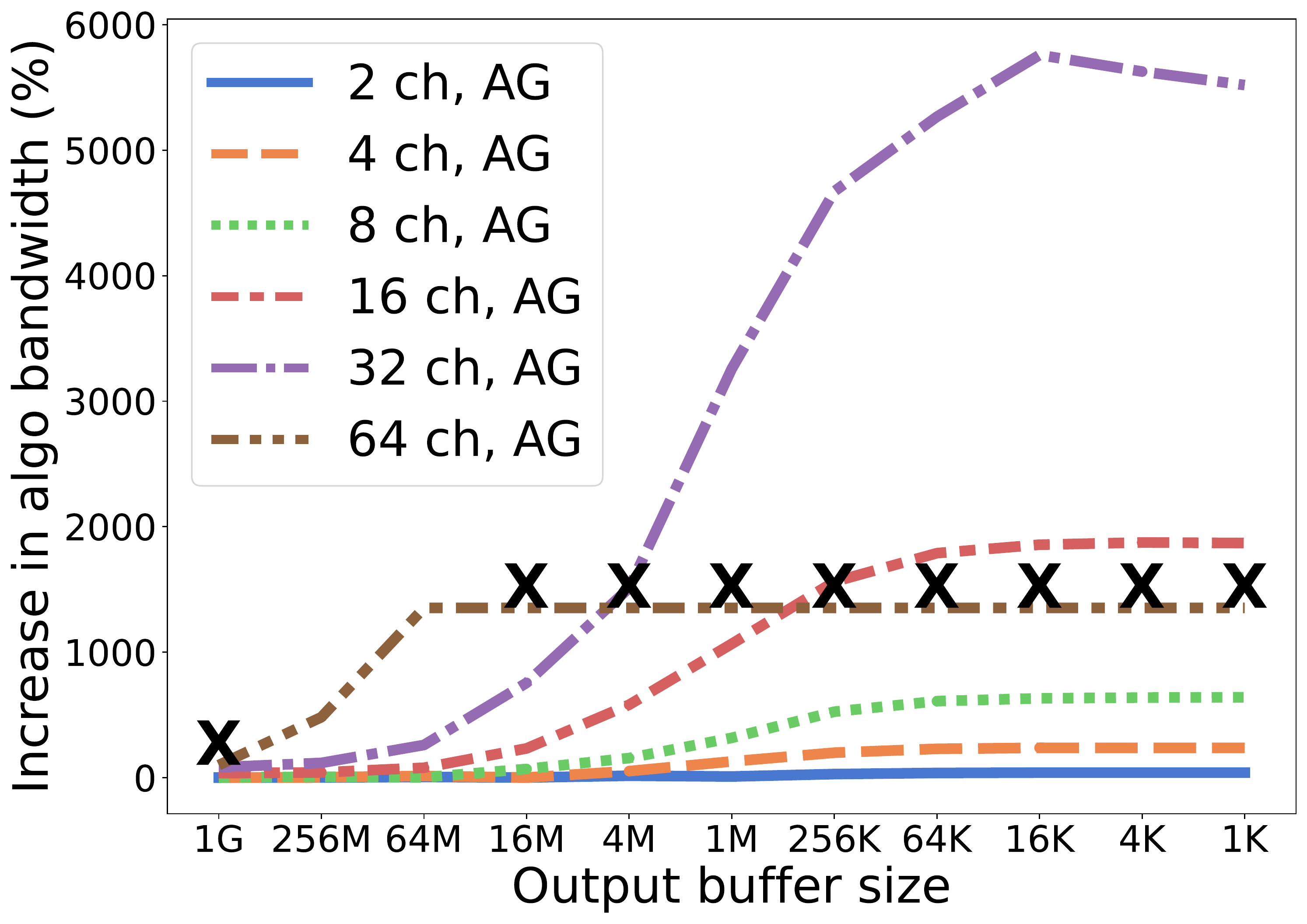}}
		\caption{\internaltwo}
	\end{subfigure}
	\caption{\label{fig:taccl-quality-allgather}Compares the algorithmic bandwidth of \sysname and TACCL ($\frac{100(TECCL - TACCL)}{TACCL}$). We mark the scenarios where TACCL is infeasible~---~which cause dips in the graph~---~using an X.}
 \vspace*{-\baselineskip}
\end{figure*}

\begin{figure*}[th]
	\centering
	\begin{subfigure}[b]{0.23\textwidth}
		\centering
		{\label{fig:ndv2_solver_time_taccl}\includegraphics{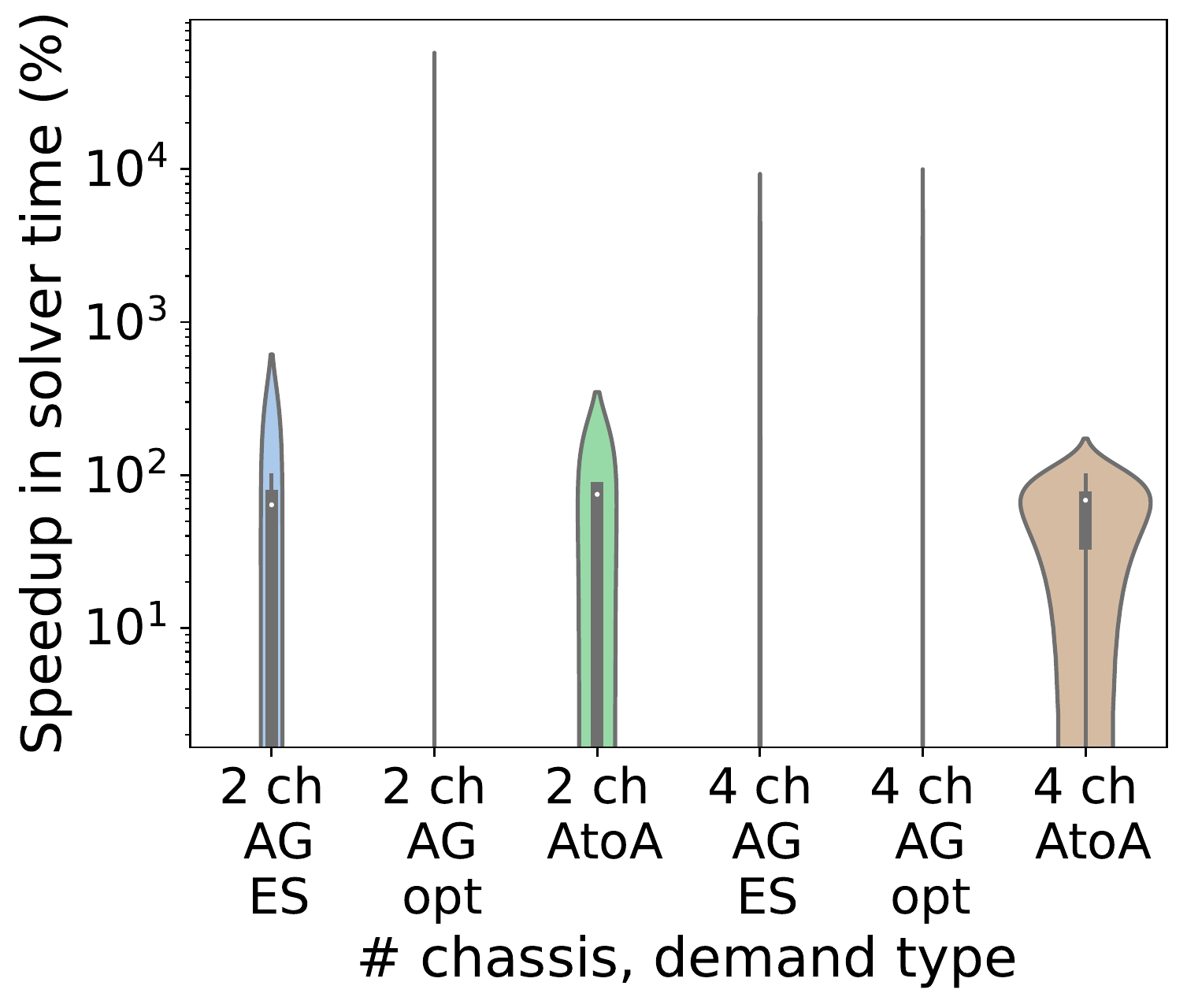}}
		\caption{\ndv}
	\end{subfigure}
	\begin{subfigure}[b]{0.23\textwidth}
		\centering
		{\label{fig:dgx2_solver_time_taccl}\includegraphics{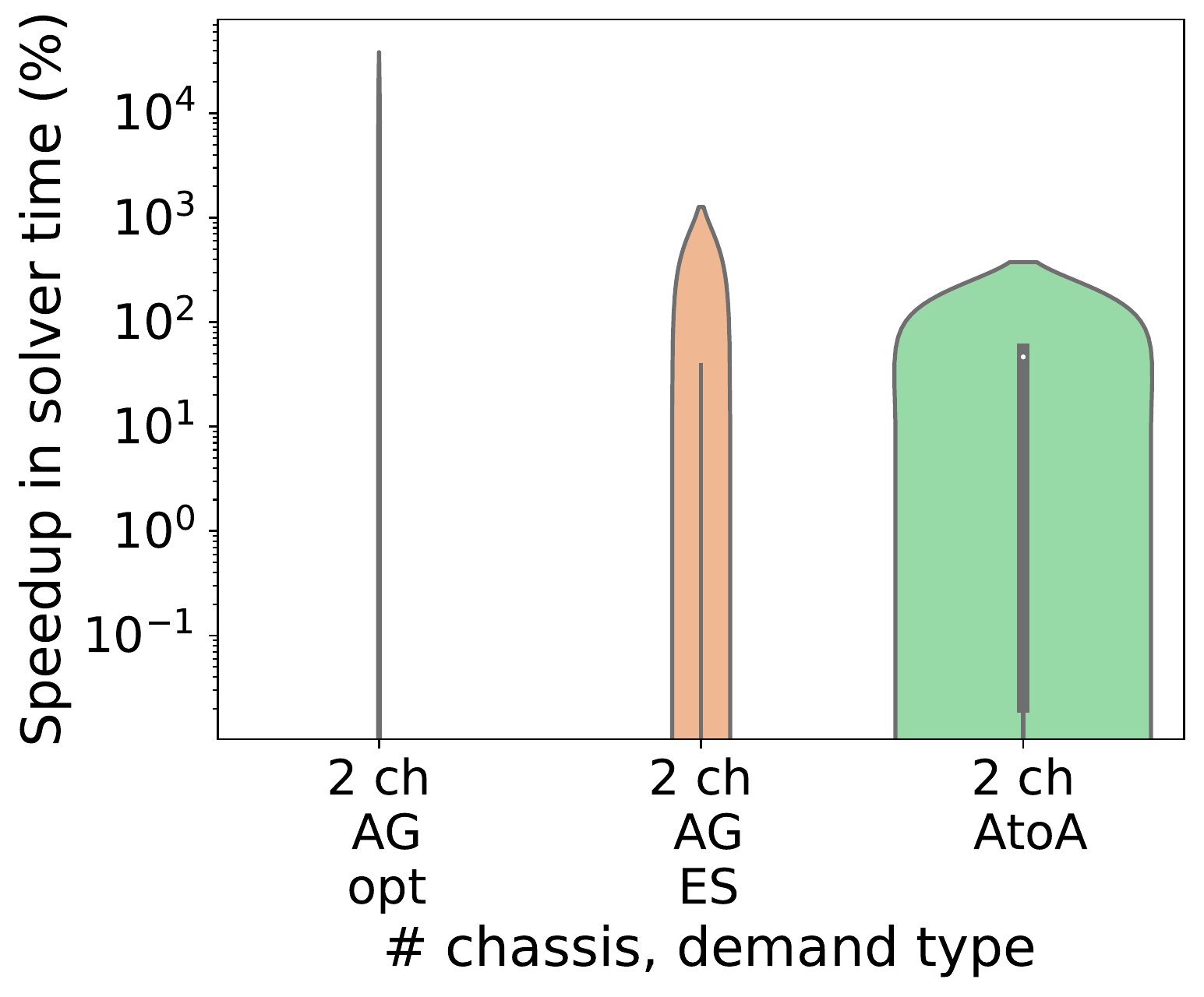}}
		\caption{\dgxtwo}
	\end{subfigure}
	\begin{subfigure}[b]{0.23\textwidth}
		\centering
		{\label{fig:internal1_solver_time_taccl}\includegraphics{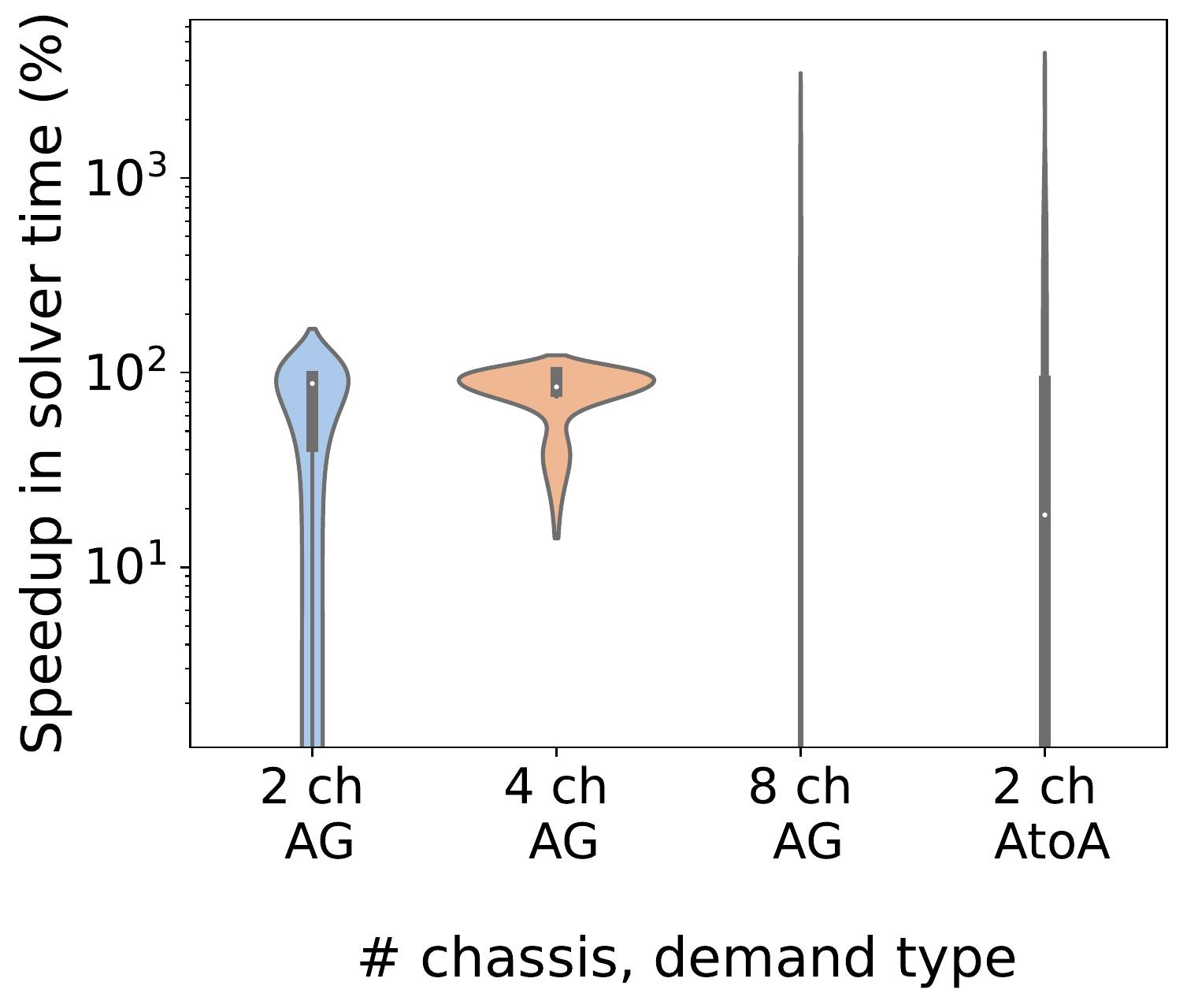}}
		\caption{\internalone}
	\end{subfigure}
	\begin{subfigure}[b]{0.23\textwidth}
		\centering
		{\label{fig:internal2_solver_time_taccl}\includegraphics{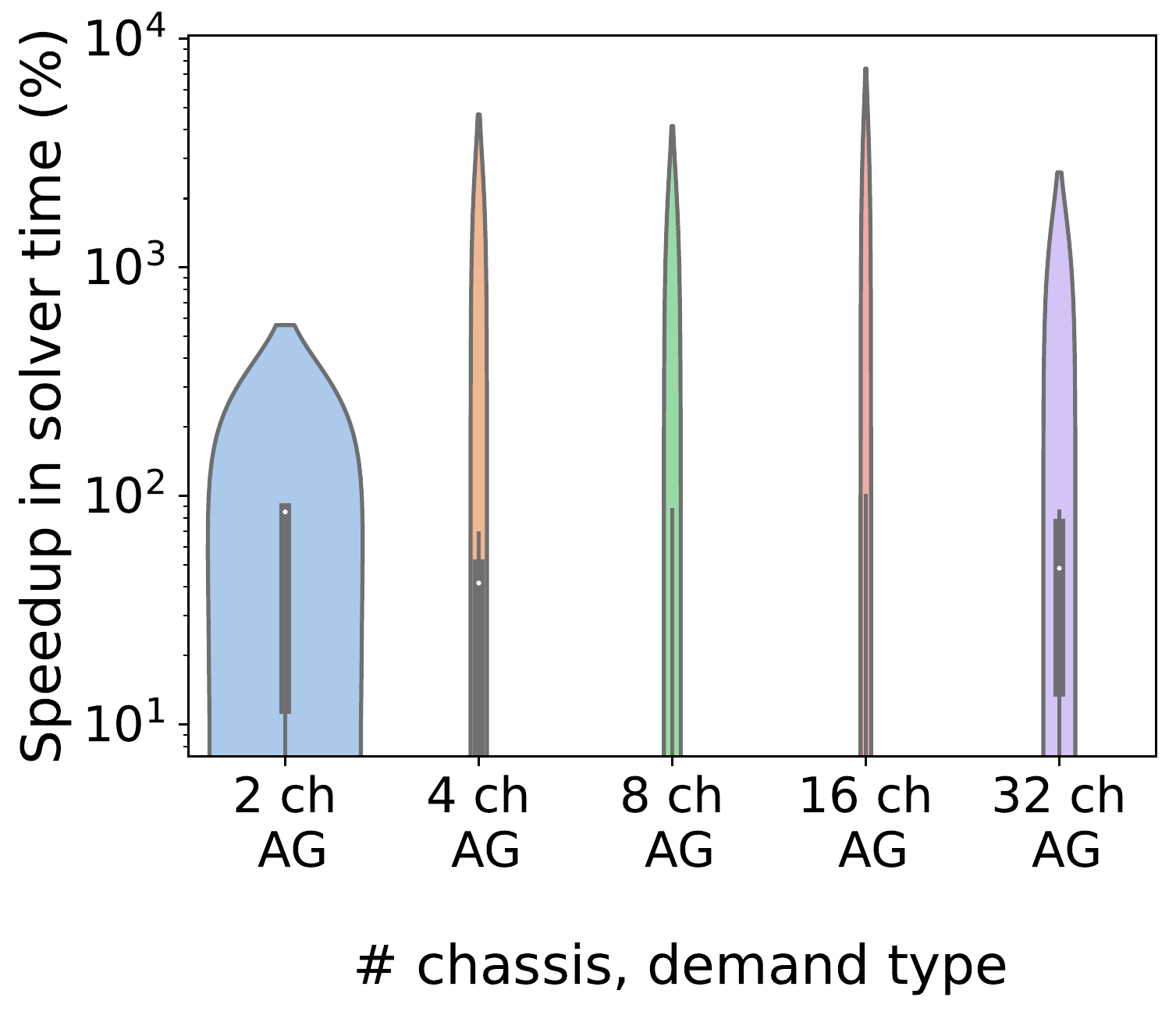}}
		\caption{\internaltwo}
	\end{subfigure}
	\caption{\label{fig:taccl-solver_time}Compares the solver time of \sysname and TACCL ($\frac{100(TECCL -  TACCL)}{TACCL}$). Ch stands for chassis, ES early stop, AG \allgather, and AtoA \alltoall. We use log scale for the y-axis to improve resolution. \sysname is faster than TACCL on $45\%$ of \alltoall scenarios and $40\%$ of \allgather scenarios (with early stop) on the \ndv topology; $72\%$ and $27\%$ for \dgxtwo; $72\%$ and $83\%$ for \internalone; and $100\%$ and $50\%$ for \internaltwo.}
 \vspace*{-\baselineskip}
\end{figure*}

We also compare with SCCL's instance solution (due to space constraints, we show the results in the \cref{sec:instance_sccl}). To create an apples-to-apples comparison, we use the number of rounds in SCCL for $\mathtt{K}$ in \sysname~---~since SCCL is no longer running an optimization~---~and use $\alpha = 0$ (this is necessary as our model will need more epochs otherwise to account for $\alpha$). We use the scenarios from Table 4 in SCCL~\cite{SCCL} and run both solvers on a desktop with 6 cores and $32$ GB RAM. SCCL failed to produce a solution for \allgather workloads with more than $1$ chunk even after 3 days. \sysname runs faster than SCCL in almost all cases and even improves SCCL's solution quality by $33\%$ in the \alltoall scenario. \sysname is slower than SCCL in one instance ($(6,7)$): this is because in \sysname we solve for the optimal number of epochs, and we use a value for $\mathtt{K}$ that is too tight~---~we can reduce the solver time to $11$ seconds by increasing $\mathtt{K}$ to $20$ (the quality of the solution does not change). We can use the $A^{*}$ technique to speed up the solution further.

To fully highlight our runtime advantage over SCCL, we ran an \alltoall demand with $8$ chunks using both solvers: SCCL timed out after $10032.7$s and did not produce a schedule, whereas ours finished in $1.88$s with a valid schedule that finished the transfer in $21\mu$s (for $25$KB chunks).

\parab{TACCL.} We compare the solver time and algorithmic bandwidth of \sysname and TACCL using \allgather and \alltoall demands and on \dgxtwo and \ndv based topologies with up to $34$ nodes (a 2-chassis \dgxtwo topology has $34$ nodes) and on both internal topologies with up to $128$ nodes.
We ran all experiments on a Linux Ubuntu 20.04 VM with two Intel Xeon(R) Platinum 8380 CPUs with a total of 80-cores/160-threads and 512 GB RAM and used Gurobi 9.5.2 version as our solver.
TACCL \alltoall does not terminate for large topologies (including the 2 chassis \dgxtwo \alltoall)~---~ we use a timeout of $2+2$ hrs or $4+4$ hrs for their routing and scheduling phases depending on the topology size. 

TACCL ran out of memory and did not produce a solution for large \internaltwo topologies (with over $64$ chassis) and for almost all \internalone topologies (with over $4$ chassis).~\cref{tab:large_topologies} reports the numbers for \sysname on $\ge 64$ nodes topologies.

TACCL scales better on the \ndv topology compared to internal topologies 1 and 2. In \ndv only 2 nodes in a chassis connect to a switch but in internal topologies 1 and 2 many nodes in a chassis are connected to a switch~---~TACCL replaces the switch with direct edges; as we increase the size of internal topologies 1 and 2 the number of such edges increases exponentially. The TACCL authors recommended we use a sketch that only uses a subset of these edges. Doing so improved the runtime for smaller topologies but TACCL still failed to produce a solution after 8 hours for larger ones.

\begin{figure}[th]
	\centering
	\begin{subfigure}[b]{0.23\textwidth}
		\centering
		{\label{fig:internal2_alltoall_solver_time_svl}\includegraphics{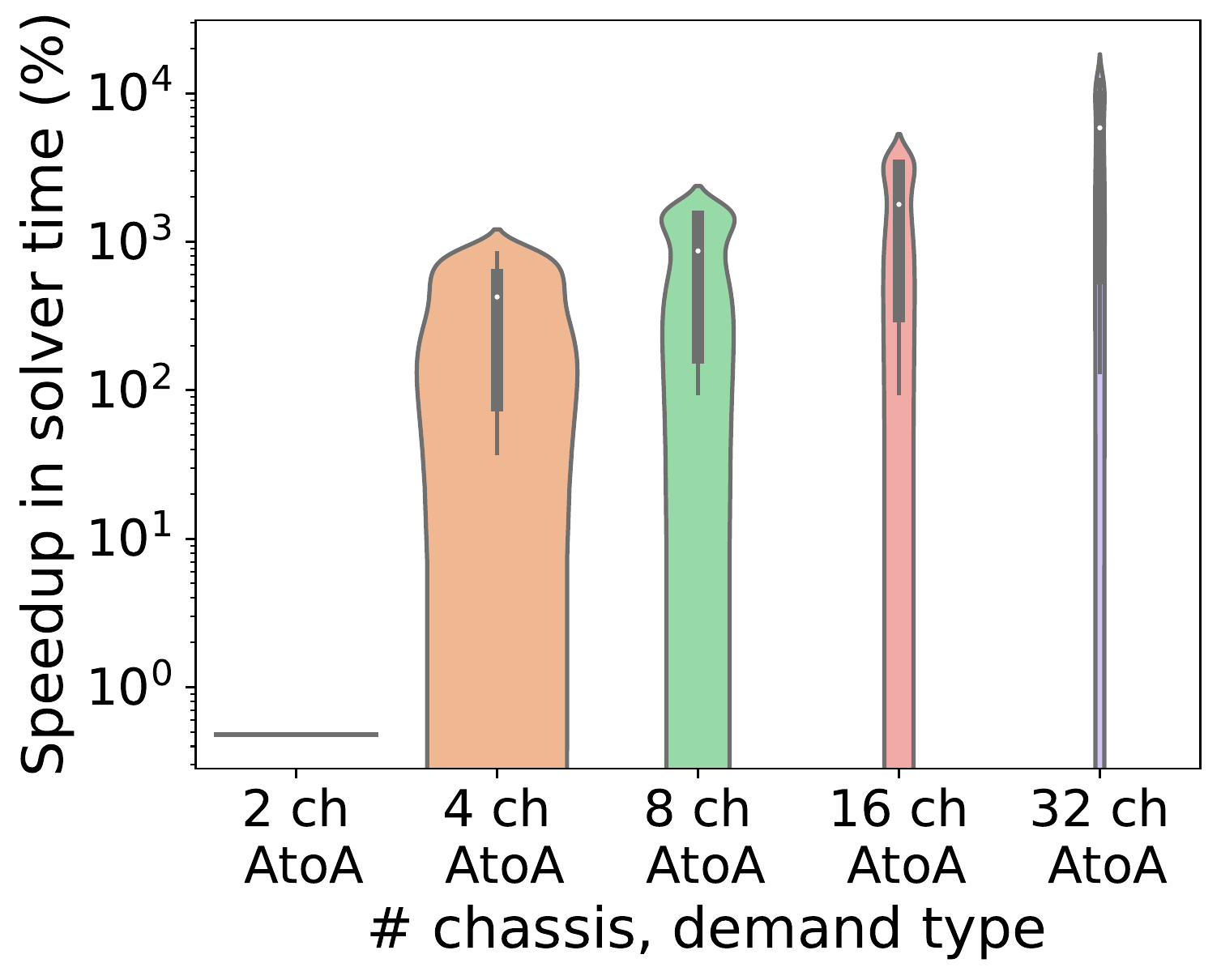}}
		\caption{Solver time}
	\end{subfigure}
	\begin{subfigure}[b]{0.23\textwidth}
		\centering
		{\label{fig:internal2_alltoall_solution_time_svl}\includegraphics{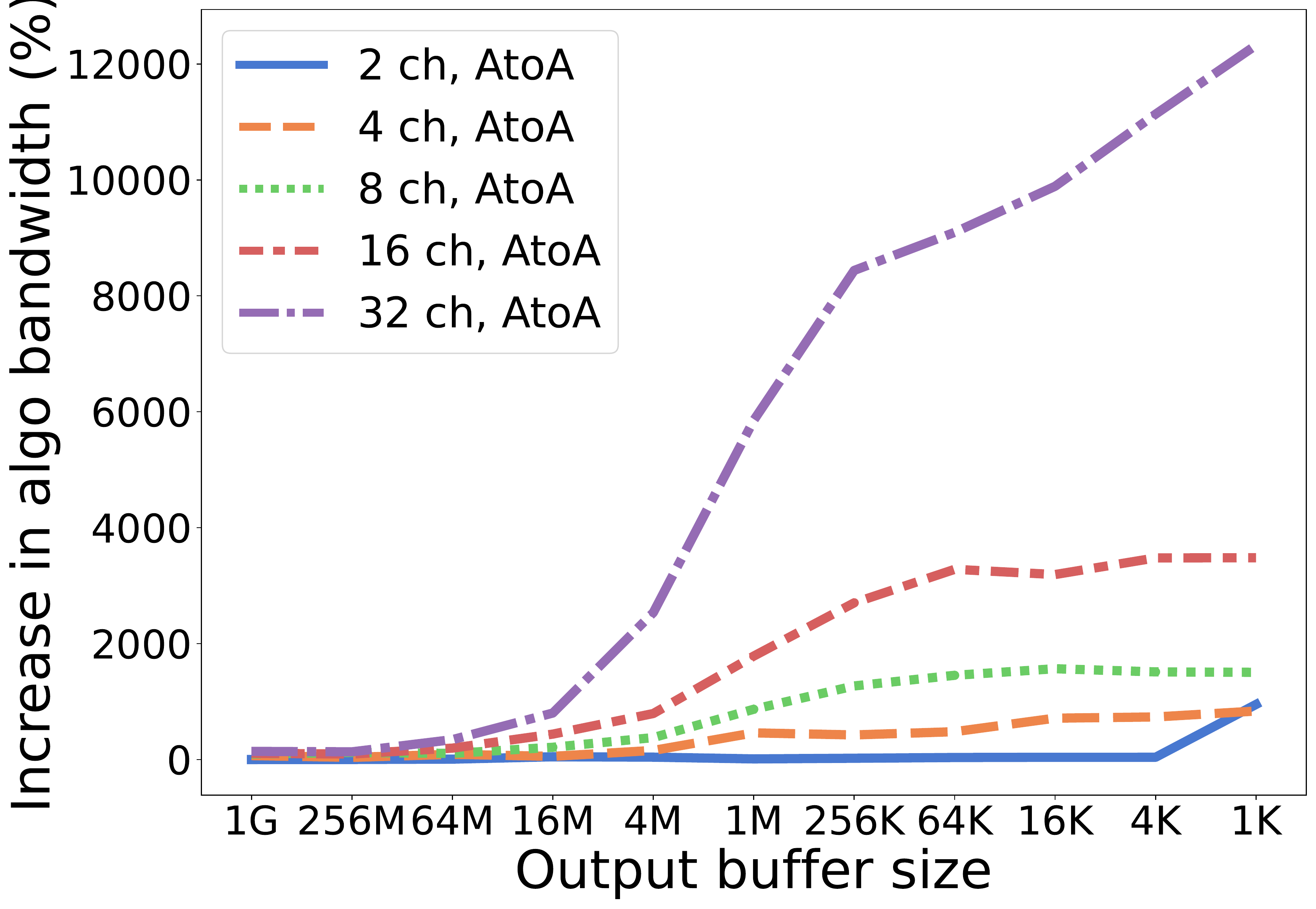}}
	   \vspace{-0.04cm}
		\caption{Transfer time}
	\end{subfigure}	
	\caption{\label{fig:internal2_alltoall} We compare TACCL and \sysname for \alltoall demands on \internaltwo with different number of chassis. \sysname is faster than TACCL in {\em all} cases and also produces higher quality solutions.}
 \vspace*{-\baselineskip}
\end{figure}

\sysname often produces higher quality solutions compared to TACCL (in some cases TACCL fails to produce a schedule and times out~---~we show those cases with an X): on \dgxtwo the improvement is at least $12\%$ and $9\%$ (maximum $471\%$ and $2979\%$) for \allgather and \alltoall respectively; on \ndv $0.36\%$ and $0.18\%$ (maximum $970\%$ and $2919\%$); on \internalone $-5\%$ and $20\%$ (maximum $689\%$ and $197\%$), and on \internaltwo, $0.33\%$ and $0.48\%$ (maximum $5759\%$ and $12322\%$). We show these results in \cref{fig:taccl-quality-allgather} and \cref{fig:internal2_alltoall} (we report \alltoall numbers for \internaltwo separately for clarity). We report the raw algorithmic bandwidths for \sysname variants in the appendix (see \cref{tab:ndv2-raw-data}) for \ndv 2 chassis as a sample.

We use Gurobi's early-stop for \allgather demands to improve \sysname's ability to scale: this does not materially impact the quality of {\sysname}'s solution~---~even with an aggressive optimality gap threshold of $30\%$~---~but allows \sysname to solve the problem faster in the \allgather scenario (we found TACCL also uses this under the hood~--~our solver time matches TACCL even when TACCL uses this feature). TACCL uses this early stop mechanism in the \alltoall case as well but we run \sysname to completion: \sysname always produces schedules that match or beat those of TACCL and in many cases it produces these schedules more quickly. We compare the two solver times in \cref{fig:taccl-solver_time}.

\subsection{Scale}
TACCL often crashes on large topologies, either due to requiring more than 400 GB RAM or memory leaks and segmentation faults. \sysname also requires a lot of memory in some cases (around $350$ GB for \alltoall on large topologies), but we can control this by changing the epoch duration to trade off the quality of the solution with the amount of memory the solver needs. \cref{tab:large_topologies} summarizes our results on large topologies and reports the scale factor (EM). We use output buffer sizes larger than 16 MB~---~as the number of GPUs increases, chunks become too small beyond this point.
We adjust the epoch size by a factor of, at most, 4 for these cases to limit memory usage.
 

\begin{table}[t]
    \renewcommand{\arraystretch}{1.05}
    \small
    \begin{tabular}{llccc}
    \textbf{Topology} & \textbf{Collective} & \multicolumn{1}{l}{\textbf{\# GPUs}} & \textbf{EM} & \textbf{Solver time} \\\hline\hline
    \rowcolor[HTML]{EFEFEF} 
    \internalone & AG (A*) & 64  & 1 & 3000 $s$ \\
    \rowcolor[HTML]{FFFFFF} 
    \internalone & AG (A*) & 128 & 1 & 7 $h$    \\
    \rowcolor[HTML]{EFEFEF} 
    \internaltwo & AG (A*) & 128 & 1 & 1300 $s$ \\
    \rowcolor[HTML]{FFFFFF} 
    \internaltwo & AG (A*) & 256 & 2 & 2.8 $h$  \\
    \rowcolor[HTML]{EFEFEF} 
    \internalone & AtoA    & 16  & 1 & 66 $s$   \\
    \rowcolor[HTML]{FFFFFF} 
    \internalone & AtoA    & 32  & 1 & 215 $s$  \\
    \rowcolor[HTML]{EFEFEF} 
    \internalone & AtoA    & 64  & 1 & 500 $s$  \\
    \rowcolor[HTML]{FFFFFF} 
    \internalone & AtoA    & 128  & 2 & 800 $s$  \\
    \rowcolor[HTML]{EFEFEF} 
    \internaltwo & AtoA    & 128 & 1 & 2600 $s$ \\
    \rowcolor[HTML]{FFFFFF} 
    \internaltwo & AtoA    & 256 & 4 & 1500 $s$
    \end{tabular}
    \caption{Large Topologies for which TACCL can't synthesize the schedule. The solver time is the average \sysname time to synthesize the schedule and EM is the epoch multiplier factor to change the epoch duration from the optimal duration for scalability.}
    \label{tab:large_topologies}
    \vspace*{-\baselineskip}
    \vspace*{-\baselineskip}
\end{table}

\subsection{Microbenchmarks}
\label{subsec:microbenchmarks}
We next evaluate our design choices:

\parab{Copy.} In-network copy is most helpful for large transfers where there is not enough capacity to transfer multiple copies directly from the source to each destination: we see in the largest transfer size ($0.21$ GB) copy reduces the transfer time by $50\%$ for \dgx, the \internalone with $\alpha = 0$ and $\alpha > 0$, and $12.5\%$ for \internaltwo. In-network copy does not help with small transfers as there is enough capacity between the source and the destinations to send multiple copies of the data directly from the source. We use $4$ chunks to complete these transfers. 

\begin{figure}[t]
	\centering
	\includegraphics[width=0.75\linewidth]{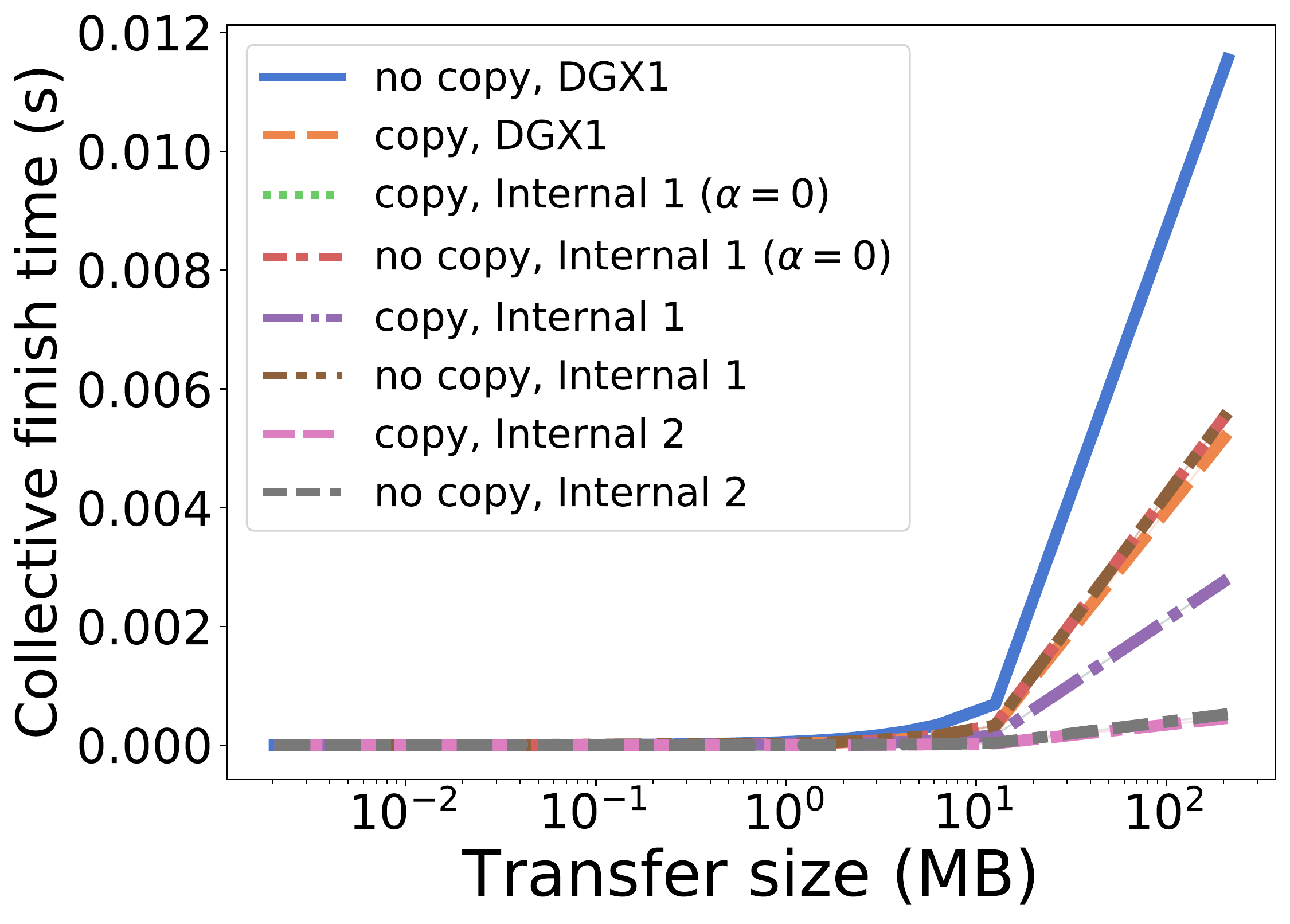}
	\caption{\label{f:nocopy} The benefit of copy: for large transfers, copy helps finish the transfer faster. }
\end{figure}

\begin{figure}[t]
	\centering
	\begin{subfigure}[b]{0.23\textwidth}
		\centering
		{\label{fig:solver_time_svl}\includegraphics{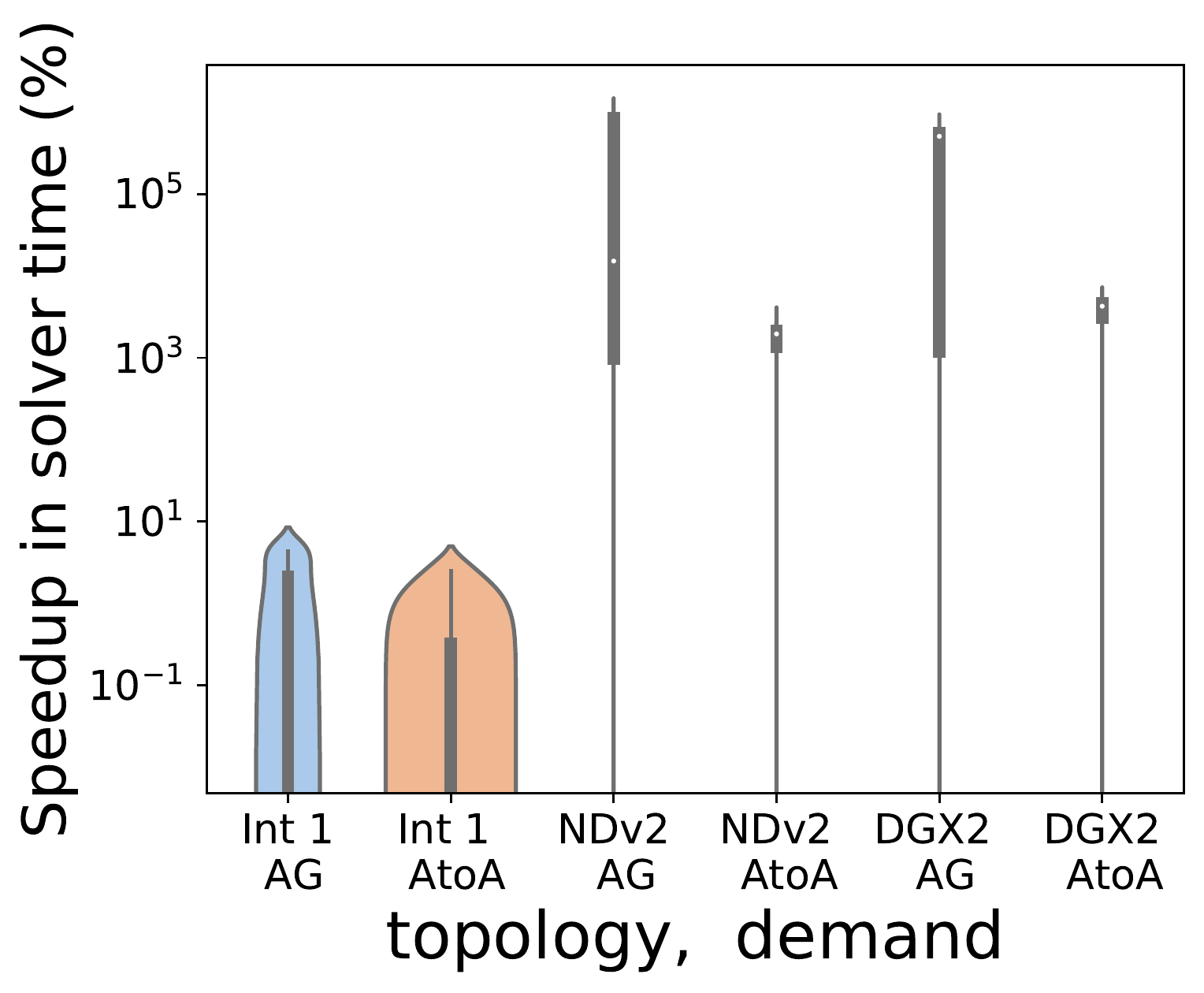}}
		\caption{Solver time}
	\end{subfigure}
	\begin{subfigure}[b]{0.23\textwidth}
		\centering
		{\label{fig:solution_time_svl}\includegraphics{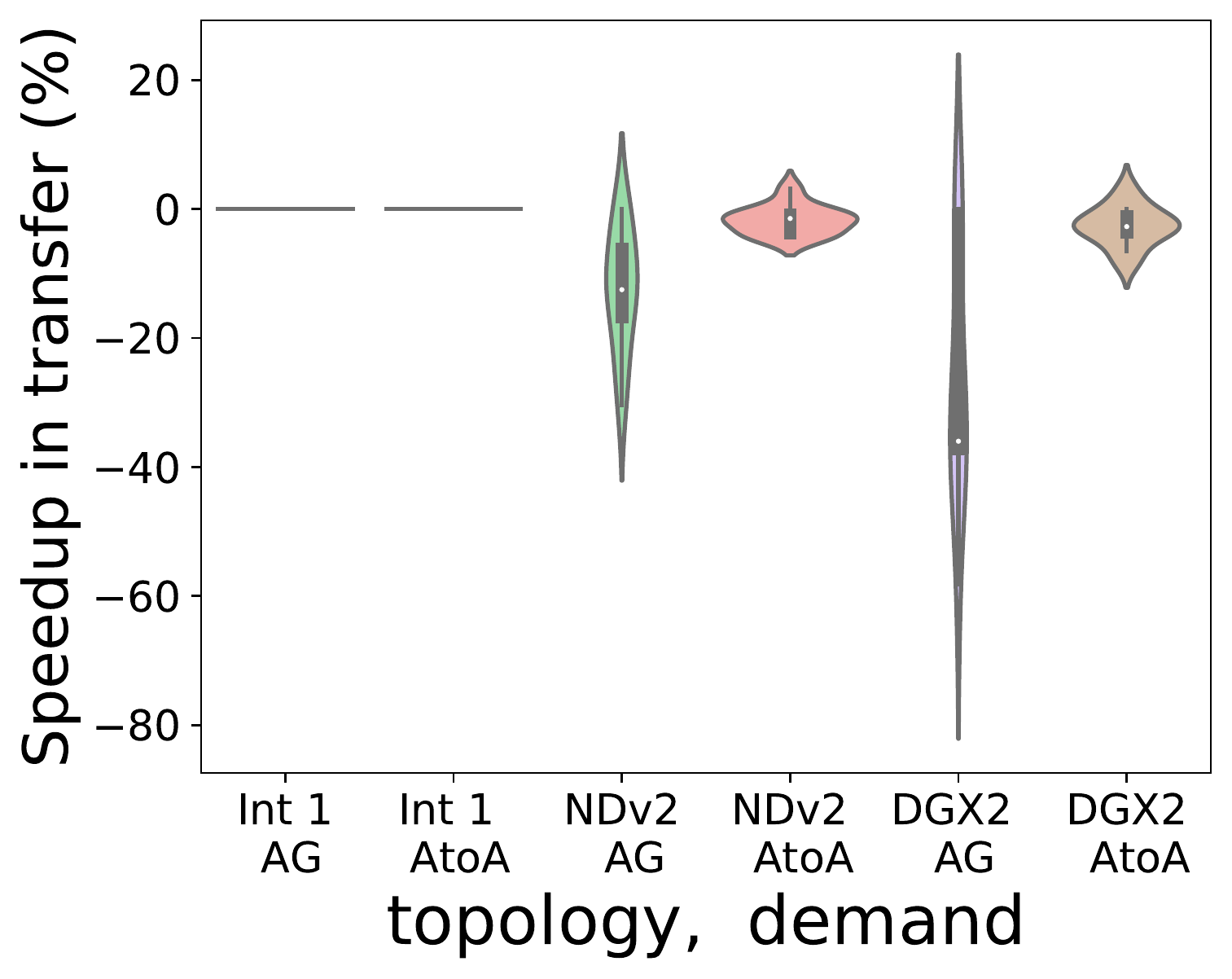}}
		\caption{Transfer time}
	\end{subfigure}
	\caption{\label{fig:epoch_duration} We compare the impact of small vs large epochs on the solver speed (a) and solution quality (b). We use 2 chassis for all topologies. Both graphs compute $\frac{100(\text{small} - \text{large})}{\text{large}}$. The solver finds a solution faster with large epochs but produces better quality solutions with small ones.}
 \vspace*{-\baselineskip}
\end{figure}

\parab{Small vs large epochs.} We investigate how the duration of epochs impacts the solver speed and the quality of the solution (\cref{fig:epoch_duration} where we use 2 chassis for each topology). In \allgather we only allow chunks to traverse one link in a single epoch: the length of the longest path dominates the transfer time when we use large epochs because the length of the epoch is too large compared to how long it takes for the chunk actually to traverse the link (on faster links). We see this more predominantly in the \ndv and \dgxtwo topology where the fast links have $4\times$ higher bandwidth (large epoch duration is, therefore, $4\times$ small epoch duration) compared to slower ones. In contrast, we do not see a difference on \internalone, where the links are mostly homogeneous.

\begin{figure}[t]
	\centering
	\begin{subfigure}[b]{0.23\textwidth}
		\centering
		{\label{fig:solver_time_withandwithoutbuffers}\includegraphics{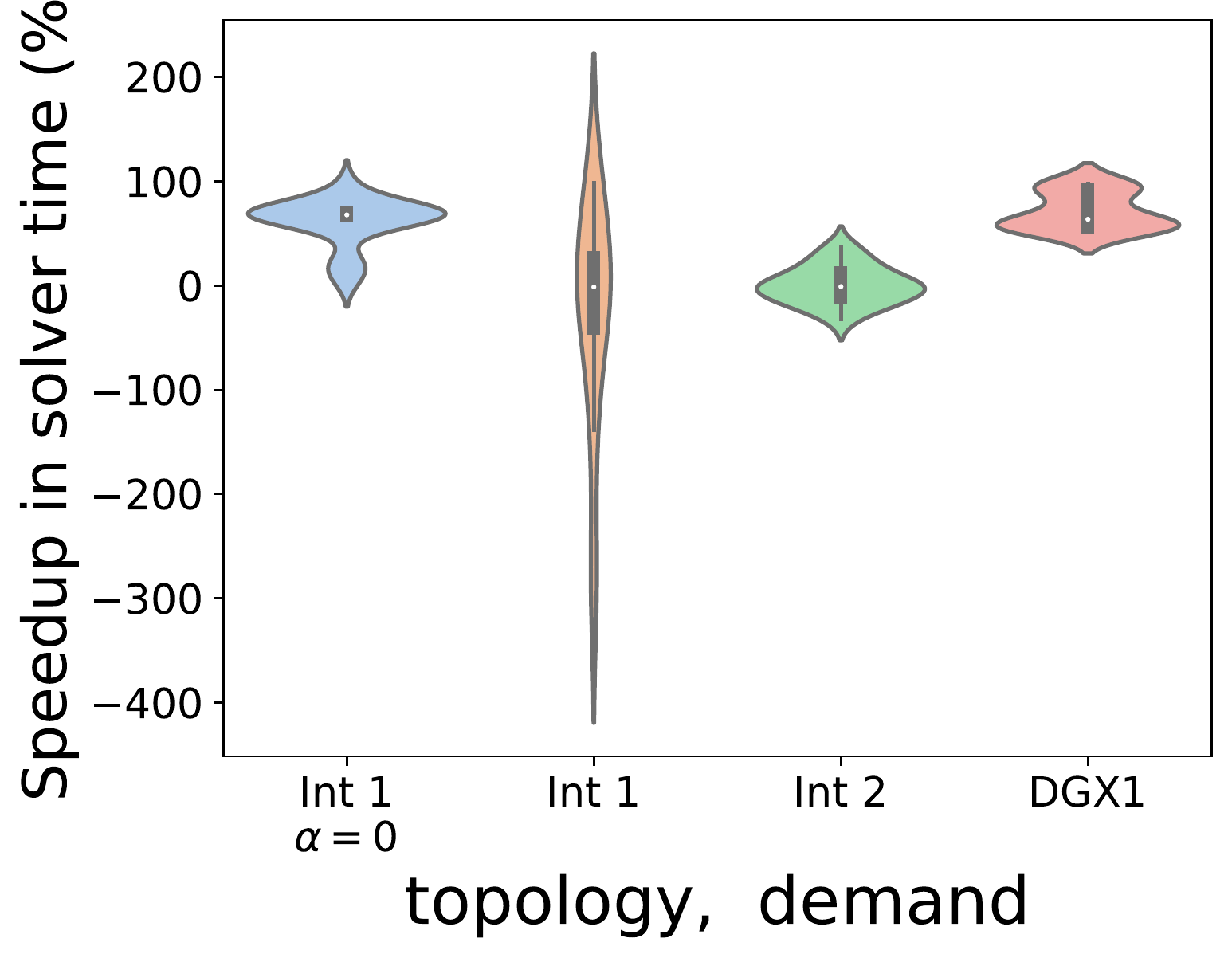}}
		\caption{Solver time}
	\end{subfigure}
	\begin{subfigure}[b]{0.23\textwidth}
		\centering
		{\label{fig:solution_time_withandwithoutbuffers}\includegraphics{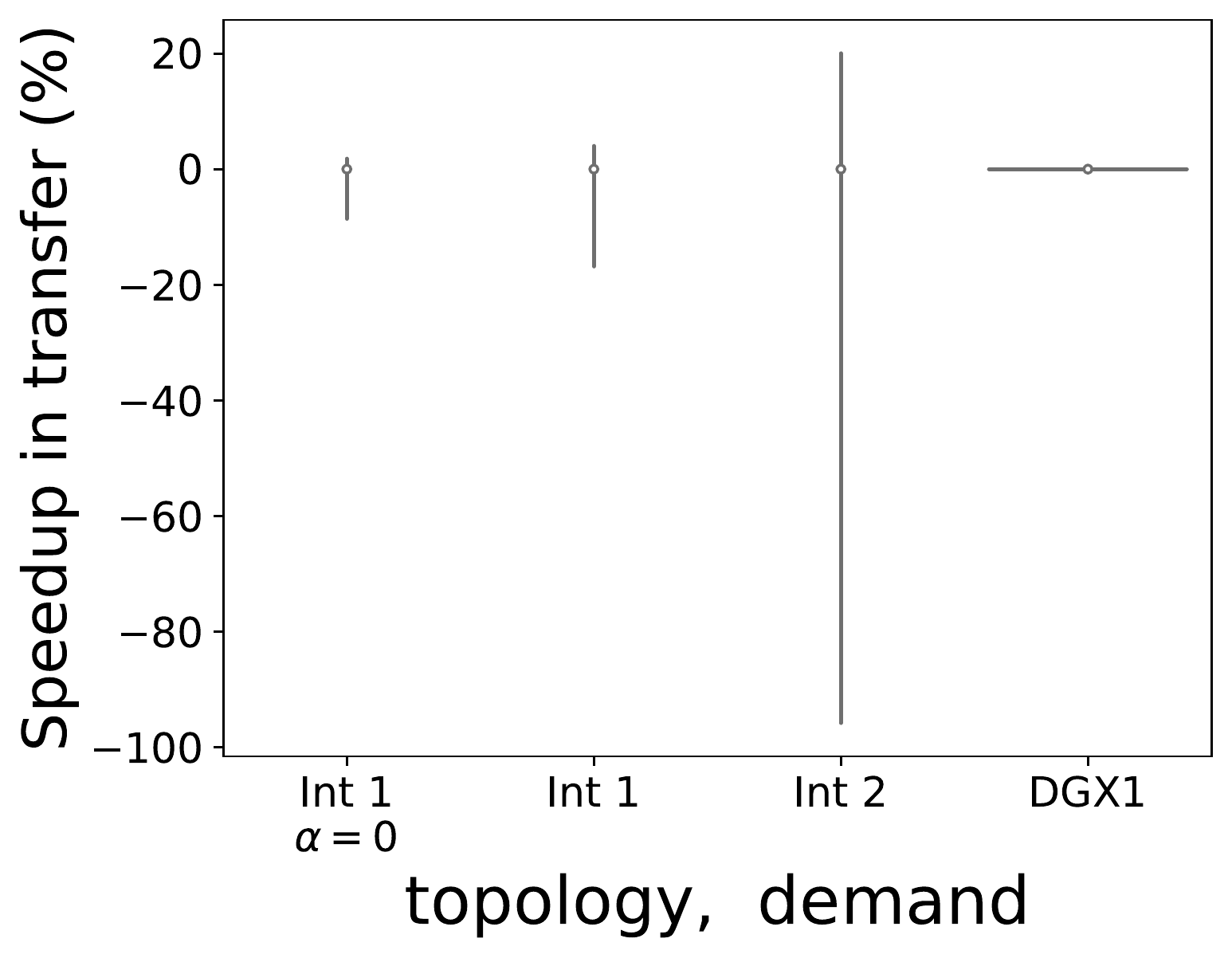}}
		\caption{Transfer time}
	\end{subfigure}
	
	\caption{\label{fig:withandwithoutbuffers} We evaluate the impact of buffers on (a) and solution quality (b) solver time. We use 2 chassis for all topologies. Both graphs compute $\frac{100(\text{without buffers} - \text{with buffers})}{\text{without buffers}}$. Buffers don't impact the solution quality in most cases, but only the solver times! The average speedups in solver time are: $61\%$, $-28.46\%$, $0.23\%$, $71\%$ for \internalone without $\alpha$, \internalone with $\alpha$, \internaltwo, and \dgx respectively.}
\end{figure}

\parab{Store and forward.} We find a somewhat surprising result. Buffers don't impact the solution quality but only the solver time (\cref{fig:withandwithoutbuffers})! This is because of the nature of collective demands such as \allgather and \alltoall: because each node needs the same amount of traffic as it has to forward it can
 interleave consuming traffic with forwarding it to compensate for the lack of buffers. But in the presence of buffers the feasible space of solutions is larger which in many cases enables the solver to find the optimal solution more quickly (the speedup is $71\%$ and $61\%$ for \internalone and \dgx respectively). We believe it is possible to formally prove this result but defer this proof to future work.

\parab{$A^{*}$ vs OPT.} We compared the quality of the $A^{*}$ technique to the optimal on a $16$-chassis \internaltwo topology with both $\alpha > 0$ and $\alpha = 0$. We used both single chunk and $2$ chunk transfers. 

When $\alpha = 0$, $A^{*}$ finished in $86.61$s ($263.29$s for $2$ chunk demands) whereas the optimal took $346$s ($4392$s for two chunks). The optimal solution was $10\%$ better than $A^{*}$ ($6\%$ in the $2$ chunk case)~---~transfer times were $3.48$s vs $3.89$s. 

The results are similar when $\alpha > 0$: $A^{*}$ finished in $137.02$s ($901.25$s for the $2$ chunk case) whereas the optimal took $363.40$s ($3047$s). The optimal solution was $20\%$ better ($8\%$ in the $2$ chunk case).

%% file: relatedwork.tex
\section{Related work}
\label{sec:relatedwork}

\sysname provides a scalable method for collective communication optimization by using a network flow-based approach. Our solution supports unsustained demands, store-and-forward, and copy. Our work builds on prior work both in network traffic engineering and in collective optimization:

\noindent\textbf{Multi-cast TE.} Prior works have looked at traffic engineering for multi-cast networks~\cite{doar1993multicast,noronha1994optimum}. Oliveira and Pardalos~\cite{oliveira2005survey} provide a comprehensive summary of these works. Blink~\cite{Blink} used these techniques to optimize collective communication but does not model delay and store-and-forward. 

\noindent \textbf{WAN TE.} Many prior works in networking use the network flow model to scalably route traffic in wide area networks~\cite{SWAN, B4, NCFlow,PoP}. However, most of these works assume sustained demands, copy, and store-and-forward. Among these works, Calendaring~\cite{kandula2014calendaring} provides a solution that models unsustained demands. NetStitcher~\cite{netstitcher} adds to this the support for store and forward but assumes flows do not compete for bandwidth. Neither of these works simultaneously model copy, store-and-forward, and delay.

\noindent \textbf{Prior work on collective communication optimization.} Many prior work have tackled the collective communication optimization problem~\cite{SCCL, TACCL, Blink, zhao2022optimal, rashidi2022themis}. We find these solutions do not scale to the topologies and data sizes we have in production today and those we anticipate for the future. TACCL is the most scalable of these solutions, but it has trouble scaling when it sends more than 1-2 chunks, and is sub-optimal. Work such as~\cite{mahajanbetter, wang23topoopt, zhao2022optimal} aims to co-optimize either topologies and parallelization strategies (\cite{wang23topoopt}) or collective scheduling and execution planning~\cite{mahajanbetter}. These works rely on collective communication optimizers as part of their search but do not provide optimal solutions to the problem themselves --- they can use \sysname as part of their search. Our work is complementary to these works.   

%% file: conclusion.tex
\section{Conclusion}
We presented \sysname: a scalable collective communication optimizer that models the problem through a TE-based approach. We provide three algorithms to solve this problem: the MILP approach which optimally solves the general collective communication optimization problem and supports multi-cast; the LP form which is also optimal and much more scalable but removes support for multi-cast; and finally the $A^{*}$-based approximation method which is much more scalable than the MILP technique and continues to support multi-cast but is no longer optimal. We show our solution outperforms prior, state-of-the-art, techniques such as SCCL and TACCL by over $2\times$.

%% file: appendix.tex
\appendix

\section{Initialization and termination constraints}
\label{sec::initialization}

We introduced the main constraints for the MILP and LP formulations in \cref{sec::solution} and \cref{sec::partial}. However, we need to add a few additional constraints to initialize and terminate these problems. 

\noindent\textbf{The first epoch.} We use buffers to indicate when the node has a specific chunk. In the first epoch of the MILP we initialize the source buffers as follows:

\begin{align*}
B_{n,n,0,c} =  \max_{d \in N} D_{n,d,c} & \quad \forall n\in N, \forall c \in C    \\
 B_{s,n,0,c} =  0  & \quad \forall s,n\in N: s\neq n, \forall c \in C 
\end{align*}

We no longer need to buffer chunks we have already sent out in the LP form and therefore these equations become:

\begin{align*}
B_{s,n,0} + \sum_{\forall j: (n,j)\in E}F_{s,n,j,0} = \sum_{\forall c \in C, \forall d \in N}D_{s,d,c} & \;\forall s,n \in N: s,n \notin S 
\end{align*}

\noindent \textbf{The last epoch.} In the LP we do not need to buffer chunks if they are not going to be forwarded. Nodes also don't need to send out any traffic after this epoch. Therefore, in the last epoch of the LP we have:

\begin{align*}
\forall s,n \in N: s\neq n, n \notin S \quad \sum_{\forall j: (j,n) \in E} F_{s,j,n, (\mathtt{K} -\lceil{\frac{\alpha_{j,n}}{\tau}}{\rceil})} =  \mathsf{R}_{s,n, \mathtt{K}} 
\end{align*}

\section{Modeling limited buffers}
\label{sec:limitedbuffers}

\noindent\textbf{In the MILP.} To model limited buffers in the MILP we need to change the buffer constraints to track which chunks to remove from the buffer and in which epoch. Hence, we introduce a new variable $X_{s,n,k,c}$ which encodes whether we should remove chunk $c$ from node $s$ from the buffer at node $n$ in epoch $k$. The buffer constraints become:

\begin{align*}
& \text{Buffer constraints} \big(s,n,k,c) \triangleq \nonumber \\
& B_{s,n, k,c } = B_{s,n,k-1,c} - X_{s,n,k - 1,c} + \sum_{\forall j \mid (j,n) \in E} F_{s,j,n,k - \lceil\delta_{jn}\rceil - 1, c}.
\end{align*}

To enforce the limit on the buffer size, we add the constraint:

\begin{align*}
\sum_{s,c} B_{s,n,k,c} \le \mathcal{L} \quad \forall n \in N, \forall k \in K,
\end{align*}

where $\mathcal{L}$ is the limit on the buffer size. We impose no limit on the auxiliary variable $X_{s,n,k-1,c}$ as the algorithm can choose to re-buffer a chunk at a node at any point in time and again remove it later.

\noindent \textbf{In the LP.} The LP removes from the buffer what it sends out on a link. Hence to use limited buffers we only have to impose an upper bound on the sum of the buffer variables at a node:

\begin{align*}
\sum_{s} B_{s,n,k} \le \mathcal{L} \quad \forall n \in N, \forall k \in K
\end{align*}

\section{Modeling legacy switches}
\label{sec:legacy}

For switches that don't support copy, we use an approach similar to TACCL's hyper-edges. We remove the switch from the topology and replace it with direct links between all pairs of GPUs that were connected through the switch. We now need to account for the capacity to and from the switch: this translates to a upper bound on the number of hyper-edges we can use simultaneously in each epoch.

We augment our notation with the variables in \cref{table:allgather_direct}. We need to add a constraint to the problem that enforces we can only use a subset of the hyper-edges: the minimum of the number of edges that come into the switch and go out of it. This constraint is as follows:

{
\small
\begin{align*}
 \sum_{\forall n \in N, \forall c \in C, \forall (i, j) \in \Omega(s)} F_{n,i,j,k,c} \leq \min(|\{(s, x) \in E \}|, |\{(y, s) \in E \}|) \nonumber \\
 \forall k \in K, \forall s \in S    
\end{align*} 
}
Each node $i$ can only send (receive) traffic on one of its  outgoing (incoming) hyper-edges:

\begin{align*}
\forall k \in K, \forall i \in N, \forall s \in S \quad    \sum_{\forall n \in N, \forall c \in C, \forall (i, j) \in \Omega(s)} F_{n,i,j,k,c} \leq 1\\
\forall k \in K, \forall i \in N, \forall s \in S \quad    \sum_{\forall n \in N, \forall c \in C, \forall (j, i) \in \Omega(s)} F_{n,j,i,k,c} \leq 1.
\end{align*}

We only need to use this model in the general MILP form to ensure the solution can scale --- the LP model already assumes none of the nodes copy traffic.

\begin{table*}[h!]
	\centering
	\begin{tabularx}{\linewidth}{ l X  }
		\textbf{Notation} & \textbf{Description}  \\ 
		\hline
		\hline
		$\Gamma$ & The function to get non-switch set of edges from the set of edges ($\Gamma: E \to E'$). Therefore, $E' \subseteq 2^{N-S \times N-S}$ and  $(i, j) \in E' \implies (i,j) \in E \land i, j \notin S.$\\
		$\Omega$ & The function from a switch node to the set of  direct-connect edges ($\Omega: S \to 2^{N-S \times N-S}$). $\Omega(s) = \{(i, j) | (i, s) \in E \land (s, j) \in E  \land (i, j) \notin E\}$ \\
		$L$ & The set of edges in the transformed graph ($L = \Gamma(E) \cup \bigcup_{s\in S}\Omega(s) $). \\
	\end{tabularx}
	\caption{Additional notation we need to model legacy switches.}\label{table:allgather_direct}
\end{table*}

\section{The $A^{*}$ technique}
\label{sec:Astar}

In the $A^{*}$ based approach we split the problem into multiple time partitions (or rounds). Our goal in each round is to get the chunks closer to the destination. We solve each of these rounds sequentially until we satisfy all the demands. 

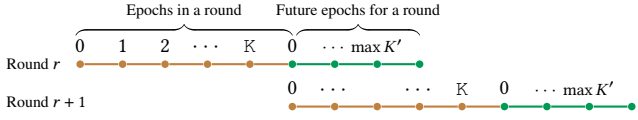
\begin{figure}
	\centering
		\begin{tikzpicture}[
                roundnode/.style={circle, fill=brown, minimum size=1mm, inner sep=0},
                futurenode/.style={circle, fill=ForestGreen, minimum size=1mm, inner sep=0},
                node distance=0.45cm and 0.45cm,
                edge/.style = {color=brown, thick},
                edge2/.style = {color=ForestGreen, thick},
                ] 
                \footnotesize
                \node[roundnode, label=above:{$0$}]        (0)                    {};
                \node[roundnode, label=above:{$1$}]        (1)    [right=of 0]    {};
                \node[roundnode, label=above:{$2$}]        (2)    [right=of 1]    {};
                \node[roundnode, label=above:{$\cdots$}]        (dot)  [right=of 2]    {};
                \node[roundnode, label=above:{$\texttt{K}$}]        (K)    [right=of dot]    {};
                \node[futurenode, label=above:{$0$}]        (f0)    [right=of K]    {};
                \node[futurenode, label=above:{\tiny$\cdots$}]        (fdots)    [right=of f0]    {};
                \node[futurenode, label=above:{\tiny$\max{K'}$}]        (fK)    [right=of fdots]    {};
                \node[futurenode]        (fl)    [right=of fK]    {};

                \node[roundnode, label=above:{$0$}]        (r0)    [below=of f0]                {};
                \node[roundnode, label=above:{$\cdots$}]        (r1)    [right=of r0]    {};
                \node[roundnode, label=above:{}]        (r2)    [right=of r1]    {};
                \node[roundnode, label=above:{$\cdots$}]        (rdot)  [right=of r2]    {};
                \node[roundnode, label=above:{$\texttt{K}$}]        (rK)    [right=of rdot]    {};
                \node[futurenode, label=above:{$0$}]        (rf0)    [right=of rK]    {};
                \node[futurenode, label=above:{\tiny$\cdots$}]        (rfdots)    [right=of rf0]    {};
                \node[futurenode, label=above:{\tiny$\max{K'}$}]        (rfK)    [right=of rfdots]    {};
                \node[futurenode]        (rfl)    [right=of rfK]    {};

                \draw[edge] (0) -- (1);
                \draw[edge] (1) -- (2);
                \draw[edge] (2) -- (dot);
                \draw[edge] (dot) -- (K);
                \draw[edge] (K) -- (f0);
                \draw[edge2] (f0) -- (fdots);
                \draw[edge2] (fdots) -- (fK);
                \draw[edge2] (fK) -- (fl);

                \draw[edge] (r0) -- (r1);
                \draw[edge] (r1) -- (r2);
                \draw[edge] (r2) -- (rdot);
                \draw[edge] (rdot) -- (rK);
                \draw[edge] (rK) -- (rf0);
                \draw[edge2] (rf0) -- (rfdots);
                \draw[edge2] (rfdots) -- (rfK);
                \draw[edge2] (rfK) -- (rfl);

                \node (rx) [left=0.1cm of 0] {\tiny Round $r$};
                \node (ry) [below=0.5cm of rx.west,anchor=west] {\tiny Round $r+1$};
                \draw [decorate,decoration = {calligraphic brace,amplitude=4pt}, line width=0.2mm] ($(0.west) + (0,11pt)$) --  ($(f0) + (0,11pt)$) node[black, midway, yshift=9pt] {\tiny Epochs in a round};
                \draw [decorate,decoration = {calligraphic brace,amplitude=4pt}, line width=0.2mm] ($(f0.east) + (0,11pt)$) --  ($(fl) + (0,11pt)$) node[black, midway, yshift=9pt] {\tiny Future epochs for a round};
		\end{tikzpicture}
	\protect\caption{$A^*$ time progression between rounds}
	\label{fig:time}
\end{figure}

The delay on each link (\ie $\alpha_{ij}$) means some chunks we send on link $(i,j)$ in a particular round may arrive at node $j$ in a subsequent round. We use the set $K'$ to denote all subsequent rounds and $Q_{s,c,i,k',r}$ to denote the chunks that arrive in these rounds to account for this (\cref{fig:time}). To keep things simple, we choose to set the number of epochs in a round in a way that ensures chunks are only delayed by a single round at most. This means the total duration of the round is greater than the largest link delay. However, users can choose to use shorter chunks --- they will have to maintain more state between rounds in that case.


\begin{table*}[h!]
	\centering
	\begin{tabularx}{\textwidth}{ l X  }
		\textbf{Variable} & \textbf{Description}  \\ 
		\hline
		\hline
		$R$ & The set of rounds ($R = \{0, 1,2, \ldots \mathtt{R}\}$) \\
		$K$ & The set of epochs in a round ($K = \{0, 1, 2, \ldots, \mathtt{K}\}$). The number of epochs in a round is constant and does not change with the round. \\
		$K'$ & The set of future epochs relevant for a round ($K' = \{0, 1, 2, \ldots, \max_{\forall  (i, j) \in E} \lceil\frac{\alpha_{i,j}}{\tau}\rceil\}$) \\
		$D$ & The demand function ($N \times N \times C \to \{0, 1\}$) where $D_{s,d,c,r}$ represents whether destination $d$ wants chunk with id $c$ from node $s$ at the start of round $r$ \\
		$F_{s,c, i, j, k, r}$ & (boolean) whether chunk $c$ of source $s$ is going over link $(i,j) \in E$ at epoch $k$ in round $r$  \\
		$B_{s,c,i,k, r}$ & (boolean) whether chunk $c$ of source $s$ is in node $i$'s buffer at the start of epoch $k$ in round $r$ \\
		$Q_{s,c,i,k, r}$ & (boolean) whether chunk $c$ of source $s$ is in node $i$'s buffer at the start of future epoch $k'$ in round $r$.\\
		$\mathcal{R}_{s,c,d,k, r}$ & whether chunk $c$ of source $s$ is delivered to node $d$ by the end of epoch $k$in round $r$\\
	\end{tabularx}
	\caption{New variables for the $A^{*}$ technique.}\label{table:A}
\end{table*}

To encode $A^{*}$ we maintain most constraints from the MILP formulation but need to modify the objective function and the buffer constraints to account for chunks arriving in future rounds. For switches, we need to modify the flow conservation constraints as they do not have enough memory for buffering.

\noindent \textbf{Look ahead constraints.} To account for chunks that will arrive in the subsequent epoch we need to maintain additional state. For none switch nodes, if the chunk arrives in the first epoch of the next round ($k' = 0$) we have:

\begin{align*}
&  Q_{s,n,c,0, r} = \nonumber \\
& B_{s,n,c, \mathtt{K}, r} + \sum_{\forall j : (j,n) \in E} F_{s,j,n,c,\mathtt{K} - \lceil\frac{\alpha_{j,n}}{\tau}\rceil, r}  \nonumber \\
&\forall s, n\in N: n \notin S, \forall c \in C
\end{align*}

and for all later arrivals we have:

\begin{align*}
 &  Q_{s,n,c,k', r} = \nonumber\\
& Q_{s,n,c,k'-1, r} + \sum_{\forall j : (j,n) \in E \land (k' - \lceil\frac{\alpha_{j,n}}{\tau}\rceil) <= 0} F_{s,j,n,c,\mathtt{K} + k' - \lceil\frac{\alpha_{j,n}}{\tau}\rceil, r} \nonumber \\
&\forall s, n\in N: n \notin S, \forall c \in C, \forall k' \in K': k'>0 .
\end{align*}

These equations allow us to store in the variables $Q$ what chunks are arriving in the next round. Notice how we also account for buffers by $B_{s,n,c, \mathtt{K}, r}$ in $k' =0$ and by $ Q_{s,n,c,k'-1, r}$ for the $k' > 0$ case. Since the switches do not have large enough buffers we use the following:

\begin{align*}
 & Q_{s,n,c,k, r} = \nonumber \\
 &   \sum_{\forall j : (j,n) \in E \land (k' - \lceil\frac{\alpha_{j,n}}{\tau}\rceil) <= 0} F_{s,j,n,c,\mathtt{K} + k' - \lceil\frac{\alpha_{j,n}}{\tau}\rceil, r} \nonumber \\
 &   \forall s, n\in N: n \in S, \forall c \in C, \forall k' \in K'. 
\end{align*}

All that we have to do now is to set the buffers at the beginning of each round $r > 0$ to $Q$ (we exclude $r = 0$ since there is no prior round, and we can use the same initialization that we had earlier):

\begin{align*}
 &  B_{s,n,c,0, r} =  Q_{s,n,c,0, r-1}\\
 & \forall s,n\in N: s\neq n \land n \notin S, \forall c \in C, r>0 
\end{align*}

For $k>0$, if $Q_{s,n,c,k-1, r-1} = 0$ and $r >0, k <= \max{K'}$ we have:

	\begin{align}
	 &\forall s, n\in N: n \notin  S, \forall c \in C, \forall k \in K: k>0 \nonumber \\
	 &B_{s,n,c,k, r} = \nonumber \\
	&B_{s,n,c,k-1, r} + \sum_{\forall j : (j,n) \in E} F_{s,j,n,c, k - \lceil\frac{\alpha_{j,n}}{\tau}\rceil -1, r} + Q_{s,n,c,k, r-1}    \nonumber
	\end{align}
	
	otherwise:
	
	\begin{align}
	&\forall s, n\in N: n \notin  S, \forall c \in C, \forall k \in K: k>0 \nonumber \\
	&B_{s,n,c,k, r} = \nonumber \\
	&B_{s,n,c,k-1} + \sum_{\forall j : (j,n) \in E} F_{s,j,n,c, k - \lceil\frac{\alpha_{j,n}}{\tau}\rceil -1} \nonumber
	\end{align}

Specifically, we are adding to the buffer what is arriving from the previous round. The two cases are there to ensure we account for each arrival only once for non-switch nodes. The equations are similar for switches:
\vspace{-8pt}

{\small
	\begin{align*}
	& \forall s,n \in N: n \in S, \forall k \in K: k>0, \forall c \in C \\
	&\max_{\forall j: (n,j) \in E} F_{s,n,j,c, k, r} \le \nonumber \\
	& \begin{array}{ccc}
	\begin{cases*}
	\sum_{\forall j : (j,n) \in E} F_{s,j,n,c, + k - \lceil\frac{\alpha_{j,n}}{\tau}\rceil -1} + Q_{s,n,c,k, r-1}  & $ r> 0, k <= \max{K'}$ \\
	\sum_{\forall j : (j,n) \in E} F_{s,j,n,c, + k - \lceil\frac{\alpha_{j,n}}{\tau}\rceil -1}  & otherwise \\
	\end{cases*} 
	\end{array}
	\end{align*}
}

but since switches don't buffer chunks we incorporate them into the flow conservation constraints.

\noindent \textbf{The objective.} We now need to motivate the optimization in each round to get the chunks closer to the destination (while making it even more profitable to satisfy the demand fully). So first, we need to automatically compute this additional payoff. To do this, we add logical edges to the graph that allow nodes to form a clique. We assign a weight to each of these edges which we calculate using the Floyd Warshall algorithm~\cite{FW} and the values for $\alpha_{ij}$. The chunks we send in this epoch that don't contribute to satisfying a demand are stored in our $Q$ variables. We now introduce a new variable: $P_{s,d,k',r}$ --- the total number of chunks coming from source $s$ and going towards destination $d$ that are currently on their way towards the destination. We have:

\begin{align*}
 & P_{s, d, k', r} \le \sum_{\forall n \in N, \forall c \in C: D_{n,d,c, r} =1} Q_{n,s,c,k', r}  \quad \forall k' \in K', \forall s, d \in N\\
 & \sum_{\forall s \in N} P_{s, d, k', r} = \sum_{\forall s \in N, \forall c \in C} D_{s,d,c, r} \quad \forall k' \in K', \forall d \in N 
\end{align*} 

we also modify the demands from round to round to remove the demands we have already satisfied. For $r > 0$ we have:

\begin{align*}
& \forall s, d \in N,  \forall c \in C & \\
& D_{s,d,c,r} = 
\begin{cases*}
0  & $D_{s,d,c,r-1} = 1, Q_{s,d,c,\max K', r-1} = 1 $ \\
D_{s,d,c,r-1}  & otherwise \\
\end{cases*} 
\end{align*}

Given these new values of $D$ and $P$ we can now add the following to our objective:

\begin{align*}
&\text{Distance Objective}(r) = \\
&\sum_{\forall k' \in K', \forall s, d \in N: s\neq d} \frac{\gamma}{(k' + 1)( 1 + FW_{s, d})} P_{s,d,k', r} +\\
& \sum_{\forall k' \in K', \forall s, d \in N: s = d} \frac{1}{(k' + 1)} P_{s,d,k', r}
\end{align*}
where the second term ensures having the chunk at the destination gives more payoff to the optimization ($\gamma < 1$).

\section{Number of epochs}
\label{sec:num_epochs}

We provide a simple algorithm for finding the number of epochs to run the optimization with. This algorithm has no bearing on the optimality of the solution as the optimization automatically identifies if less epochs are sufficient. 

\RestyleAlgo{ruled}
\begin{algorithm}[h]
	\DontPrintSemicolon
	\LinesNumbered
	\caption{This algorithm identifies the number of epochs we need to run the optimization with. We use the resulting $n_e$ to instantiate the general optimization~---~this is an upper bound on the number of epochs we need, and the optimization can automatically discover if a smaller number of epochs is sufficient.\label{algo:finding_num_epochs}}
	\label{Alg:finding-best-chunk-size} 
	\KwIn{$\mathcal{D}$. The demand matrix.}
	\KwIn{$G(N,E)$. The topology.}
	\KwIn{$\tau_{opt}$}
	\KwIn{$\alpha_{ij}$. The latency cost of each link $(i,j) \in E$.}
	\KwIn{$C_{ij}$. The capacity of each link $(i,j) \in E$.}
	\KwIn{$C_\tau$. A set of candidate completion times.}
	\KwOut{$n_e$. The upper bound on the number of epochs we need.}
	\For{$\text{total\_time} \in C_\tau$}
	{
		\For{$n_e \in \{4, 8, 12\}$}
		{
			$\tau \gets \frac{\text{total\_time}}{n_e}$ \;
			$Opt, \text{status} \gets \text{general\_form}(\mathcal{D}, \tau, \alpha, C, n_e, G(N,E))$ \;
			\If{$\text{status} = \text{feasible}$}
			{
				$\text{feasible\_time} \gets \text{total\_time}$\;
				break\;
			}
			
		}
	}
	$n_e \gets \frac{\text{feasible\_time}}{\tau_{opt}}$\;
	\Return{$n_e$}\;
\end{algorithm}

{\small
	\begin{table*}[!htbp]
		\centering
		\begin{tabularx}{\textwidth}{ l p{3cm} c c c}
			\textbf{Collective} & \textbf{(\# chunks, \#epochs)} & \textbf{SCCL solver time (s)} & \textbf{{\sysname} solver time (s)} & \textbf{Diff in transfer time (\%)}\\ 
			\hline
			\hline
			\allgather  & $(1, 2)$ & $0.3$ & $0.09$ & $0$\\
			& $(2, 3)$ & $0.7$ & $0.07$ & $0$\\ 
			& $(3, 4)$ & $1.8$ & $0.19$ & $0$\\
			& $(4, 5)$ & $4.1$ & $1.45$ & $0$\\
			& $(5, 6)$ & $11.2$& $8.96$ & $0$ \\
			& $(6, 7)$ & $27.7$& $50.57$ (11s)& $0$ \\
			\hline
			\alltoall &  $(1, 3)$ & $8.8$ & $0.11$ & $33\%$ \\
			&  $(3, 8)$ & NA & $0.18$ & NA\\
			&  $(8, 30)$ & NA & $1.88$ & NA\\
		\end{tabularx}
		\caption{Comparing {\sysname}'s runtime to SCCL. We use $25$ KB chunks for these experiments and $\alpha=0$. The difference in transfer time is $\frac{100(SCCL - TECCL)}{SCCL}$. For all-to-all we use our notation~---~the number of chunks represents the number of chunks the sender wants to send to each destination (SCCL's notation uses the number of chunks to mean the total number of chunks the source needs to send).}\label{table:comparison_sccl_runtime}
	\end{table*}
}


\section{Epoch duration set based on the fastest link}
\label{sec:fastlink}

To set the epoch duration based on the speed of the fastest link in the LP we do not need to change anything: the LP supports fractional chunks and handles this automatically.The MILP only allows us to send whole chunks~---~if we set the epoch duration to be lower than the transmission time of the chunk on the slowest link we can never use that link: we need to modify both the flow conservation constraint and the capacity constraints to address this issue.

We can model the flow conservation constraints similar to how we model $\alpha$: we account for how many epochs it takes a chunk to traverse the slowest link and change the value of $\delta_{ij}$ accordingly. 

To model the capacity constraint, we need to ensure the number of chunks on a link never exceed its capacity. We first calculate how many epochs we need to transmit the chunk over a link ($\kappa$) and modify the capacity constraints to: 

\begin{align}
&\text{Capacity Constraint} \big(i,j,k\big)  \triangleq \nonumber\\
&\sum_{k-\kappa \le k' \le k}\sum_{s \in N}\sum_{c \in C} F_{s,i,j,k',c} \le \kappa T_{ij} \tau \nonumber
\end{align}

Notice this capacity constraint ensures the same behavior we had when we used the larger epoch duration.

\section{Comparing to SCCL instance}
\label{sec:instance_sccl}

SCCL has two modes: the \texttt{least-steps} and \texttt{instance}. We compare \sysname to SCCL \texttt{instance} in \cref{table:comparison_sccl_runtime}.

\section{Details of each topology}
\label{sec:topology details}

\begin{figure*}
	\centering
	\begin{tikzpicture}[
	node distance=7mm,
	gpu_icon/.style={circle, draw=green!50, fill=green!5, very thick, minimum size=6mm},
	gpu_icon_two/.style={circle, draw=blue!50, fill=blue!5, very thick, minimum size=6mm},
        gpu_icon_three/.style={circle, draw=brown!70, fill=brown!10, very thick, minimum size=6mm},
        gpu_icon_four/.style={circle, draw=violet!50, fill=violet!5, very thick, minimum size=6mm},
	]
	\node[gpu_icon]  at (0,-7)    (gpu_1_0)  {\includegraphics[width=24pt]{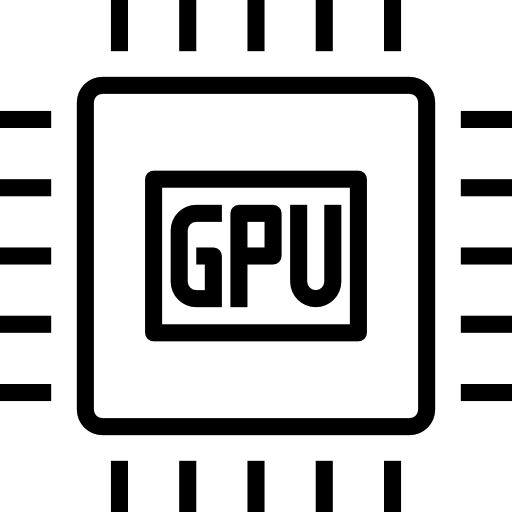}};
	\node[gpu_icon]      (gpu_1_1)       [above=of gpu_1_0]{\includegraphics[width=24pt]{Figures/gpu.png}};
	\node[gpu_icon]      (gpu_1_2)       [left=of gpu_1_1]{\includegraphics[width=24pt]{Figures/gpu.png}};
	\node[gpu_icon]      (gpu_1_3)       [left=of gpu_1_0]{\includegraphics[width=24pt]{Figures/gpu.png}};
	\node[gpu_icon]      (gpu_1_4)       [above=of gpu_1_1]{\includegraphics[width=24pt]{Figures/gpu.png}};
	\node[gpu_icon]      (gpu_1_5)       [above=of gpu_1_4]{\includegraphics[width=24pt]{Figures/gpu.png}};
	\node[gpu_icon]      (gpu_1_6)       [left=of gpu_1_5]{\includegraphics[width=24pt]{Figures/gpu.png}};
	\node[gpu_icon]      (gpu_1_7)       [left=of gpu_1_4]{\includegraphics[width=24pt]{Figures/gpu.png}};
		
	\draw[<->, thick] (gpu_1_0) -- (gpu_1_1);
	\draw[<->, thick, dashed] (gpu_1_0) -- (gpu_1_2);
	\draw[<->, thick, dashed] (gpu_1_0) -- (gpu_1_3);
	\draw[<->, thick, dashed] (gpu_1_1) -- (gpu_1_2);
	\draw[<->, thick] (gpu_1_2) -- (gpu_1_3);
	\draw[<->, thick] (gpu_1_3) -- (gpu_1_1);
	
	\draw[<->, thick] (gpu_1_4) -- (gpu_1_5);
	\draw[<->, thick, dashed] (gpu_1_4) -- (gpu_1_6);
	\draw[<->, thick, dashed] (gpu_1_4) -- (gpu_1_7);
	\draw[<->, thick, dashed] (gpu_1_5) -- (gpu_1_6);
	\draw[<->, thick] (gpu_1_6) -- (gpu_1_7);
	\draw[<->, thick] (gpu_1_7) -- (gpu_1_5);
	
	\draw[<->, thick] (gpu_1_0) -- ++(1.35,0) -- ($ (gpu_1_4) + (1.35, 0) $) -- (gpu_1_4);
	\draw[<->, thick, dashed] (gpu_1_1) -- ++(1.15, 0) -- ($ (gpu_1_5) + (1.15, 0) $) -- (gpu_1_5);
	\draw[<->, thick] (gpu_1_2) -- ++(-1.35, 0) -- ($ (gpu_1_6) + (-1.35, 0) $) -- (gpu_1_6);
	\draw[<->, thick, dashed] (gpu_1_3) -- ++(-1.15, 0) -- ($ (gpu_1_7) + (-1.15, 0) $) -- (gpu_1_7);

	\node[gpu_icon_two]    at (7, -7)    (gpu_2_0)  {\includegraphics[width=24pt]{Figures/gpu.png}};
	\node[gpu_icon_two]      (gpu_2_1)       [above=of gpu_2_0]{\includegraphics[width=24pt]{Figures/gpu.png}};
	\node[gpu_icon_two]      (gpu_2_2)       [right=of gpu_2_1]{\includegraphics[width=24pt]{Figures/gpu.png}};
	\node[gpu_icon_two]      (gpu_2_3)       [right=of gpu_2_0]{\includegraphics[width=24pt]{Figures/gpu.png}};
	\node[gpu_icon_two]      (gpu_2_4)       [above=of gpu_2_1]{\includegraphics[width=24pt]{Figures/gpu.png}};
	\node[gpu_icon_two]      (gpu_2_5)       [above=of gpu_2_4]{\includegraphics[width=24pt]{Figures/gpu.png}};
	\node[gpu_icon_two]      (gpu_2_6)       [right=of gpu_2_5]{\includegraphics[width=24pt]{Figures/gpu.png}};
	\node[gpu_icon_two]      (gpu_2_7)       [right=of gpu_2_4]{\includegraphics[width=24pt]{Figures/gpu.png}};

        \draw[<->, thick] (gpu_2_0) -- (gpu_2_1);
	\draw[<->, thick, dashed] (gpu_2_0) -- (gpu_2_2);
	\draw[<->, thick, dashed] (gpu_2_0) -- (gpu_2_3);
	\draw[<->, thick, dashed] (gpu_2_1) -- (gpu_2_2);
	\draw[<->, thick] (gpu_2_2) -- (gpu_2_3);
	\draw[<->, thick] (gpu_2_3) -- (gpu_2_1);
	
	\draw[<->, thick] (gpu_2_4) -- (gpu_2_5);
	\draw[<->, thick, dashed] (gpu_2_4) -- (gpu_2_6);
	\draw[<->, thick, dashed] (gpu_2_4) -- (gpu_2_7);
	\draw[<->, thick, dashed] (gpu_2_5) -- (gpu_2_6);
	\draw[<->, thick] (gpu_2_6) -- (gpu_2_7);
	\draw[<->, thick] (gpu_2_7) -- (gpu_2_5);
	
	\draw[<->, thick] (gpu_2_0) -- ++(-1.35, 0) -- ($ (gpu_2_4) + (-1.35, 0) $) -- (gpu_2_4);
	\draw[<->, thick, dashed] (gpu_2_1) -- ++(-1.15, 0) -- ($ (gpu_2_5) + (-1.15, 0) $) -- (gpu_2_5);
	\draw[<->, thick] (gpu_2_2) -- ++(1.35, 0) -- ($ (gpu_2_6) + (1.35, 0) $) -- (gpu_2_6);
	\draw[<->, thick, dashed] (gpu_2_3) -- ++(1.15, 0) -- ($ (gpu_2_7) + (1.15, 0) $) -- (gpu_2_7);
	
	\node[gpu_icon_three]  at (0, -11)    (gpu_3_0)  {\includegraphics[width=24pt]{Figures/gpu.png}};
	\node[gpu_icon_three]      (gpu_3_1)       [below=of gpu_3_0]{\includegraphics[width=24pt]{Figures/gpu.png}};
	\node[gpu_icon_three]      (gpu_3_2)       [left=of gpu_3_1]{\includegraphics[width=24pt]{Figures/gpu.png}};
	\node[gpu_icon_three]      (gpu_3_3)       [left=of gpu_3_0]{\includegraphics[width=24pt]{Figures/gpu.png}};
	\node[gpu_icon_three]      (gpu_3_4)       [below=of gpu_3_1]{\includegraphics[width=24pt]{Figures/gpu.png}};
	\node[gpu_icon_three]      (gpu_3_5)       [below=of gpu_3_4]{\includegraphics[width=24pt]{Figures/gpu.png}};
	\node[gpu_icon_three]      (gpu_3_6)       [left=of gpu_3_5]{\includegraphics[width=24pt]{Figures/gpu.png}};
	\node[gpu_icon_three]      (gpu_3_7)       [left=of gpu_3_4]{\includegraphics[width=24pt]{Figures/gpu.png}};

        \draw[<->, thick] (gpu_3_0) -- (gpu_3_1);
	\draw[<->, thick, dashed] (gpu_3_0) -- (gpu_3_2);
	\draw[<->, thick, dashed] (gpu_3_0) -- (gpu_3_3);
	\draw[<->, thick, dashed] (gpu_3_1) -- (gpu_3_2);
	\draw[<->, thick] (gpu_3_2) -- (gpu_3_3);
	\draw[<->, thick] (gpu_3_3) -- (gpu_3_1);
	
	\draw[<->, thick] (gpu_3_4) -- (gpu_3_5);
	\draw[<->, thick, dashed] (gpu_3_4) -- (gpu_3_6);
	\draw[<->, thick, dashed] (gpu_3_4) -- (gpu_3_7);
	\draw[<->, thick, dashed] (gpu_3_5) -- (gpu_3_6);
	\draw[<->, thick] (gpu_3_6) -- (gpu_3_7);
	\draw[<->, thick] (gpu_3_7) -- (gpu_3_5);
	
	\draw[<->, thick] (gpu_3_0) -- ++(1.35, 0) -- ($ (gpu_3_4) + (1.35, 0) $) -- (gpu_3_4);
	\draw[<->, thick, dashed] (gpu_3_1) -- ++(1.15, 0) -- ($ (gpu_3_5) + (1.15, 0) $) -- (gpu_3_5);
	\draw[<->, thick] (gpu_3_2) -- ++(-1.35, 0) -- ($ (gpu_3_6) + (-1.35, 0) $) -- (gpu_3_6);
	\draw[<->, thick, dashed] (gpu_3_3) -- ++(-1.15, 0) -- ($ (gpu_3_7) + (-1.15, 0) $) -- (gpu_3_7);
	
	\node[gpu_icon_four]  at (7, -11)    (gpu_4_0)  {\includegraphics[width=24pt]{Figures/gpu.png}};
	\node[gpu_icon_four]      (gpu_4_1)       [below=of gpu_4_0]{\includegraphics[width=24pt]{Figures/gpu.png}};
	\node[gpu_icon_four]      (gpu_4_2)       [right=of gpu_4_1]{\includegraphics[width=24pt]{Figures/gpu.png}};
	\node[gpu_icon_four]      (gpu_4_3)       [right=of gpu_4_0]{\includegraphics[width=24pt]{Figures/gpu.png}};
	\node[gpu_icon_four]      (gpu_4_4)       [below=of gpu_4_1]{\includegraphics[width=24pt]{Figures/gpu.png}};
	\node[gpu_icon_four]      (gpu_4_5)       [below=of gpu_4_4]{\includegraphics[width=24pt]{Figures/gpu.png}};
	\node[gpu_icon_four]      (gpu_4_6)       [right=of gpu_4_5]{\includegraphics[width=24pt]{Figures/gpu.png}};
	\node[gpu_icon_four]      (gpu_4_7)       [right=of gpu_4_4]{\includegraphics[width=24pt]{Figures/gpu.png}};

        \draw[<->, thick] (gpu_4_0) -- (gpu_4_1);
	\draw[<->, thick, dashed] (gpu_4_0) -- (gpu_4_2);
	\draw[<->, thick, dashed] (gpu_4_0) -- (gpu_4_3);
	\draw[<->, thick, dashed] (gpu_4_1) -- (gpu_4_2);
	\draw[<->, thick] (gpu_4_2) -- (gpu_4_3);
	\draw[<->, thick] (gpu_4_3) -- (gpu_4_1);
	
	\draw[<->, thick] (gpu_4_4) -- (gpu_4_5);
	\draw[<->, thick, dashed] (gpu_4_4) -- (gpu_4_6);
	\draw[<->, thick, dashed] (gpu_4_4) -- (gpu_4_7);
	\draw[<->, thick, dashed] (gpu_4_5) -- (gpu_4_6);
	\draw[<->, thick] (gpu_4_6) -- (gpu_4_7);
	\draw[<->, thick] (gpu_4_7) -- (gpu_4_5);
	
	\draw[<->, thick] (gpu_4_0) -- ++(-1.35, 0) -- ($ (gpu_4_4) + (-1.35, 0) $) -- (gpu_4_4);
	\draw[<->, thick, dashed] (gpu_4_1) -- ++(-1.15, 0) -- ($ (gpu_4_5) + (-1.15, 0) $) -- (gpu_4_5);
	\draw[<->, thick] (gpu_4_2) -- ++(1.35, 0) -- ($ (gpu_4_6) + (1.35, 0) $) -- (gpu_4_6);
	\draw[<->, thick, dashed] (gpu_4_3) -- ++(1.15, 0) -- ($ (gpu_4_7) + (1.15, 0) $) -- (gpu_4_7);

        \node (switch) at (3.5,-9) {\includegraphics[width=64pt]{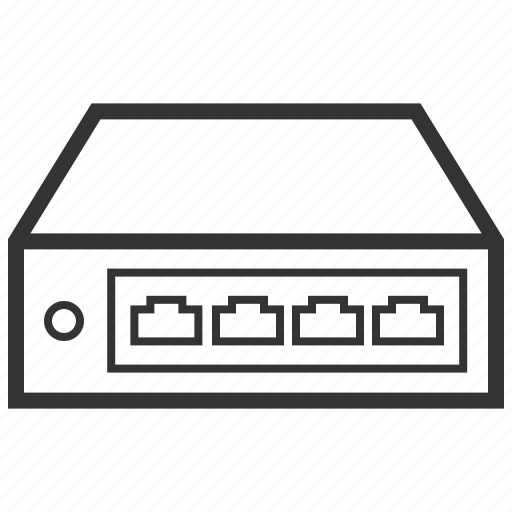}};
	\draw[->, very thick, dotted] (gpu_1_0) -- (switch);
        \draw[->, very thick, dotted] (gpu_2_0) -- (switch);
        \draw[->, very thick, dotted] (gpu_3_0) -- (switch);
        \draw[->, very thick, dotted] (gpu_4_0) -- (switch);

        \draw[->, very thick, dotted] (switch) -- (gpu_1_1);
        \draw[->, very thick, dotted] (switch) -- (gpu_2_1);
        \draw[->, very thick, dotted] (switch) -- (gpu_3_1);
        \draw[->, very thick, dotted] (switch) -- (gpu_4_1);

        \draw[<->, thick] (1.7,-19) -- (2.7,-19) node [pos=1, right] {$50$ GBps with $\alpha = 0.7 \mu s$ in each direction};
	\draw[<->, dashed, thick] (1.7,-19.5) -- (2.7,-19.5) node [pos=1, right] {$25$ GBps with $\alpha = 0.7 \mu s$ in each direction};
        \draw[->, dotted, very thick] (1.7,-20) -- (2.7,-20) node [pos=1, right] {$12.5$ GBps with $\alpha = 1.3 \mu s$};
        \end{tikzpicture}
	\caption{Four chassis \ndv topology used by \sysname. Each chassis has $8$ GPUs connected with $50$ GBps and $25$ GBps links. TACCL replaces the switch by connecting GPU $0$ of a chassis to GPU $1$ of all other chassis and constraints that only one of the three links can be used at a given time.}
 \label{fig:ndv2_topo}
\end{figure*}

We use \dgx, \dgxtwo, \ndv, and internal topologies 1 and 2 for our evaluation.
\cref{fig:ndv2_topo} and  \ref{fig:dgx_two_topo} shows the topologies, capacity and $\alpha$ we used for \ndv and \dgxtwo respectively.
\dgx has $8$ GPUs and is similar to a single chassis \ndv.
Internal topologies 1 and 2 are proprietary, and we cannot report numbers for those.

\begin{landscape}
   \begin{figure}
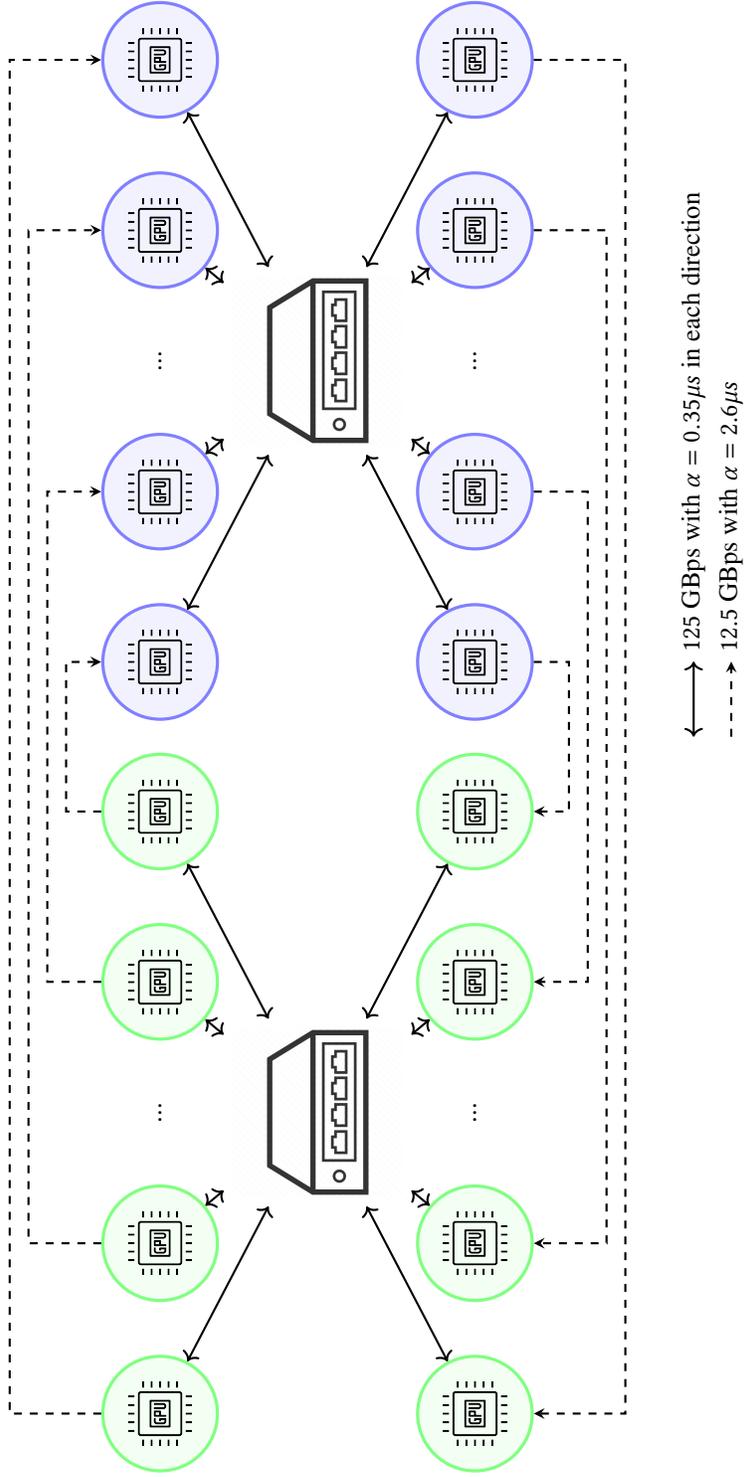

	\vspace{4cm}
	\centering
	\begin{tikzpicture}[
	node distance=7mm,
	gpu_icon/.style={circle, draw=green!50, fill=green!5, very thick, minimum size=6mm},
	gpu_icon_two/.style={circle, draw=blue!50, fill=blue!5, very thick, minimum size=6mm},
	]
	
	
	\node (leftswitch) at (0,0) {\includegraphics[width=64pt]{Figures/switch-512.png}};
	\node (rightswitch) at (10,0) {\includegraphics[width=64pt]{Figures/switch-512.png}};
	
	\node (uncertain_NW) [above=of leftswitch]{...};
	\node (uncertain_NE) [above=of rightswitch]{...};
	\node (uncertain_SW) [below=of leftswitch]{...};
	\node (uncertain_SE) [below=of rightswitch]{...};
	
	\node[gpu_icon]      (gpu_NW_b)       [left=of uncertain_NW]{\includegraphics[width=24pt]{Figures/gpu.png}};
	\node[gpu_icon]      (gpu_NW_a)       [left=of gpu_NW_b]{\includegraphics[width=24pt]{Figures/gpu.png}};
	\node[gpu_icon]      (gpu_NW_c)       [right=of uncertain_NW]{\includegraphics[width=24pt]{Figures/gpu.png}};
	\node[gpu_icon]      (gpu_NW_d)       [right=of gpu_NW_c]{\includegraphics[width=24pt]{Figures/gpu.png}};
	
	\node[gpu_icon]      (gpu_SW_b)       [left=of uncertain_SW]{\includegraphics[width=24pt]{Figures/gpu.png}};
	\node[gpu_icon]      (gpu_SW_a)       [left=of gpu_SW_b]{\includegraphics[width=24pt]{Figures/gpu.png}};
	\node[gpu_icon]      (gpu_SW_c)       [right=of uncertain_SW]{\includegraphics[width=24pt]{Figures/gpu.png}};
	\node[gpu_icon]      (gpu_SW_d)       [right=of gpu_SW_c]{\includegraphics[width=24pt]{Figures/gpu.png}};
	
	\node[gpu_icon_two]      (gpu_NE_b)       [left=of uncertain_NE]{\includegraphics[width=24pt]{Figures/gpu.png}};
	\node[gpu_icon_two]      (gpu_NE_a)       [left=of gpu_NE_b]{\includegraphics[width=24pt]{Figures/gpu.png}};
	\node[gpu_icon_two]      (gpu_NE_c)       [right=of uncertain_NE]{\includegraphics[width=24pt]{Figures/gpu.png}};
	\node[gpu_icon_two]      (gpu_NE_d)       [right=of gpu_NE_c]{\includegraphics[width=24pt]{Figures/gpu.png}};
	
	\node[gpu_icon_two]      (gpu_SE_b)       [left=of uncertain_SE]{\includegraphics[width=24pt]{Figures/gpu.png}};
	\node[gpu_icon_two]      (gpu_SE_a)       [left=of gpu_SE_b]{\includegraphics[width=24pt]{Figures/gpu.png}};
	\node[gpu_icon_two]      (gpu_SE_c)       [right=of uncertain_SE]{\includegraphics[width=24pt]{Figures/gpu.png}};
	\node[gpu_icon_two]      (gpu_SE_d)       [right=of gpu_SE_c]{\includegraphics[width=24pt]{Figures/gpu.png}};
	
	\draw[<->, thick] (leftswitch) -- (gpu_NW_a);
	\draw[<->, thick] (leftswitch) -- (gpu_NW_b);
	\draw[<->, thick] (leftswitch) -- (gpu_NW_c);
	\draw[<->, thick] (leftswitch) -- (gpu_NW_d);
	\draw[<->, thick] (leftswitch) -- (gpu_SW_a);
	\draw[<->, thick] (leftswitch) -- (gpu_SW_b);
	\draw[<->, thick] (leftswitch) -- (gpu_SW_c);
	\draw[<->, thick] (leftswitch) -- (gpu_SW_d);
	
	\draw[<->, thick](rightswitch) -- (gpu_NE_a);
	\draw[<->, thick](rightswitch) -- (gpu_NE_b);
	\draw[<->, thick](rightswitch) -- (gpu_NE_c);
	\draw[<->, thick](rightswitch) -- (gpu_NE_d);
	\draw[<->, thick](rightswitch) -- (gpu_SE_a);
	\draw[<->, thick](rightswitch) -- (gpu_SE_b);
	\draw[<->, thick](rightswitch) -- (gpu_SE_c);
	\draw[<->, thick](rightswitch) -- (gpu_SE_d);

	\draw[->,dashed, -stealth, thick]  (gpu_NW_a) -- ++(0,2) -- ($ (gpu_NE_d) + (0,2) $) -- (gpu_NE_d);
	\draw[->,dashed, -stealth, thick]  (gpu_NW_b) -- ++(0,1.75) -- ($ (gpu_NE_c) + (0,1.75) $) -- (gpu_NE_c);
	\draw[->,dashed, -stealth, thick]  (gpu_NW_c) -- ++(0,1.5) -- ($ (gpu_NE_b) + (0,1.5) $) -- (gpu_NE_b);
	\draw[->,dashed, -stealth, thick]  (gpu_NW_d) -- ++(0,1.25) -- ($ (gpu_NE_a) + (0,1.25) $) -- (gpu_NE_a);
	
	\draw[->,dashed, -stealth, thick]  (gpu_SE_a) -- ++(0,-1.25) -- ($ (gpu_SW_d) - (0,1.25) $) -- (gpu_SW_d);
	\draw[->,dashed, -stealth, thick]  (gpu_SE_b) -- ++(0,-1.5) -- ($ (gpu_SW_c) - (0,1.5) $) -- (gpu_SW_c);
	\draw[->,dashed, -stealth, thick]  (gpu_SE_c) -- ++(0,-1.75) -- ($ (gpu_SW_b) - (0,1.75) $) -- (gpu_SW_b);
	\draw[->,dashed, -stealth, thick]  (gpu_SE_d) -- ++(0,-2) -- ($ (gpu_SW_a) - (0,2) $) -- (gpu_SW_a);
	
	\draw[<->, thick] (5,-5) -- (6,-5) node [pos=1, right] {$125$ GBps with $\alpha = 0.35 \mu s$ in each direction};
	\draw[<->, dashed, -stealth, thick] (5,-5.5) -- (6,-5.5) node [pos=1, right] {$12.5$ GBps with $\alpha = 2.6 \mu s$};

	\end{tikzpicture}
	\vspace{1cm}
	\caption{Two chassis \dgxtwo topology used by \sysname. Each chassis has $16$ GPUs ($8$ GPUs are used for sending chunks to another chassis, and $8$ GPUs are used for receiving chunks from the other chassis). Each dashed link is $12.5$ GBps with $\alpha = 2.6 \mu s$, and each thick straight link is $125$ GBps with $\alpha = 0.35 \mu s$ in each direction. TACCL replaces the switch in each chassis and connects each GPU in a chassis to every other GPU, effectively forming a clique and uses its \texttt{uc-min} strategy to minimize the number of edges used.}
 \end{figure}
 \customlabel{fig:dgx_two_topo}{Figure 12}{} 
\end{landscape}

\renewcommand{\arraystretch}{1.3}

\onecolumn
\begin{longtable}{|ccccccccc|}
\caption{Experimental results for \sysname and comparison to TACCL on \ndv 2 chassis topology.}
\label{tab:ndv2-raw-data}\\
\hline
\multicolumn{1}{|c|}{\begin{tabular}[c]{@{}c@{}}Output\\ Buffer Size\end{tabular}} &
  \multicolumn{1}{c|}{\begin{tabular}[c]{@{}c@{}}ED\\ ($\mu$s)\end{tabular}} &
  \multicolumn{1}{c|}{\begin{tabular}[c]{@{}c@{}}CT \\ ($\mu$s)\end{tabular}} &
  \multicolumn{1}{c|}{\begin{tabular}[c]{@{}c@{}}ST\\ (s)\end{tabular}} &
  \multicolumn{1}{c|}{{\color[HTML]{333333} \textbf{\begin{tabular}[c]{@{}c@{}}AB\\ (GB/s)\end{tabular}}}} &
  \multicolumn{1}{c|}{\begin{tabular}[c]{@{}c@{}}TACCL \\ CT ($\mu$s)\end{tabular}} &
  \multicolumn{1}{c|}{\begin{tabular}[c]{@{}c@{}}TACCL \\ ST (s)\end{tabular}} &
  \multicolumn{1}{c|}{{\color[HTML]{333333} \textbf{\begin{tabular}[c]{@{}c@{}}TACCL \\ AB (GB/s)\end{tabular}}}} &
  {\color[HTML]{036400} \textbf{Improvement \%}} \\ \hline\hline
\endfirsthead
\multicolumn{9}{c}%
{{\bfseries Table \thetable\ continued from previous page}} \\
\endhead
\multicolumn{3}{|c|}{ED - Epoch Duration} &
  \multicolumn{3}{c|}{CT - Collective finish Time} &
  \multicolumn{3}{c|}{ST- Solver Time} \\ \hline
\multicolumn{9}{|c|}{AB - Algorithmic Bandwidth = output buffer size / collective time} \\ \hline\hline
\multicolumn{9}{|c|}{\textbf{\ndv 2 chassis \alltoall optimal epoch duration}} \\ \hline
\multicolumn{1}{|c|}{1 GB} &
  \multicolumn{1}{c|}{1250} &
  \multicolumn{1}{c|}{320235.81} &
  \multicolumn{1}{c|}{336.50} &
  \multicolumn{1}{c|}{{\color[HTML]{333333} \textbf{3.123}}} &
  \multicolumn{1}{c|}{320049.4} &
  \multicolumn{1}{c|}{1214.69} &
  \multicolumn{1}{c|}{{\color[HTML]{333333} \textbf{3.125}}} &
  {\color[HTML]{036400} \textbf{-0.058}} \\ \hline
\multicolumn{1}{|c|}{256 MB} &
  \multicolumn{1}{c|}{320} &
  \multicolumn{1}{c|}{82000.00} &
  \multicolumn{1}{c|}{307.33} &
  \multicolumn{1}{c|}{{\color[HTML]{333333} \textbf{3.122}}} &
  \multicolumn{1}{c|}{81964.2} &
  \multicolumn{1}{c|}{1217.56} &
  \multicolumn{1}{c|}{{\color[HTML]{333333} \textbf{3.123}}} &
  {\color[HTML]{036400} \textbf{-0.044}} \\ \hline
\multicolumn{1}{|c|}{64 MB} &
  \multicolumn{1}{c|}{80} &
  \multicolumn{1}{c|}{20495.09} &
  \multicolumn{1}{c|}{339.92} &
  \multicolumn{1}{c|}{{\color[HTML]{333333} \textbf{3.123}}} &
  \multicolumn{1}{c|}{20532} &
  \multicolumn{1}{c|}{1220.6} &
  \multicolumn{1}{c|}{{\color[HTML]{333333} \textbf{3.117}}} &
  {\color[HTML]{036400} \textbf{0.180}} \\ \hline
\multicolumn{1}{|c|}{16 MB} &
  \multicolumn{1}{c|}{20} &
  \multicolumn{1}{c|}{5123.77} &
  \multicolumn{1}{c|}{280.82} &
  \multicolumn{1}{c|}{{\color[HTML]{333333} \textbf{3.123}}} &
  \multicolumn{1}{c|}{5164.4} &
  \multicolumn{1}{c|}{1213.9} &
  \multicolumn{1}{c|}{{\color[HTML]{333333} \textbf{3.098}}} &
  {\color[HTML]{036400} \textbf{0.793}} \\ \hline
\multicolumn{1}{|c|}{4 MB} &
  \multicolumn{1}{c|}{5} &
  \multicolumn{1}{c|}{1296.25} &
  \multicolumn{1}{c|}{165.63} &
  \multicolumn{1}{c|}{{\color[HTML]{333333} \textbf{3.086}}} &
  \multicolumn{1}{c|}{1324.2} &
  \multicolumn{1}{c|}{1214.51} &
  \multicolumn{1}{c|}{{\color[HTML]{333333} \textbf{3.021}}} &
  {\color[HTML]{036400} \textbf{2.156}} \\ \hline
\multicolumn{1}{|c|}{1 MB} &
  \multicolumn{1}{c|}{1.25} &
  \multicolumn{1}{c|}{325.28} &
  \multicolumn{1}{c|}{189.47} &
  \multicolumn{1}{c|}{{\color[HTML]{333333} \textbf{3.074}}} &
  \multicolumn{1}{c|}{359} &
  \multicolumn{1}{c|}{1213.52} &
  \multicolumn{1}{c|}{{\color[HTML]{333333} \textbf{2.786}}} &
  {\color[HTML]{036400} \textbf{10.366}} \\ \hline
\multicolumn{1}{|c|}{256 KB} &
  \multicolumn{1}{c|}{0.32} &
  \multicolumn{1}{c|}{85.52} &
  \multicolumn{1}{c|}{218.50} &
  \multicolumn{1}{c|}{{\color[HTML]{333333} \textbf{2.993}}} &
  \multicolumn{1}{c|}{115.72} &
  \multicolumn{1}{c|}{1221.78} &
  \multicolumn{1}{c|}{{\color[HTML]{333333} \textbf{2.212}}} &
  {\color[HTML]{036400} \textbf{35.313}} \\ \hline
\multicolumn{1}{|c|}{64 KB} &
  \multicolumn{1}{c|}{0.08} &
  \multicolumn{1}{c|}{23.30} &
  \multicolumn{1}{c|}{161.99} &
  \multicolumn{1}{c|}{{\color[HTML]{333333} \textbf{2.747}}} &
  \multicolumn{1}{c|}{50.34} &
  \multicolumn{1}{c|}{860.88} &
  \multicolumn{1}{c|}{{\color[HTML]{333333} \textbf{1.271}}} &
  {\color[HTML]{036400} \textbf{116.052}} \\ \hline
\multicolumn{1}{|c|}{16 KB} &
  \multicolumn{1}{c|}{0.02} &
  \multicolumn{1}{c|}{7.27} &
  \multicolumn{1}{c|}{182.08} &
  \multicolumn{1}{c|}{{\color[HTML]{333333} \textbf{2.202}}} &
  \multicolumn{1}{c|}{35.76} &
  \multicolumn{1}{c|}{86.03} &
  \multicolumn{1}{c|}{{\color[HTML]{333333} \textbf{0.447}}} &
  {\color[HTML]{036400} \textbf{392.223}} \\ \hline
\multicolumn{1}{|c|}{4 KB} &
  \multicolumn{1}{c|}{0.02} &
  \multicolumn{1}{c|}{4.64} &
  \multicolumn{1}{c|}{69.58} &
  \multicolumn{1}{c|}{{\color[HTML]{333333} \textbf{0.862}}} &
  \multicolumn{1}{c|}{32.16} &
  \multicolumn{1}{c|}{31.14} &
  \multicolumn{1}{c|}{{\color[HTML]{333333} \textbf{0.125}}} &
  {\color[HTML]{036400} \textbf{592.134}} \\ \hline
\multicolumn{1}{|c|}{1 KB} &
  \multicolumn{1}{c|}{0.005} &
  \multicolumn{1}{c|}{4.24} &
  \multicolumn{1}{c|}{196.72} &
  \multicolumn{1}{c|}{{\color[HTML]{333333} \textbf{0.236}}} &
  \multicolumn{1}{c|}{36.8} &
  \multicolumn{1}{c|}{27.66} &
  \multicolumn{1}{c|}{{\color[HTML]{333333} \textbf{0.027}}} &
  {\color[HTML]{036400} \textbf{768.920}} \\ \hline\hline
\multicolumn{9}{|c|}{\textbf{\ndv 2 chassis \alltoall max epoch duration}} \\ \hline
\multicolumn{1}{|c|}{1 GB} &
  \multicolumn{1}{c|}{5000} &
  \multicolumn{1}{c|}{325000} &
  \multicolumn{1}{c|}{14.82} &
  \multicolumn{1}{c|}{\textbf{3.077}} &
  \multicolumn{1}{c|}{320049.400} &
  \multicolumn{1}{c|}{1214.692} &
  \multicolumn{1}{c|}{{\color[HTML]{333333} \textbf{3.125}}} &
  {\color[HTML]{036400} \textbf{-1.52}} \\ \hline
\multicolumn{1}{|c|}{256 MB} &
  \multicolumn{1}{c|}{1280.41} &
  \multicolumn{1}{c|}{83226.63} &
  \multicolumn{1}{c|}{14.36} &
  \multicolumn{1}{c|}{\textbf{3.076}} &
  \multicolumn{1}{c|}{81964.200} &
  \multicolumn{1}{c|}{1217.557} &
  \multicolumn{1}{c|}{{\color[HTML]{333333} \textbf{3.123}}} &
  {\color[HTML]{036400} \textbf{-1.52}} \\ \hline
\multicolumn{1}{|c|}{64 MB} &
  \multicolumn{1}{c|}{320.10} &
  \multicolumn{1}{c|}{20806.66} &
  \multicolumn{1}{c|}{11.01} &
  \multicolumn{1}{c|}{\textbf{3.076}} &
  \multicolumn{1}{c|}{20532.000} &
  \multicolumn{1}{c|}{1220.602} &
  \multicolumn{1}{c|}{{\color[HTML]{333333} \textbf{3.117}}} &
  {\color[HTML]{036400} \textbf{-1.32}} \\ \hline
\multicolumn{1}{|c|}{16 MB} &
  \multicolumn{1}{c|}{80.01} &
  \multicolumn{1}{c|}{5200.42} &
  \multicolumn{1}{c|}{9.96} &
  \multicolumn{1}{c|}{\textbf{3.077}} &
  \multicolumn{1}{c|}{5164.400} &
  \multicolumn{1}{c|}{1213.903} &
  \multicolumn{1}{c|}{{\color[HTML]{333333} \textbf{3.098}}} &
  {\color[HTML]{036400} \textbf{-0.69}} \\ \hline
\multicolumn{1}{|c|}{4 MB} &
  \multicolumn{1}{c|}{20} &
  \multicolumn{1}{c|}{1300.03} &
  \multicolumn{1}{c|}{11.81} &
  \multicolumn{1}{c|}{\textbf{3.077}} &
  \multicolumn{1}{c|}{1324.200} &
  \multicolumn{1}{c|}{1214.507} &
  \multicolumn{1}{c|}{{\color[HTML]{333333} \textbf{3.021}}} &
  {\color[HTML]{036400} \textbf{1.86}} \\ \hline
\multicolumn{1}{|c|}{1 MB} &
  \multicolumn{1}{c|}{5} &
  \multicolumn{1}{c|}{340} &
  \multicolumn{1}{c|}{10.85} &
  \multicolumn{1}{c|}{\textbf{2.941}} &
  \multicolumn{1}{c|}{359.000} &
  \multicolumn{1}{c|}{1213.521} &
  \multicolumn{1}{c|}{{\color[HTML]{333333} \textbf{2.786}}} &
  {\color[HTML]{036400} \textbf{5.59}} \\ \hline
\multicolumn{1}{|c|}{256 KB} &
  \multicolumn{1}{c|}{1.28} &
  \multicolumn{1}{c|}{88.32} &
  \multicolumn{1}{c|}{9.97} &
  \multicolumn{1}{c|}{\textbf{2.899}} &
  \multicolumn{1}{c|}{115.720} &
  \multicolumn{1}{c|}{1221.779} &
  \multicolumn{1}{c|}{{\color[HTML]{333333} \textbf{2.212}}} &
  {\color[HTML]{036400} \textbf{31.02}} \\ \hline
\multicolumn{1}{|c|}{64 KB} &
  \multicolumn{1}{c|}{0.32} &
  \multicolumn{1}{c|}{24.32} &
  \multicolumn{1}{c|}{10.46} &
  \multicolumn{1}{c|}{\textbf{2.632}} &
  \multicolumn{1}{c|}{50.340} &
  \multicolumn{1}{c|}{860.875} &
  \multicolumn{1}{c|}{{\color[HTML]{333333} \textbf{1.271}}} &
  {\color[HTML]{036400} \textbf{106.99}} \\ \hline
\multicolumn{1}{|c|}{16 KB} &
  \multicolumn{1}{c|}{0.08} &
  \multicolumn{1}{c|}{7.6} &
  \multicolumn{1}{c|}{8.83} &
  \multicolumn{1}{c|}{\textbf{2.105}} &
  \multicolumn{1}{c|}{35.760} &
  \multicolumn{1}{c|}{86.034} &
  \multicolumn{1}{c|}{{\color[HTML]{333333} \textbf{0.447}}} &
  {\color[HTML]{036400} \textbf{370.53}} \\ \hline
\multicolumn{1}{|c|}{4 KB} &
  \multicolumn{1}{c|}{0.02} &
  \multicolumn{1}{c|}{4.5} &
  \multicolumn{1}{c|}{20.90} &
  \multicolumn{1}{c|}{\textbf{0.889}} &
  \multicolumn{1}{c|}{32.115} &
  \multicolumn{1}{c|}{31.139} &
  \multicolumn{1}{c|}{{\color[HTML]{333333} \textbf{0.125}}} &
  {\color[HTML]{036400} \textbf{613.67}} \\ \hline
\multicolumn{1}{|c|}{1 KB} &
  \multicolumn{1}{c|}{0.01} &
  \multicolumn{1}{c|}{4.235} &
  \multicolumn{1}{c|}{276.47} &
  \multicolumn{1}{c|}{{\color[HTML]{333333} \textbf{0.236}}} &
  \multicolumn{1}{c|}{36.799} &
  \multicolumn{1}{c|}{27.660} &
  \multicolumn{1}{c|}{{\color[HTML]{333333} \textbf{0.027}}} &
  {\color[HTML]{036400} \textbf{768.92}} \\ \hline\hline
\multicolumn{9}{|c|}{\textbf{\ndv2 chassis \allgather optimal epoch duration}} \\ \hline
\multicolumn{1}{|c|}{1 GB} &
  \multicolumn{1}{c|}{1250} &
  \multicolumn{1}{c|}{43750} &
  \multicolumn{1}{c|}{7201.05} &
  \multicolumn{1}{c|}{\textbf{22.86}} &
  \multicolumn{1}{c|}{53766.70} &
  \multicolumn{1}{c|}{7.01} &
  \multicolumn{1}{c|}{\textbf{18.60}} &
  {\color[HTML]{036400} \textbf{22.90}} \\ \hline
\multicolumn{1}{|c|}{256 MB} &
  \multicolumn{1}{c|}{320} &
  \multicolumn{1}{c|}{11200} &
  \multicolumn{1}{c|}{7214.16} &
  \multicolumn{1}{c|}{\textbf{22.86}} &
  \multicolumn{1}{c|}{12494.60} &
  \multicolumn{1}{c|}{6.56} &
  \multicolumn{1}{c|}{\textbf{20.49}} &
  {\color[HTML]{036400} \textbf{11.56}} \\ \hline
\multicolumn{1}{|c|}{64 MB} &
  \multicolumn{1}{c|}{80} &
  \multicolumn{1}{c|}{2800} &
  \multicolumn{1}{c|}{7209.46} &
  \multicolumn{1}{c|}{\textbf{22.86}} &
  \multicolumn{1}{c|}{3133.20} &
  \multicolumn{1}{c|}{8.27} &
  \multicolumn{1}{c|}{\textbf{20.43}} &
  {\color[HTML]{036400} \textbf{11.90}} \\ \hline
\multicolumn{1}{|c|}{16 MB} &
  \multicolumn{1}{c|}{20} &
  \multicolumn{1}{c|}{700} &
  \multicolumn{1}{c|}{7208.70} &
  \multicolumn{1}{c|}{\textbf{22.86}} &
  \multicolumn{1}{c|}{-} &
  \multicolumn{1}{c|}{-} &
  \multicolumn{1}{c|}{\textbf{-}} &
  {\color[HTML]{036400} \textbf{-}} \\ \hline
\multicolumn{1}{|c|}{4 MB} &
  \multicolumn{1}{c|}{5} &
  \multicolumn{1}{c|}{190} &
  \multicolumn{1}{c|}{152.60} &
  \multicolumn{1}{c|}{\textbf{21.05}} &
  \multicolumn{1}{c|}{216.50} &
  \multicolumn{1}{c|}{8.37} &
  \multicolumn{1}{c|}{\textbf{18.48}} &
  {\color[HTML]{036400} \textbf{13.95}} \\ \hline
\multicolumn{1}{|c|}{1 MB} &
  \multicolumn{1}{c|}{1.25} &
  \multicolumn{1}{c|}{48.75} &
  \multicolumn{1}{c|}{160.10} &
  \multicolumn{1}{c|}{\textbf{20.51}} &
  \multicolumn{1}{c|}{62.15} &
  \multicolumn{1}{c|}{62.65} &
  \multicolumn{1}{c|}{\textbf{16.09}} &
  {\color[HTML]{036400} \textbf{27.49}} \\ \hline
\multicolumn{1}{|c|}{256 KB} &
  \multicolumn{1}{c|}{0.32} &
  \multicolumn{1}{c|}{14.72} &
  \multicolumn{1}{c|}{59.55} &
  \multicolumn{1}{c|}{\textbf{17.39}} &
  \multicolumn{1}{c|}{25.26} &
  \multicolumn{1}{c|}{11.17} &
  \multicolumn{1}{c|}{\textbf{10.13}} &
  {\color[HTML]{036400} \textbf{71.60}} \\ \hline
\multicolumn{1}{|c|}{64 KB} &
  \multicolumn{1}{c|}{0.08} &
  \multicolumn{1}{c|}{6.08} &
  \multicolumn{1}{c|}{27.61} &
  \multicolumn{1}{c|}{\textbf{10.53}} &
  \multicolumn{1}{c|}{13.08} &
  \multicolumn{1}{c|}{3.66} &
  \multicolumn{1}{c|}{\textbf{4.89}} &
  {\color[HTML]{036400} \textbf{115.13}} \\ \hline
\multicolumn{1}{|c|}{16 KB} &
  \multicolumn{1}{c|}{0.02} &
  \multicolumn{1}{c|}{4.44} &
  \multicolumn{1}{c|}{18.80} &
  \multicolumn{1}{c|}{\textbf{3.60}} &
  \multicolumn{1}{c|}{12.68} &
  \multicolumn{1}{c|}{6.34} &
  \multicolumn{1}{c|}{\textbf{1.26}} &
  {\color[HTML]{036400} \textbf{185.59}} \\ \hline
\multicolumn{1}{|c|}{4 KB} &
  \multicolumn{1}{c|}{0.02} &
  \multicolumn{1}{c|}{4.24} &
  \multicolumn{1}{c|}{12.26} &
  \multicolumn{1}{c|}{\textbf{0.94}} &
  \multicolumn{1}{c|}{11.85} &
  \multicolumn{1}{c|}{4.30} &
  \multicolumn{1}{c|}{\textbf{0.34}} &
  {\color[HTML]{036400} \textbf{179.48}} \\ \hline
\multicolumn{1}{|c|}{1 KB} &
  \multicolumn{1}{c|}{0.005} &
  \multicolumn{1}{c|}{4.135} &
  \multicolumn{1}{c|}{50.28} &
  \multicolumn{1}{c|}{\textbf{0.24}} &
  \multicolumn{1}{c|}{10.16} &
  \multicolumn{1}{c|}{3.02} &
  \multicolumn{1}{c|}{\textbf{0.1}} &
  {\color[HTML]{036400} \textbf{145.68}} \\ \hline\hline
\multicolumn{9}{|c|}{\textbf{\ndv 2 chassis \allgather early stop at 30\% using optimal epoch duration}} \\ \hline
\multicolumn{1}{|c|}{1 GB} &
  \multicolumn{1}{c|}{1250} &
  \multicolumn{1}{c|}{47500} &
  \multicolumn{1}{c|}{2.66} &
  \multicolumn{1}{c|}{\textbf{21.05}} &
  \multicolumn{1}{c|}{53766.70} &
  \multicolumn{1}{c|}{7.01} &
  \multicolumn{1}{c|}{\textbf{18.60}} &
  {\color[HTML]{036400} \textbf{13.19}} \\ \hline
\multicolumn{1}{|c|}{256 MB} &
  \multicolumn{1}{c|}{320} &
  \multicolumn{1}{c|}{12163.89} &
  \multicolumn{1}{c|}{2.37} &
  \multicolumn{1}{c|}{\textbf{21.05}} &
  \multicolumn{1}{c|}{12494.60} &
  \multicolumn{1}{c|}{6.56} &
  \multicolumn{1}{c|}{\textbf{20.49}} &
  {\color[HTML]{036400} \textbf{2.72}} \\ \hline
\multicolumn{1}{|c|}{64 MB} &
  \multicolumn{1}{c|}{80} &
  \multicolumn{1}{c|}{3920.31} &
  \multicolumn{1}{c|}{2.45} &
  \multicolumn{1}{c|}{\textbf{16.33}} &
  \multicolumn{1}{c|}{3133.20} &
  \multicolumn{1}{c|}{8.27} &
  \multicolumn{1}{c|}{\textbf{20.43}} &
  {\color[HTML]{036400} \textbf{-20.08}} \\ \hline
\multicolumn{1}{|c|}{16 MB} &
  \multicolumn{1}{c|}{20} &
  \multicolumn{1}{c|}{980.02} &
  \multicolumn{1}{c|}{2.42} &
  \multicolumn{1}{c|}{\textbf{16.33}} &
  \multicolumn{1}{c|}{-} &
  \multicolumn{1}{c|}{-} &
  \multicolumn{1}{c|}{\textbf{-}} &
  {\color[HTML]{036400} \textbf{-}} \\ \hline
\multicolumn{1}{|c|}{4 MB} &
  \multicolumn{1}{c|}{5} &
  \multicolumn{1}{c|}{240} &
  \multicolumn{1}{c|}{2.40} &
  \multicolumn{1}{c|}{\textbf{16.67}} &
  \multicolumn{1}{c|}{216.50} &
  \multicolumn{1}{c|}{8.37} &
  \multicolumn{1}{c|}{\textbf{18.48}} &
  {\color[HTML]{036400} \textbf{-9.79}} \\ \hline
\multicolumn{1}{|c|}{1 MB} &
  \multicolumn{1}{c|}{1.25} &
  \multicolumn{1}{c|}{63.75} &
  \multicolumn{1}{c|}{4.32} &
  \multicolumn{1}{c|}{\textbf{15.69}} &
  \multicolumn{1}{c|}{62.15} &
  \multicolumn{1}{c|}{62.65} &
  \multicolumn{1}{c|}{\textbf{16.09}} &
  {\color[HTML]{036400} \textbf{-2.51}} \\ \hline
\multicolumn{1}{|c|}{256 KB} &
  \multicolumn{1}{c|}{0.32} &
  \multicolumn{1}{c|}{16.96} &
  \multicolumn{1}{c|}{2.83} &
  \multicolumn{1}{c|}{\textbf{15.09}} &
  \multicolumn{1}{c|}{25.26} &
  \multicolumn{1}{c|}{11.17} &
  \multicolumn{1}{c|}{\textbf{10.13}} &
  {\color[HTML]{036400} \textbf{48.94}} \\ \hline
\multicolumn{1}{|c|}{64 KB} &
  \multicolumn{1}{c|}{0.08} &
  \multicolumn{1}{c|}{6.32} &
  \multicolumn{1}{c|}{3.94} &
  \multicolumn{1}{c|}{\textbf{10.13}} &
  \multicolumn{1}{c|}{13.08} &
  \multicolumn{1}{c|}{3.66} &
  \multicolumn{1}{c|}{\textbf{4.89}} &
  {\color[HTML]{036400} \textbf{106.96}} \\ \hline
\multicolumn{1}{|c|}{16 KB} &
  \multicolumn{1}{c|}{0.02} &
  \multicolumn{1}{c|}{4.44} &
  \multicolumn{1}{c|}{12.98} &
  \multicolumn{1}{c|}{\textbf{3.60}} &
  \multicolumn{1}{c|}{12.68} &
  \multicolumn{1}{c|}{6.34} &
  \multicolumn{1}{c|}{\textbf{1.26}} &
  {\color[HTML]{036400} \textbf{185.59}} \\ \hline
\multicolumn{1}{|c|}{4 KB} &
  \multicolumn{1}{c|}{0.02} &
  \multicolumn{1}{c|}{4.24} &
  \multicolumn{1}{c|}{10.17} &
  \multicolumn{1}{c|}{\textbf{0.94}} &
  \multicolumn{1}{c|}{11.85} &
  \multicolumn{1}{c|}{4.30} &
  \multicolumn{1}{c|}{\textbf{0.34}} &
  {\color[HTML]{036400} \textbf{179.48}} \\ \hline
\multicolumn{1}{|c|}{1 KB} &
  \multicolumn{1}{c|}{0.005} &
  \multicolumn{1}{c|}{4.135} &
  \multicolumn{1}{c|}{42.94} &
  \multicolumn{1}{c|}{\textbf{0.24}} &
  \multicolumn{1}{c|}{10.16} &
  \multicolumn{1}{c|}{3.02} &
  \multicolumn{1}{c|}{\textbf{0.1}} &
  {\color[HTML]{036400} \textbf{145.68}} \\ \hline\hline
\multicolumn{9}{|c|}{\textbf{\ndv 2 Chassis \allgather max epoch duration}} \\ \hline
\multicolumn{1}{|c|}{1 GB} &
  \multicolumn{1}{c|}{5000} &
  \multicolumn{1}{c|}{50000} &
  \multicolumn{1}{c|}{0.94} &
  \multicolumn{1}{c|}{\textbf{20}} &
  \multicolumn{1}{c|}{53766.70} &
  \multicolumn{1}{c|}{7.01} &
  \multicolumn{1}{c|}{\textbf{18.60}} &
  {\color[HTML]{036400} \textbf{7.53}} \\ \hline
\multicolumn{1}{|c|}{256 MB} &
  \multicolumn{1}{c|}{1280.41} &
  \multicolumn{1}{c|}{12804.10} &
  \multicolumn{1}{c|}{0.77} &
  \multicolumn{1}{c|}{\textbf{19.99}} &
  \multicolumn{1}{c|}{12494.60} &
  \multicolumn{1}{c|}{6.56} &
  \multicolumn{1}{c|}{\textbf{20.49}} &
  {\color[HTML]{036400} \textbf{-2.42}} \\ \hline
\multicolumn{1}{|c|}{64 MB} &
  \multicolumn{1}{c|}{320.10} &
  \multicolumn{1}{c|}{3201.02} &
  \multicolumn{1}{c|}{0.78} &
  \multicolumn{1}{c|}{\textbf{19.99}} &
  \multicolumn{1}{c|}{3133.20} &
  \multicolumn{1}{c|}{8.27} &
  \multicolumn{1}{c|}{\textbf{20.43}} &
  {\color[HTML]{036400} \textbf{-2.12}} \\ \hline
\multicolumn{1}{|c|}{16 MB} &
  \multicolumn{1}{c|}{80.01} &
  \multicolumn{1}{c|}{800.06} &
  \multicolumn{1}{c|}{0.77} &
  \multicolumn{1}{c|}{\textbf{20}} &
  \multicolumn{1}{c|}{-} &
  \multicolumn{1}{c|}{-} &
  \multicolumn{1}{c|}{\textbf{-}} &
  {\color[HTML]{036400} \textbf{-}} \\ \hline
\multicolumn{1}{|c|}{4 MB} &
  \multicolumn{1}{c|}{20} &
  \multicolumn{1}{c|}{200} &
  \multicolumn{1}{c|}{0.77} &
  \multicolumn{1}{c|}{\textbf{20}} &
  \multicolumn{1}{c|}{216.50} &
  \multicolumn{1}{c|}{8.37} &
  \multicolumn{1}{c|}{\textbf{18.48}} &
  {\color[HTML]{036400} \textbf{8.25}} \\ \hline
\multicolumn{1}{|c|}{1 MB} &
  \multicolumn{1}{c|}{5} &
  \multicolumn{1}{c|}{70} &
  \multicolumn{1}{c|}{1.04} &
  \multicolumn{1}{c|}{\textbf{14.29}} &
  \multicolumn{1}{c|}{62.15} &
  \multicolumn{1}{c|}{62.65} &
  \multicolumn{1}{c|}{\textbf{16.09}} &
  {\color[HTML]{036400} \textbf{-11.21}} \\ \hline
\multicolumn{1}{|c|}{256 KB} &
  \multicolumn{1}{c|}{1.28} &
  \multicolumn{1}{c|}{19.20} &
  \multicolumn{1}{c|}{1.09} &
  \multicolumn{1}{c|}{\textbf{13.33}} &
  \multicolumn{1}{c|}{25.26} &
  \multicolumn{1}{c|}{11.17} &
  \multicolumn{1}{c|}{\textbf{10.13}} &
  {\color[HTML]{036400} \textbf{31.56}} \\ \hline
\multicolumn{1}{|c|}{64 KB} &
  \multicolumn{1}{c|}{0.32} &
  \multicolumn{1}{c|}{7.68} &
  \multicolumn{1}{c|}{1.74} &
  \multicolumn{1}{c|}{\textbf{8.33}} &
  \multicolumn{1}{c|}{13.08} &
  \multicolumn{1}{c|}{3.66} &
  \multicolumn{1}{c|}{\textbf{4.89}} &
  {\color[HTML]{036400} \textbf{70.31}} \\ \hline
\multicolumn{1}{|c|}{16 KB} &
  \multicolumn{1}{c|}{0.08} &
  \multicolumn{1}{c|}{4.80} &
  \multicolumn{1}{c|}{3.35} &
  \multicolumn{1}{c|}{\textbf{3.33}} &
  \multicolumn{1}{c|}{12.68} &
  \multicolumn{1}{c|}{6.34} &
  \multicolumn{1}{c|}{\textbf{1.26}} &
  {\color[HTML]{036400} \textbf{164.17}} \\ \hline
\multicolumn{1}{|c|}{4 KB} &
  \multicolumn{1}{c|}{0.02} &
  \multicolumn{1}{c|}{4.24} &
  \multicolumn{1}{c|}{21.56} &
  \multicolumn{1}{c|}{\textbf{0.94}} &
  \multicolumn{1}{c|}{11.85} &
  \multicolumn{1}{c|}{4.30} &
  \multicolumn{1}{c|}{\textbf{0.34}} &
  {\color[HTML]{036400} \textbf{179.48}} \\ \hline
\multicolumn{1}{|c|}{1 KB} &
  \multicolumn{1}{c|}{0.01} &
  \multicolumn{1}{c|}{4.14} &
  \multicolumn{1}{c|}{89.07} &
  \multicolumn{1}{c|}{\textbf{0.24}} &
  \multicolumn{1}{c|}{10.16} &
  \multicolumn{1}{c|}{3.02} &
  \multicolumn{1}{c|}{\textbf{0.1}} &
  {\color[HTML]{036400} \textbf{145.68}} \\ \hline
\end{longtable}

%% file: main.bbl

\begin{thebibliography}{31}


\ifx \showCODEN    \undefined \def \showCODEN     #1{\unskip}     \fi
\ifx \showDOI      \undefined \def \showDOI       #1{#1}\fi
\ifx \showISBNx    \undefined \def \showISBNx     #1{\unskip}     \fi
\ifx \showISBNxiii \undefined \def \showISBNxiii  #1{\unskip}     \fi
\ifx \showISSN     \undefined \def \showISSN      #1{\unskip}     \fi
\ifx \showLCCN     \undefined \def \showLCCN      #1{\unskip}     \fi
\ifx \shownote     \undefined \def \shownote      #1{#1}          \fi
\ifx \showarticletitle \undefined \def \showarticletitle #1{#1}   \fi
\ifx \showURL      \undefined \def \showURL       {\relax}        \fi
\providecommand\bibfield[2]{#2}
\providecommand\bibinfo[2]{#2}
\providecommand\natexlab[1]{#1}
\providecommand\showeprint[2][]{arXiv:#2}

\bibitem[\protect\citeauthoryear{Abuzaid, Kandula, Arzani, Menache, Zaharia,
  and Bailis}{Abuzaid et~al\mbox{.}}{2021}]%
        {NCFlow}
\bibfield{author}{\bibinfo{person}{Firas Abuzaid}, \bibinfo{person}{Srikanth
  Kandula}, \bibinfo{person}{Behnaz Arzani}, \bibinfo{person}{Ishai Menache},
  \bibinfo{person}{Matei Zaharia}, {and} \bibinfo{person}{Peter Bailis}.}
  \bibinfo{year}{2021}\natexlab{}.
\newblock \showarticletitle{Contracting wide-area network topologies to solve
  flow problems quickly}. In \bibinfo{booktitle}{{\em 18th USENIX Symposium on
  Networked Systems Design and Implementation (NSDI 21)}}.
  \bibinfo{pages}{175--200}.
\newblock


\bibitem[\protect\citeauthoryear{Azure NDv2-series}{Azure NDv2-series}{2021}]%
        {ndv2}
Azure NDv2-series \bibinfo{year}{2021}\natexlab{}.
\newblock   (\bibinfo{year}{2021}).
\newblock
\showURL{%
\url{https://learn.microsoft.com/en-us/azure/virtual-machines/ndv2-series}}


\bibitem[\protect\citeauthoryear{Bertsimas and Tsitsiklis}{Bertsimas and
  Tsitsiklis}{1997}]%
        {bertsimas1997introduction}
\bibfield{author}{\bibinfo{person}{Dimitris Bertsimas} {and}
  \bibinfo{person}{John~N Tsitsiklis}.} \bibinfo{year}{1997}\natexlab{}.
\newblock \bibinfo{booktitle}{{\em Introduction to linear optimization}}.
  Vol.~\bibinfo{volume}{6}.
\newblock \bibinfo{publisher}{Athena Scientific Belmont, MA}.
\newblock


\bibitem[\protect\citeauthoryear{Boyd and Vandenberghe}{Boyd and
  Vandenberghe}{2004}]%
        {boyd_co}
\bibfield{author}{\bibinfo{person}{Stephen Boyd} {and} \bibinfo{person}{Lieven
  Vandenberghe}.} \bibinfo{year}{2004}\natexlab{}.
\newblock \bibinfo{booktitle}{{\em {Convex Optimization}}}.
\newblock \bibinfo{publisher}{Cambridge University Press}.
\newblock


\bibitem[\protect\citeauthoryear{Cai, Liu, Maleki, Musuvathi, Mytkowicz,
  Nelson, and Saarikivi}{Cai et~al\mbox{.}}{2021}]%
        {SCCL}
\bibfield{author}{\bibinfo{person}{Zixian Cai}, \bibinfo{person}{Zhengyang
  Liu}, \bibinfo{person}{Saeed Maleki}, \bibinfo{person}{Madanlal Musuvathi},
  \bibinfo{person}{Todd Mytkowicz}, \bibinfo{person}{Jacob Nelson}, {and}
  \bibinfo{person}{Olli Saarikivi}.} \bibinfo{year}{2021}\natexlab{}.
\newblock \showarticletitle{Synthesizing Optimal Collective Algorithms}. In
  \bibinfo{booktitle}{{\em Proceedings of the 26th ACM SIGPLAN Symposium on
  Principles and Practice of Parallel Programming}} {\em
  (\bibinfo{series}{PPoPP '21})}. \bibinfo{publisher}{Association for Computing
  Machinery}, \bibinfo{address}{New York, NY, USA}, \bibinfo{pages}{62–75}.
\newblock
\showISBNx{9781450382946}
\showDOI{%
\url{https://doi.org/10.1145/3437801.3441620}}


\bibitem[\protect\citeauthoryear{ChatGPT runs 10K Nvidia training GPUs with
  potential for thousands more}{ChatGPT runs 10K Nvidia training GPUs with
  potential for thousands more}{2023}]%
        {chatgpt}
ChatGPT runs 10K Nvidia training GPUs with potential for thousands more
  \bibinfo{year}{2023}\natexlab{}.
\newblock   (\bibinfo{year}{2023}).
\newblock
\showURL{%
\url{https://www.fierceelectronics.com/sensors/chatgpt-runs-10k-nvidia-training-gpus-potential-thousands-more}}


\bibitem[\protect\citeauthoryear{Deng, Pan, Zhou, Kong, Flores, and Lin}{Deng
  et~al\mbox{.}}{2021}]%
        {deng2021deeplight}
\bibfield{author}{\bibinfo{person}{Wei Deng}, \bibinfo{person}{Junwei Pan},
  \bibinfo{person}{Tian Zhou}, \bibinfo{person}{Deguang Kong},
  \bibinfo{person}{Aaron Flores}, {and} \bibinfo{person}{Guang Lin}.}
  \bibinfo{year}{2021}\natexlab{}.
\newblock \showarticletitle{DeepLight: Deep Lightweight Feature Interactions
  for Accelerating CTR Predictions in Ad Serving}. In \bibinfo{booktitle}{{\em
  Proceedings of the 14th ACM International Conference on Web Search and Data
  Mining}} {\em (\bibinfo{series}{WSDM '21})}. \bibinfo{publisher}{Association
  for Computing Machinery}, \bibinfo{address}{New York, NY, USA},
  \bibinfo{pages}{922–930}.
\newblock
\showISBNx{9781450382977}
\showDOI{%
\url{https://doi.org/10.1145/3437963.3441727}}


\bibitem[\protect\citeauthoryear{Devlin, Chang, Lee, and Toutanova}{Devlin
  et~al\mbox{.}}{2018}]%
        {devlin2018bert}
\bibfield{author}{\bibinfo{person}{Jacob Devlin}, \bibinfo{person}{Ming-Wei
  Chang}, \bibinfo{person}{Kenton Lee}, {and} \bibinfo{person}{Kristina
  Toutanova}.} \bibinfo{year}{2018}\natexlab{}.
\newblock \showarticletitle{Bert: Pre-training of deep bidirectional
  transformers for language understanding}.
\newblock \bibinfo{journal}{{\em arXiv preprint arXiv:1810.04805\/}}
  (\bibinfo{year}{2018}).
\newblock


\bibitem[\protect\citeauthoryear{Doar}{Doar}{1993}]%
        {Sharp}
\bibfield{author}{\bibinfo{person}{John Matthew~Simon Doar}.}
  \bibinfo{year}{1993}\natexlab{}.
\newblock \bibinfo{booktitle}{{\em Multicast in the asynchronous transfer mode
  environment}}.
\newblock \bibinfo{type}{{T}echnical {R}eport}.
  \bibinfo{institution}{University of Cambridge, Computer Laboratory}.
\newblock


\bibitem[\protect\citeauthoryear{Doar and Leslie}{Doar and Leslie}{1993}]%
        {doar1993multicast}
\bibfield{author}{\bibinfo{person}{M. Doar} {and} \bibinfo{person}{I. Leslie}.}
  \bibinfo{year}{1993}\natexlab{}.
\newblock \showarticletitle{How bad is naive multicast routing?}. In
  \bibinfo{booktitle}{{\em IEEE INFOCOM '93 The Conference on Computer
  Communications, Proceedings}}. \bibinfo{pages}{82--89 vol.1}.
\newblock
\showDOI{%
\url{https://doi.org/10.1109/INFCOM.1993.253246}}


\bibitem[\protect\citeauthoryear{Gurobi Algorithm used to solve continuous
  models}{Gurobi Algorithm used to solve continuous models}{2023}]%
        {method2}
Gurobi Algorithm used to solve continuous models
  \bibinfo{year}{2023}\natexlab{}.
\newblock   (\bibinfo{year}{2023}).
\newblock
\showURL{%
\url{https://www.gurobi.com/documentation/9.1/refman/method.html}}


\bibitem[\protect\citeauthoryear{Hart, Nilsson, and Raphael}{Hart
  et~al\mbox{.}}{1968}]%
        {Astar}
\bibfield{author}{\bibinfo{person}{Peter Hart}, \bibinfo{person}{Nils Nilsson},
  {and} \bibinfo{person}{Bertram Raphael}.} \bibinfo{year}{1968}\natexlab{}.
\newblock \showarticletitle{A Formal Basis for the Heuristic Determination of
  Minimum Cost Paths}.
\newblock \bibinfo{journal}{{\em {IEEE} Transactions on Systems Science and
  Cybernetics\/}} \bibinfo{volume}{4}, \bibinfo{number}{2}
  (\bibinfo{year}{1968}), \bibinfo{pages}{100--107}.
\newblock
\showDOI{%
\url{https://doi.org/10.1109/tssc.1968.300136}}


\bibitem[\protect\citeauthoryear{Hockney}{Hockney}{1994}]%
        {alphabeta}
\bibfield{author}{\bibinfo{person}{Roger~W Hockney}.}
  \bibinfo{year}{1994}\natexlab{}.
\newblock \showarticletitle{The communication challenge for MPP: Intel Paragon
  and Meiko CS-2}.
\newblock \bibinfo{journal}{{\em Parallel computing\/}} \bibinfo{volume}{20},
  \bibinfo{number}{3} (\bibinfo{year}{1994}), \bibinfo{pages}{389--398}.
\newblock


\bibitem[\protect\citeauthoryear{Hong, Kandula, Mahajan, Zhang, Gill, Nanduri,
  and Wattenhofer}{Hong et~al\mbox{.}}{2013}]%
        {SWAN}
\bibfield{author}{\bibinfo{person}{Chi-Yao Hong}, \bibinfo{person}{Srikanth
  Kandula}, \bibinfo{person}{Ratul Mahajan}, \bibinfo{person}{Ming Zhang},
  \bibinfo{person}{Vijay Gill}, \bibinfo{person}{Mohan Nanduri}, {and}
  \bibinfo{person}{Roger Wattenhofer}.} \bibinfo{year}{2013}\natexlab{}.
\newblock \showarticletitle{Achieving High Utilization with Software-Driven
  WAN}. In \bibinfo{booktitle}{{\em Proceedings of the ACM SIGCOMM 2013
  Conference on SIGCOMM}} {\em (\bibinfo{series}{SIGCOMM '13})}.
  \bibinfo{publisher}{Association for Computing Machinery},
  \bibinfo{address}{New York, NY, USA}, \bibinfo{pages}{15–26}.
\newblock
\showISBNx{9781450320566}
\showDOI{%
\url{https://doi.org/10.1145/2486001.2486012}}


\bibitem[\protect\citeauthoryear{Hougardy}{Hougardy}{2010}]%
        {FW}
\bibfield{author}{\bibinfo{person}{Stefan Hougardy}.}
  \bibinfo{year}{2010}\natexlab{}.
\newblock \showarticletitle{The Floyd--Warshall algorithm on graphs with
  negative cycles}.
\newblock \bibinfo{journal}{{\it Inform. Process. Lett.}}
  \bibinfo{volume}{110}, \bibinfo{number}{8-9} (\bibinfo{year}{2010}),
  \bibinfo{pages}{279--281}.
\newblock


\bibitem[\protect\citeauthoryear{Jain, Kumar, Mandal, Ong, Poutievski, Singh,
  Venkata, Wanderer, Zhou, Zhu, et~al\mbox{.}}{Jain et~al\mbox{.}}{2013}]%
        {B4}
\bibfield{author}{\bibinfo{person}{Sushant Jain}, \bibinfo{person}{Alok Kumar},
  \bibinfo{person}{Subhasree Mandal}, \bibinfo{person}{Joon Ong},
  \bibinfo{person}{Leon Poutievski}, \bibinfo{person}{Arjun Singh},
  \bibinfo{person}{Subbaiah Venkata}, \bibinfo{person}{Jim Wanderer},
  \bibinfo{person}{Junlan Zhou}, \bibinfo{person}{Min Zhu}, {et~al\mbox{.}}}
  \bibinfo{year}{2013}\natexlab{}.
\newblock \showarticletitle{B4: Experience with a globally-deployed software
  defined WAN}.
\newblock \bibinfo{journal}{{\em ACM SIGCOMM Computer Communication Review\/}}
  \bibinfo{volume}{43}, \bibinfo{number}{4} (\bibinfo{year}{2013}),
  \bibinfo{pages}{3--14}.
\newblock


\bibitem[\protect\citeauthoryear{Kandula, Menache, Schwartz, and
  Babbula}{Kandula et~al\mbox{.}}{2014}]%
        {kandula2014calendaring}
\bibfield{author}{\bibinfo{person}{Srikanth Kandula}, \bibinfo{person}{Ishai
  Menache}, \bibinfo{person}{Roy Schwartz}, {and} \bibinfo{person}{Spandana~Raj
  Babbula}.} \bibinfo{year}{2014}\natexlab{}.
\newblock \showarticletitle{Calendaring for Wide Area Networks}. In
  \bibinfo{booktitle}{{\em Proceedings of the 2014 ACM Conference on SIGCOMM}}
  {\em (\bibinfo{series}{SIGCOMM '14})}. \bibinfo{publisher}{Association for
  Computing Machinery}, \bibinfo{address}{New York, NY, USA},
  \bibinfo{pages}{515–526}.
\newblock
\showISBNx{9781450328364}
\showDOI{%
\url{https://doi.org/10.1145/2619239.2626336}}


\bibitem[\protect\citeauthoryear{Laoutaris, Sirivianos, Yang, and
  Rodriguez}{Laoutaris et~al\mbox{.}}{2011}]%
        {netstitcher}
\bibfield{author}{\bibinfo{person}{Nikolaos Laoutaris},
  \bibinfo{person}{Michael Sirivianos}, \bibinfo{person}{Xiaoyuan Yang}, {and}
  \bibinfo{person}{Pablo Rodriguez}.} \bibinfo{year}{2011}\natexlab{}.
\newblock \showarticletitle{Inter-Datacenter Bulk Transfers with Netstitcher}.
  In \bibinfo{booktitle}{{\em Proceedings of the ACM SIGCOMM 2011 Conference}}
  {\em (\bibinfo{series}{SIGCOMM '11})}. \bibinfo{publisher}{Association for
  Computing Machinery}, \bibinfo{address}{New York, NY, USA},
  \bibinfo{pages}{74–85}.
\newblock
\showISBNx{9781450307970}
\showDOI{%
\url{https://doi.org/10.1145/2018436.2018446}}


\bibitem[\protect\citeauthoryear{Mahajan, Chu, Sridharan, and Akella}{Mahajan
  et~al\mbox{.}}{[n. d.]}]%
        {mahajanbetter}
\bibfield{author}{\bibinfo{person}{Kshiteej Mahajan},
  \bibinfo{person}{Ching-Hsiang Chu}, \bibinfo{person}{Srinivas Sridharan},
  {and} \bibinfo{person}{Aditya Akella}.} \bibinfo{year}{[n. d.]}\natexlab{}.
\newblock \showarticletitle{Better Together: Jointly Optimizing ML Collective
  Scheduling and Execution Planning using SYNDICATE}.
\newblock  (\bibinfo{year}{[n. d.]}).
\newblock


\bibitem[\protect\citeauthoryear{MSCCL codebase}{MSCCL codebase}{[n. d.]}]%
        {msccl}
MSCCL codebase \bibinfo{year}{[n. d.]}\natexlab{}.
\newblock   (\bibinfo{year}{[n. d.]}).
\newblock
\showURL{%
\url{https://github.com/microsoft/msccl}}


\bibitem[\protect\citeauthoryear{Narayanan, Kazhamiaka, Abuzaid, Kraft,
  Agrawal, Kandula, Boyd, and Zaharia}{Narayanan et~al\mbox{.}}{2021}]%
        {PoP}
\bibfield{author}{\bibinfo{person}{Deepak Narayanan}, \bibinfo{person}{Fiodar
  Kazhamiaka}, \bibinfo{person}{Firas Abuzaid}, \bibinfo{person}{Peter Kraft},
  \bibinfo{person}{Akshay Agrawal}, \bibinfo{person}{Srikanth Kandula},
  \bibinfo{person}{Stephen Boyd}, {and} \bibinfo{person}{Matei Zaharia}.}
  \bibinfo{year}{2021}\natexlab{}.
\newblock \showarticletitle{Solving Large-Scale Granular Resource Allocation
  Problems Efficiently with POP}. In \bibinfo{booktitle}{{\em Proceedings of
  the ACM SIGOPS 28th Symposium on Operating Systems Principles}} {\em
  (\bibinfo{series}{SOSP '21})}. \bibinfo{publisher}{Association for Computing
  Machinery}, \bibinfo{address}{New York, NY, USA}, \bibinfo{pages}{521–537}.
\newblock
\showISBNx{9781450387095}
\showDOI{%
\url{https://doi.org/10.1145/3477132.3483588}}


\bibitem[\protect\citeauthoryear{Noronha and Tobagi}{Noronha and
  Tobagi}{1994}]%
        {noronha1994optimum}
\bibfield{author}{\bibinfo{person}{C.A. Noronha} {and} \bibinfo{person}{F.A.
  Tobagi}.} \bibinfo{year}{1994}\natexlab{}.
\newblock \showarticletitle{Optimum routing of multicast streams}. In
  \bibinfo{booktitle}{{\em Proceedings of INFOCOM '94 Conference on Computer
  Communications}}. \bibinfo{pages}{865--873 vol.2}.
\newblock
\showDOI{%
\url{https://doi.org/10.1109/INFCOM.1994.337651}}


\bibitem[\protect\citeauthoryear{Nvidia DGX System}{Nvidia DGX System}{2021}]%
        {dgx2}
Nvidia DGX System \bibinfo{year}{2021}\natexlab{}.
\newblock   (\bibinfo{year}{2021}).
\newblock
\showURL{%
\url{https://www.nvidia.com/en-us/data-center/dgx-systems/}}


\bibitem[\protect\citeauthoryear{Oliveira and Pardalos}{Oliveira and
  Pardalos}{2005}]%
        {oliveira2005survey}
\bibfield{author}{\bibinfo{person}{Carlos~AS Oliveira} {and}
  \bibinfo{person}{Panos~M Pardalos}.} \bibinfo{year}{2005}\natexlab{}.
\newblock \showarticletitle{A survey of combinatorial optimization problems in
  multicast routing}.
\newblock \bibinfo{journal}{{\em Computers \& Operations Research\/}}
  \bibinfo{volume}{32}, \bibinfo{number}{8} (\bibinfo{year}{2005}),
  \bibinfo{pages}{1953--1981}.
\newblock


\bibitem[\protect\citeauthoryear{Pedroso}{Pedroso}{2011}]%
        {gurobi}
\bibfield{author}{\bibinfo{person}{Joo~Pedro Pedroso}.}
  \bibinfo{year}{2011}\natexlab{}.
\newblock \showarticletitle{Optimization with gurobi and python}.
\newblock \bibinfo{journal}{{\em INESC Porto and Universidade do Porto,, Porto,
  Portugal\/}}  \bibinfo{volume}{1} (\bibinfo{year}{2011}).
\newblock


\bibitem[\protect\citeauthoryear{Rashidi, Won, Srinivasan, Sridharan, and
  Krishna}{Rashidi et~al\mbox{.}}{2022}]%
        {rashidi2022themis}
\bibfield{author}{\bibinfo{person}{Saeed Rashidi}, \bibinfo{person}{William
  Won}, \bibinfo{person}{Sudarshan Srinivasan}, \bibinfo{person}{Srinivas
  Sridharan}, {and} \bibinfo{person}{Tushar Krishna}.}
  \bibinfo{year}{2022}\natexlab{}.
\newblock \showarticletitle{Themis: A Network Bandwidth-Aware Collective
  Scheduling Policy for Distributed Training of DL Models}. In
  \bibinfo{booktitle}{{\em Proceedings of the 49th Annual International
  Symposium on Computer Architecture}} {\em (\bibinfo{series}{ISCA '22})}.
  \bibinfo{publisher}{Association for Computing Machinery},
  \bibinfo{address}{New York, NY, USA}, \bibinfo{pages}{581–596}.
\newblock
\showISBNx{9781450386104}
\showDOI{%
\url{https://doi.org/10.1145/3470496.3527382}}


\bibitem[\protect\citeauthoryear{Shah, Chidambaram, Cowan, Maleki, Musuvathi,
  Mytkowicz, Nelson, Saarikivi, and Singh}{Shah et~al\mbox{.}}{2021}]%
        {TACCL}
\bibfield{author}{\bibinfo{person}{Aashaka Shah}, \bibinfo{person}{Vijay
  Chidambaram}, \bibinfo{person}{Meghan Cowan}, \bibinfo{person}{Saeed Maleki},
  \bibinfo{person}{Madan Musuvathi}, \bibinfo{person}{Todd Mytkowicz},
  \bibinfo{person}{Jacob Nelson}, \bibinfo{person}{Olli Saarikivi}, {and}
  \bibinfo{person}{Rachee Singh}.} \bibinfo{year}{2021}\natexlab{}.
\newblock \bibinfo{title}{TACCL: Guiding Collective Algorithm Synthesis using
  Communication Sketches}.
\newblock   (\bibinfo{year}{2021}).
\newblock
\showDOI{%
\url{https://doi.org/10.48550/ARXIV.2111.04867}}


\bibitem[\protect\citeauthoryear{Tang, Kakarla, Beckett, Zhai, Brown,
  Millstein, Tamir, and Varghese}{Tang et~al\mbox{.}}{2021}]%
        {campion}
\bibfield{author}{\bibinfo{person}{Alan Tang}, \bibinfo{person}{Siva
  Kesava~Reddy Kakarla}, \bibinfo{person}{Ryan Beckett}, \bibinfo{person}{Ennan
  Zhai}, \bibinfo{person}{Matt Brown}, \bibinfo{person}{Todd Millstein},
  \bibinfo{person}{Yuval Tamir}, {and} \bibinfo{person}{George Varghese}.}
  \bibinfo{year}{2021}\natexlab{}.
\newblock \showarticletitle{Campion: Debugging Router Configuration
  Differences}. In \bibinfo{booktitle}{{\em Proceedings of the 2021 ACM SIGCOMM
  2021 Conference}} {\em (\bibinfo{series}{SIGCOMM '21})}.
  \bibinfo{publisher}{Association for Computing Machinery},
  \bibinfo{address}{New York, NY, USA}, \bibinfo{pages}{748–761}.
\newblock
\showISBNx{9781450383837}
\showDOI{%
\url{https://doi.org/10.1145/3452296.3472925}}


\bibitem[\protect\citeauthoryear{Wang, Venkataraman, Phanishayee, Devanur,
  Thelin, and Stoica}{Wang et~al\mbox{.}}{2020}]%
        {Blink}
\bibfield{author}{\bibinfo{person}{Guanhua Wang}, \bibinfo{person}{Shivaram
  Venkataraman}, \bibinfo{person}{Amar Phanishayee}, \bibinfo{person}{Nikhil
  Devanur}, \bibinfo{person}{Jorgen Thelin}, {and} \bibinfo{person}{Ion
  Stoica}.} \bibinfo{year}{2020}\natexlab{}.
\newblock \showarticletitle{Blink: Fast and generic collectives for distributed
  ml}.
\newblock \bibinfo{journal}{{\em Proceedings of Machine Learning and
  Systems\/}}  \bibinfo{volume}{2} (\bibinfo{year}{2020}),
  \bibinfo{pages}{172--186}.
\newblock


\bibitem[\protect\citeauthoryear{Wang, Khazraee, Zhong, Ghobadi, Jia, Mudigere,
  Zhang, and Kewitsch}{Wang et~al\mbox{.}}{2022}]%
        {wang23topoopt}
\bibfield{author}{\bibinfo{person}{Weiyang Wang}, \bibinfo{person}{Moein
  Khazraee}, \bibinfo{person}{Zhizhen Zhong}, \bibinfo{person}{Manya Ghobadi},
  \bibinfo{person}{Zhihao Jia}, \bibinfo{person}{Dheevatsa Mudigere},
  \bibinfo{person}{Ying Zhang}, {and} \bibinfo{person}{Anthony Kewitsch}.}
  \bibinfo{year}{2022}\natexlab{}.
\newblock \bibinfo{title}{TopoOpt: Co-optimizing Network Topology and
  Parallelization Strategy for Distributed Training Jobs}.
\newblock   (\bibinfo{year}{2022}).
\newblock
\showDOI{%
\url{https://doi.org/10.48550/ARXIV.2202.00433}}


\bibitem[\protect\citeauthoryear{Zhao, Pal, Chugh, Wang, Basu, Khoury, and
  Krishnamurthy}{Zhao et~al\mbox{.}}{2022}]%
        {zhao2022optimal}
\bibfield{author}{\bibinfo{person}{Liangyu Zhao}, \bibinfo{person}{Siddharth
  Pal}, \bibinfo{person}{Tapan Chugh}, \bibinfo{person}{Weiyang Wang},
  \bibinfo{person}{Prithwish Basu}, \bibinfo{person}{Joud Khoury}, {and}
  \bibinfo{person}{Arvind Krishnamurthy}.} \bibinfo{year}{2022}\natexlab{}.
\newblock \bibinfo{title}{Optimal Direct-Connect Topologies for Collective
  Communications}.
\newblock   (\bibinfo{year}{2022}).
\newblock
\showDOI{%
\url{https://doi.org/10.48550/ARXIV.2202.03356}}


\end{thebibliography}
